%% file: lmcs_comp.tex
\documentclass{LMCS}

\usepackage{amsmath}
\usepackage{amsfonts}
\usepackage{amssymb}
\usepackage{color}
\usepackage{epsfig}
\usepackage{enumerate,hyperref}

\newtheorem{theorem}{Theorem}[section]
\newtheorem{xdefinition}[theorem]{Definition}

\newtheorem{xexample}[theorem]{Example}
\newenvironment{example}{\begin{xexample}}{\ebox\end{xexample}}

\newtheorem{corollary}[theorem]{Corollary}

\newtheorem{claim}[theorem]{Claim}

\def\dom{{\rm dom}}
\def\M{\mathcal{M}}
\def\inst{{\rm Inst}}
\def\bs{{\mathbf S}}
\def\bt{{\mathbf T}}

\def\FO{\ensuremath{\mathrm{FO}}}

\def\ebox{{\hfill$\Box$}}
\newcommand{\aproof}[2]{\noindent {\it Proof of #1.} #2\qed}

\newcommand{\f}{\bar{f}}

\def\a{\bar a}
\def\b{\bar b}

\def\e{\bar e}

\def\x{\bar x}
\def\y{\bar y}

\def\FF{\mathfrak{F}}
\def\CUS{\FF_{\rm univ}}

\def\G{\mathcal{G}}

\newcommand{\E}{\mathcal{E}}

\newcommand{\sk}{\text{Sk}\forall\text{CQ}^=}
\newcommand{\nat}{\text{non-atomic}}
\newcommand{\lv}{\text{list-var}}

\newcommand{\sig}{\text{skel}}
\newcommand{\C}{{\bf C}}
\newcommand{\N}{{\bf N}}

\newcommand{\certain}[3]{\mbox{{\it certain}}_{#1}(#2,#3)}

\newcommand{\R}{{\bf R}}
\newcommand{\D}{{\bf D}}

\def\doi{7 (3:13) 2011}
\lmcsheading%
{\doi}
{1--38}
{}
{}
{Nov.~\phantom08, 2010}
{Sep.~\phantom08, 2011}
{}

\begin{document}

\title[Composition with Target Constraints]{Composition with Target
Constraints\rsuper*}

\author[M.~Arenas]{Marcelo Arenas\rsuper a}
\address{{\lsuper a}Pontificia Universidad Cat\'olica de Chile}
\email{marenas@ing.puc.cl}
\thanks{{\lsuper{a,c}}Most of the work on this paper was done while Alan
Nash was at IBM Research--Almaden and Marcelo Arenas was a visitor at
IBM Research--Almaden.}

\author[R.~Fagin]{Ronald Fagin\rsuper b}
\address{{\lsuper b}IBM Research--Almaden}
\email{fagin@almaden.ibm.com}

\author[A.~Nash]{Alan Nash\rsuper c}
\address{{\lsuper c}Aleph One LLC}

\keywords{Metadata management, schema mapping, data exchange,
composition, target constraint}
\subjclass{H.2.5}
\titlecomment{{\lsuper*}This is an extended version of
\cite{AFN10}. The first two authors dedicate this paper to the memory
of our co-author Alan Nash, who died tragically, before this final
version of the paper was prepared.}

\begin{abstract}
It is known that the composition of schema mappings, each specified by
source-to-target tgds (st-tgds), can be specified by a second-order
tgd (SO tgd). We consider the question of what happens when target
constraints are allowed.  Specifically, we consider the question of
specifying the composition of {\em standard schema mappings\/} (those
specified by st-tgds, target egds, and a weakly acyclic set of target
tgds).  We show that SO tgds, even with the assistance of arbitrary
source constraints and target constraints, cannot specify in general
the composition of two standard schema mappings.  Therefore, we
introduce source-to-target second-order dependencies (st-SO
dependencies), which are similar to SO tgds, but allow equations in
the conclusion.  We show that st-SO dependencies (along with target
egds and target tgds) are sufficient to express the composition of
every finite sequence of standard schema mappings, and further, every
st-SO dependency specifies such a composition.  In addition to this
expressive power, we show that st-SO dependencies enjoy other
desirable properties.  In particular, they have a polynomial-time
chase that generates a universal solution. This universal solution can
be used to find the certain answers to unions of conjunctive queries
in polynomial time.

It is easy to show that the composition of an arbitrary number of
standard schema mappings is equivalent to the composition of only two
standard schema mappings.  We show that surprisingly, the analogous
result holds also for schema mappings specified by just st-tgds (no
target constraints). That is, the composition of an arbitrary number
of such schema mappings is equivalent to the composition of only two
such schema mappings.  This is proven by showing that every SO tgd is
equivalent to an unnested SO tgd (one where there is no nesting of
function symbols).  The language of unnested SO tgds is quite natural,
and we show that unnested SO tgds are capable of specifying the
composition of an arbitrary number of schema mappings, each specified
by st-tgds.  Similarly, we prove unnesting results for st-SO
dependencies, with the same types of consequences.

\end{abstract}

\maketitle

\section{Introduction}
\label{sec-intro}

Schema mappings are high-level specifications that describe the
relationship between two database schemas, a {\em source schema\/} and a
{\em target schema}.
Because of the crucial importance of schema mappings for
data integration and data exchange
(see the surveys
\cite{Ko05,Len02}), several different operators on schema
mappings have been singled out as important objects of study
\cite{Bernstein03}.
One of the most fundamental is the composition operator,
which combines successive
schema mappings into a single schema mapping.
The composition operator can play a useful role each time the target
of a schema mapping is also  the source of another schema mapping.
This scenario occurs, for instance, in schema evolution, where a
schema
may undergo several successive changes.  It also occurs in
extract-transform-load (ETL) processes in which the output of a transformation
may be the input to another \cite{VSS02}.
The composition operator has been
studied in depth
\cite{FKPT05,MH03,Me04,NBM05}.

One of the most basic questions is: what is the language needed
to express the composition of schema mappings?
For example, if the schema mapping $\M_{12}$ is
an {\em st-tgd mapping}, that is, a mapping
specified by a finite
set
of the widely-studied {\em source-to-target
tuple-generating dependencies (st-tgds)},
and the schema mapping $\M_{23}$ is also an st-tgd mapping,
is the composition $\M_{12} \circ \M_{23}$ also an st-tgd mapping?
Fagin et al.\
\cite{FKPT05} showed that
surprisingly, the answer is ``No.''
In fact, they showed that it is necessary to pass to existential
second-order logic to express this composition in general.
Specifically, they defined a class of dependencies,
which they call {\em second-order tgds (SO tgds)},
which are source-to-target,
with existentially-quantified function symbols,
and they showed that this is the
``language of composition''.
That is, they showed that the composition
of any number of st-tgd mappings
can be specified by an SO tgd.
They also showed that
every SO tgd specifies the composition of a finite number of
st-tgd mappings.
Thus, SO tgds are exactly the right language.

What happens if we allow not only source-to-target constraints,
but also target constraints?
Target constraints are important in practice;
examples of important target constraints are those that
specify the keys of target relations,
and referential integrity constraints (or inclusion dependencies
\cite{CFP84}).
This paper is motivated by the
question of how to express the compositions of schema mappings
that have target constraints.
This question was first explored by Nash et al.\ \cite{NBM05}, where an
even more general
class of constraints was studied: constraints expressed over the joint
source and target schemas without any restrictions.
Here we study a case intermediate between that studied by Fagin et al.
in \cite{FKPT05} and
that studied by Nash et al.\ in \cite{NBM05}.
Specifically, we study
{\em standard schema mappings}, where
the source-to-target constraints are st-tgds, and the target constraints
consist of
target equality-generating dependencies (t-egds) and a weakly acyclic
set \cite{FKMP05} of target tuple-generating dependencies (t-tgds).
Standard schema mappings have a chase that is guaranteed to
terminate in polynomial time.
In fact, weak acyclicity was introduced in
\cite{FKMP05} in order to provide a fairly general sufficient condition
for the chase to terminate in polynomial time (a slightly less
general class was introduced in \cite{DT03}, under the name
{\em constraints with stratified witness},
for the same purpose).

Standard schema mappings are a natural ``sweet spot'' between the
schema mappings studied by Fagin et al.\ \cite{FKMP05}
(with only source-to-target constraints)
and the schema mappings studied
by Nash et al.\  \cite{NBM05}  (with general constraints),
for two reasons.
The first reason is the
importance of standard schema mappings.
Source-to-target tgds are the natural and common backbone language of
data exchange systems \cite{FHHMPV09}.
Furthermore, even though the notion of
weakly acyclic sets of tgds was introduced only recently, it has now
been studied extensively
\cite{AK08,AK09,ALP08a,ALP08b,APR08,Ba09,CDO09,CG09,DNR08,DPT06,FKNP08,%
FKMP05,FKMT06,GLLR07,GN08,GS08,GKIT07,HS07,KT08,Ko05,Me08,PT09,tK09}.
Among the important special cases of weakly acyclic sets of tgds
are sets of full tgds
(those with no existential quantifiers) and
acyclic sets of inclusion dependencies \cite{CK86}, a large class
that is common in practice.
The second reason for our interest in standard schema mappings is that
as we shall see, compositions of standard
schema mappings have especially nice properties.
Thus, the language of standard schema mappings is expressive enough to
be useful in practice, and yet simple enough to allow nice properties,
such as
having a polynomial-time chase.

There are various inexpressibility results in \cite{FKPT05}
and~\cite{NBM05} that show the inability of first-order logic to express
compositions.
Thus, each of these results says that there is a pair
of schema mappings that
are each specified by simple formulas in first-order logic, but where
the composition
cannot be expressed in first-order logic.
In this paper, we show
that some compositions
cannot be expressed even in
certain fragments of
second-order logic.
First, we show
that SO tgds are not adequate to
express
the composition of an arbitrary pair of standard schema mappings.
It turns out that this is quite easy to show.
But what if we allow not only SO tgds, but also arbitrary
source constraints and target constraints?
This is a more delicate problem.
By making use of a notion of locality from \cite{ABFL04}, we show that
even these are not adequate to
express
the composition of an arbitrary pair of standard schema mappings.

Therefore, we introduce a richer class of dependencies, which we call
{\em source-to-target second-order
dependencies (st-SO dependencies)}.
This class of dependencies is the source-to-target restriction of the
class $\sk$ of dependencies introduced in \cite{NBM05}.
Our st-SO dependencies differ from SO tgds in that st-SO dependencies
may have not only relational atomic formulas $R(t_1, \ldots, t_n)$
in the conclusions, but also
equalities $t_1 = t_2$.
We show that st-SO dependencies are
exactly the right extension of SO tgds for the purpose
of expressing the composition of standard
schema mappings.
Specifically, we show that (1) the composition of standard schema
mappings
can be expressed by an st-SO dependency
(along with target constraints), and (2) every st-SO dependency
specifies the composition of some finite sequence of standard schema
mappings.
We note that a result analogous to~(1), but for schema
mappings that are not necessarily source-to-target, was obtained in
\cite{NBM05} by using their class $\sk$ of dependencies.
In fact, our proof of~(1) is simply a variation of the proof in~\cite{NBM05}.

In addition,
we
show that st-SO dependencies enjoy other desirable properties.
In particular, we show that they have a polynomial-time chase
procedure.
This chase procedure is novel, in that it has to keep track of
constantly changing values of functions.
As usual, the chase generates not just a solution, but a
{\em universal solution} \cite{FKMP05}.
(Recall that a {\em solution} for a source instance $I$ with respect to
a schema mapping $\M$ is a target instance $J$ where the pair $(I,J)$
satisfies
the constraints of $\M$, and a universal solution is a solution with a
homomorphism to every solution.)
The fact that the chase is guaranteed to terminate
(whether in polynomial time
or otherwise) implies that
if there is a solution for a given source instance $I$, then there is a
universal solution.
The fact that the chase runs in polynomial time guarantees that there is
a polynomial-time algorithm for deciding if there is a solution, and, if
so, for producing a universal solution.

Let $q$ be a query posed against the target schema.
The {\em certain answers\/} for $q$ on a source instance $I$,
with respect to a schema mapping $\M$, are those tuples that appear in
the answer $q(J)$ for every solution $J$ for $I$.
It is shown in
\cite{FKMP05} that if $q$ is a union of conjunctive queries, and
$J^*$ is a universal solution for $I$, then the certain answers for $q$
on
$I$ can be obtained by evaluating
$q$ on $J^*$ and then keeping only those tuples formed entirely of
values from $I$.
Since the chase using an st-SO dependency can be carried out in
polynomial time, it
follows that we can obtain a universal solution in polynomial time, and
so we can compute the certain answers to unions of conjunctive queries
in polynomial time.

In addition to our results about st-SO dependencies, we also have some
results directly about compositions of schema mappings.
It is easy to show that the composition of an arbitrary number
of standard schema mappings is equivalent to the composition of only two
standard schema mappings. We show the surprising result that a similar result holds also
for st-tgd mappings (no target
constraints). That is, the composition of an arbitrary number of st-tgd
mappings is equivalent to the composition of only two st-tgd mappings.
This is proven by showing that every SO tgd is equivalent to an unnested SO tgd
(one where there is no nesting of function symbols).  We also prove a similar denesting result for
st-SO dependencies.
These denesting results are the most difficult results technically in
the paper.

We feel that unnested dependencies are more natural, more readable, and
easier to understand than nested dependencies.
They are probably easier to use in practice.
For example,
it is easy to see that the ``nested mappings'' in \cite{FHHMPP06}
can be expressed by unnested SO tgds.
We show that
unnested SO tgds are also
expressive enough to specify the composition of an arbitrary number of
st-tgd mappings.
This was not
known even
for the composition of two st-tgd mappings.
Thus, although it was shown in \cite{FKPT05} that each unnested SO tgd
specifies the composition of
a
pair of st-tgd mappings, the converse
was not shown.
In fact, for the composition of two st-tgd mappings, the composition
construction in
\cite{FKPT05} can produce an SO tgd with nesting depth 2, not 1.

We close by discussing an application of our results.
In practice, a composition of many schema mappings may arise (say, as
the result of many steps of schema evolution).
If these are st-tgd  mappings, then
there are
several approaches towards ``simplifying''
this composition.
One approach
is to replace
the composition of many st-tgd mappings
by a single schema mapping, specified by an unnested SO tgd.
For another approach,
we can remain within the language of st-tgds by
replacing the composition of
many st-tgd mappings by the
composition of only two st-tgd mappings.
A similar comment applies to the composition of many standard schema
mappings.



\section{Preliminaries}
\label{sec-prelim}

A {\em schema} ${\bf R}$ is a finite set $\{ R_1, \ldots, R_k \}$ of
relation symbols, with each $R_i$ having a fixed arity $n_i > 0$. Let
$\D$ be a countably infinite domain. An {\em instance} $I$ of $\R$
assigns to each relation symbol $R_i$ of $\R$ a finite $n_i$-ary
relation $R_i^I \subseteq \D^{n_i}$.
We let $\inst(\R)$ be the set of instances of $\R$.
The {\em domain} (or
{\em active domain\/})
$\dom(I)$ of
instance $I$ is the set of all elements that occur in any of the
relations $R_i^I$.
We say that $R(a_1, \ldots, a_n)$ is a {\em fact\/} of $I$ if
$(a_1, \ldots, a_n) \in R^I$.
We sometimes denote an instance by its set of facts.

As is customary in the data exchange literature, we consider instances
with two types of values: constants and nulls \cite{FKMP05}. More
precisely, let $\C$ and $\N$ be infinite and disjoint sets of
constants and nulls, respectively, and
take the domain $\D$ to be $\C \cup \N$.
If we refer to a schema $\bs$ as a {\em source} schema, then we
assume that for every instance $I$ of $\bs$, it holds that $\dom(I)
\subseteq \C$. On the other hand, if we refer to a schema $\bt$ as a
{\em target} schema, then for every instance $J$ of $\bs$, it holds
that $\dom(J) \subseteq \C \cup \N$.
The distinction between constants and nulls is important in the
definition of a homomorphism
(which we give later).


\subsection{Source-to-target and target dependencies}
Fix a source schema $\bs$ and a target schema $\bt$, and assume that
$\bs$ and $\bt$ do not have predicate symbols in common. Then a
{\em source-to-target tuple-generating dependency (st-tgd)\/} is a
first-order sentence of the form:
\begin{eqnarray*}
\forall \bar x \, (\varphi(\bar x) \rightarrow \exists \bar y \,
\psi(\bar x,\bar y)),
\end{eqnarray*}
where $\varphi(\bar x)$ is a conjunction of relational atoms over $\bs$
and $\psi(\bar x,\bar y)$ is a conjunction of relational atoms over
$\bt$. We assume a safety condition, that every member of $\bar x$
actually appears in a relational atom in $\varphi(\bar x)$.
A {\em target equality-generating dependency (t-egd)\/} is a
first-order sentence of the form:
\begin{eqnarray*}
\forall \bar x \, (\varphi(\bar x) \rightarrow u = v),
\end{eqnarray*}
where $\varphi(\bar x)$ is a conjunction of relational atoms over
$\bt$ and $u$, $v$ are among the variables mentioned in $\bar
x$.
We again assume the
same
safety condition.
In several of the examples we give in this paper, we shall make use of
special t-egds called {\em key dependencies}, which say that one
attribute of a binary relation is a key for that relation (of course, we
could define more general key dependencies if we wanted).
The key dependencies we consider are either of the form $R(x,y) \land
R(x,z) \to y=z$ (which says that the first attribute is a key) or
$S(y,x) \land S(z,x) \to y = z$ (which says that the second attribute
is a key).
Finally, a {\em target tuple-generating dependency (t-tgd)\/} is a
first-order sentence of the form:
\begin{eqnarray*}
\forall \bar x \, (\varphi(\bar x) \rightarrow \exists \bar y \,
\psi(\bar x,\bar y)),
\end{eqnarray*}
where both $\varphi(\bar x)$ and $\psi(\bar x,\bar y)$ are conjunctions
of relational atoms over $\bt$, and
where we again assume the
same
safety condition.

The notion of satisfaction of a t-egd
$\alpha$ by a target instance $J$, denoted by $J \models \alpha$, is
defined as the standard notion of satisfaction in first-order logic,
and likewise for t-tgds. For the case of an st-tgd $\alpha$, a
source instance $I$ and a target instance $J$, the pair $(I,J)$ is
said to
satisfy
$\alpha$, denoted by $(I,J) \models \alpha$, if the
following instance $K$ of $\bs \cup \bt$ satisfies $\alpha$
in the standard first-order logic sense. For every relation symbol $S \in
\bs$, relation $S^K$ is defined as $S^I$, and for every relation symbol
$T \in \bt$, relation $T^K$ is defined as $T^J$. As usual, a set
$\Sigma_{st}$ of st-tgds is said to be satisfied by a pair $(I,J)$,
denoted by $(I,J) \models \Sigma_{st}$, if
$(I,J) \models \alpha$ for every $\alpha \in \Sigma_{st}$ (and
likewise for a set of t-egds and t-tgds).

\subsection{Schema mappings}
In general, a {\em schema mapping} from a source schema $\bs$ to a
target schema $\bt$ is a set of pairs $(I,J)$, where $I$ is an
instance of $\bs$ and $J$ is an instance of $\bt$. In this paper, we
restrict our attention to some classes of schema mappings that are
{\em specified} in some logical formalisms.
We may sometimes refer to two schema mappings with the same set of
$(I,J)$ pairs as {\em equivalent}, to capture the idea that the
formulas that specify them are logically equivalent.
A schema mapping
$\M$
from $\bs$ to $\bt$ is said to be an {\em st-tgd mapping}
if there exists a
finite
set $\Sigma_{st}$ of st-tgds such that $(I,J)$
belongs to $\M$ if and only if $(I,J) \models \Sigma_{st}$, for every
pair $I,J$ of instances of $\bs$ and $\bt$, respectively. We use
notation $\M = (\bs, \bt, \Sigma_{st})$ to indicate that $\M$ is
specified by $\Sigma_{st}$. Moreover, a schema mapping $\M$ from $\bs$
to $\bt$ is said to be a {\em standard schema mapping} if there exists
a
finite
set $\Sigma_{st}$ of st-tgds and a
finite
set $\Sigma_t$ consisting of a
set of t-egds and a {\em weakly acyclic} set of t-tgds, such that
$(I,J)$ belongs to $\M$ if and only if $(I,J) \models \Sigma_{st}$ and
$J \models \Sigma_t$, for every pair $I,J$ of instances of $\bs$ and $\bt$,
respectively;
notation $\M = (\bs, \bt, \Sigma_{st}, \Sigma_t)$ is used in this case
to indicate that $\M$ is specified by $\Sigma_{st}$
and $\Sigma_t$.
We occasionally allow a
finite
set $\Sigma_s$ of source constraints in some of
our schema mappings:
we then use the notation $\M = (\bs, \bt, \Sigma_s,\Sigma_{st},
\Sigma_t)$.

To define the widely used notion of weak acyclicity, we
need to introduce some terminology. For a set $\Gamma$ of t-tgds over
$\bt$, define the dependency graph $G_\Gamma$ of $\Gamma$ as follows.
\begin{iteMize}{$\bullet$}\itemsep=2pt
\item For every relation name $T$ in $\bt$ of arity $n$, and for
every $i \in \{1, \ldots, n\}$, include a node $(T, i)$ in $G_\Gamma$.

\item Include an edge $(T_1, i) \to (T_2, j)$ in $G_\Gamma$ if there exist
a t-tgd $\forall \bar x (\varphi(\bar x) \to \exists \bar y \, \psi(\bar
x,\bar y))$ in $\Gamma$ and a variable $x$ in $\bar
x$ such that, $x$ occurs in the $i$-th attribute of $T_1$ in a
conjunct of $\varphi$ and in the $j$-th attribute of $T_2$ in
a conjunct of $\psi$.

\item Include a {\em special} edge $(T_1, i) \to^* (T_2, j)$ in
$G_\Gamma$ if there exist a t-tgd $\forall \bar x (\varphi(\bar x)
\to \exists \bar y \, \psi(\bar x, \bar y))$ in $\Gamma$ and
variables $x$, $y$ in $\bar x$ and $\bar y$, respectively, such that
$x$ occurs in the $i$-th attribute of $T_1$ in a conjunct of
$\varphi$ and $y$ occurs in the $j$-th attribute of $T_2$ in a
conjunct of $\psi$.
\end{iteMize}
Then set $\Gamma$ of t-tgds is said to be {\em weakly acyclic} if its
dependency graph $G_\Gamma$ has no cycle through a special edge
\cite{FKMP05}.

Given a schema mapping $\M$, if a pair $(I,J)$ belongs to it, then $J$
is said to be a {\em solution} for $I$
with respect to $\M$.
A {\em universal solution\/} \cite{FKMP05} for $I$ is a solution with a
homomorphism to every solution for $I$.
A {\em homomorphism\/} from instance $J_1$ to instance $J_2$
is
a function $h$ from $\C \cup \N$ to $\C \cup \N$ such that
(1) for each $c$ in $\C$, we have that $h(c) =
c$, and
(2) whenever $R(a_1, \ldots, a_n)$ is a fact of $J_1$, then
$R(h(a_1),\ldots,h(a_n))$ is a fact of $J_2$.

\subsection{Second-order dependencies}
In this paper, we also consider schema mappings that are specified by
second-order dependencies. In the definition of these dependencies,
the following terminology is used. Given a collection $\x$ of
variables and a collection $\f$ of function symbols,
a {\em term (based on ${\bar x}$ and ${\bar f}$) with depth of nesting
$d$}
is defined recursively as follows:
\begin{enumerate}[(1)]\itemsep=1pt
\item Every member of ${\bar x}$
and every 0-ary function symbol
(constant symbol) of ${\bar f}$
is a term with depth of nesting  0.
\item If $f$ is a $k$-ary function symbol in ${\bar f}$
with $k \geq 1$,
and if $t_1, \ldots, t_k$ are
terms, with maximum depth of nesting $d-1$, then $f(t_1, \ldots, t_k)$
is a term with depth of nesting $d$.
\end{enumerate}
\begin{defi}{\bf (\cite{FKPT05})}\label{def:sotgd}
Given a source schema $\bs$ and a target schema
$\bt$, a {\em second-order source-to-target tuple-generating dependency
or SO tgd (from $\bs$ to $\bt$)\/} is a second-order formula of the form
\begin{eqnarray*}
\exists \f \, (\forall \x_1 (\varphi_1 \rightarrow \psi_1) \wedge
\cdots \wedge \forall \x_n (\varphi_n \rightarrow \psi_n)),
\end{eqnarray*}
where
\begin{enumerate}[(1)]\itemsep=1pt
\item Each member of $\f$ is a function symbol.

\item Each $\varphi_i$ is a conjunction of
\begin{iteMize}{$\bullet$}\itemsep=1pt
\item relational atomic formulas of the form $S(y_1, \ldots, y_k)$,
where $S$ is a $k$-ary relation symbol of $\bs$ and $y_1$, $\ldots$,
$y_k$ are (not necessarily distinct) variables in $\x_i$, and

\item equality atoms of the form $t = t'$, where $t$ and $t'$ are
terms based on $\x_i$ and $\f$.
\end{iteMize}

\item Each $\psi_i$ is a conjunction of relational atomic formulas of
the form $T(t_1, \ldots, t_\ell)$, where $T$ is an $\ell$-ary relation
symbol of $\bt$ and $t_1$, $\ldots,$ $t_\ell$ are terms based on $\x_i$
and $\f$.

\item Each variable in $\x_i$ appears in some relational atomic formula of
$\varphi_i$.\ebox
\end{enumerate}
\end{defi}

The fourth condition is the safety condition for SO tgds.
Note that it is ``built into'' SO tgds that they are source-to-target.
The depth of nesting of an SO tgd is the maximal depth of nesting of the
terms that appear in it.
We say that the SO tgd is {\em unnested\/} if its depth of nesting is at
most 1.
Thus, an unnested SO tgd can contain terms like $f(x)$, but not terms
like
$f(g(x))$.

As was noted in \cite{FKPT05,NBM05}, there is a subtlety in the
semantics
of SO tgds, namely, the semantics of existentially quantified
function symbols.
In particular,
in  deciding whether $(I,J) \models \sigma$,
for an SO tgd $\sigma$,
what should the domain and range of
the functions instantiating the  existentially quantified function symbols be?
The obvious choice is to let the domain and range be the active domain
of $(I,J)$, but it is shown in \cite{FKPT05,NBM05} that this does not work
properly.
Instead, the solution in \cite{FKPT05,NBM05} is as
follows.
Let $\sigma$ be an SO
tgd from a source schema $\bs$ to a target schema $\bt$. Then
given an instance $I$ of $\bs$ and an instance $J$ of $\bt$, instance
$(I,J)$ is converted into a structure $( U; I,J )$, which is just like
$( I,J )$ except that it has a {\em universe\/} $U$.  The domain and
range of the functions in $\sigma$ is then taken to be $U$. The
universe $U$ is taken to be a countably infinite set that includes
$\dom(I) \cup \dom(J)$.  The intuition is that the universe contains
the active domain along with an infinite set of nulls.  Then $(I,J)$
is said to satisfy $\sigma$, denoted by $( I,J ) \models
\sigma$, if $( U; I,J ) \models \sigma$ under the standard notion of
satisfaction in second-order logic (see, for example, \cite{En01}).
It should be noticed that it is proven in \cite{FKPT05} that in the
case of SO tgds, instead of taking the universe $U$ to be infinite,
one can take it to be finite and ``sufficiently large'',
whereas in \cite{NBM05} this is shown to be
insufficient in the presence of unrestricted target constraints.


The class of SO tgds was introduced in \cite{FKPT05} to deal with the
problem of composing schema mappings. More specifically, given a
schema mapping $\M_{12}$ from a schema $\bs_1$ to a schema $\bs_2$ and
a schema mapping $\M_{23}$ from $\bs_2$ to a schema
$\bs_3$, the composition of
these two schemas, denoted by $\M_{12} \circ \M_{23}$, is defined as
the schema mapping consisting of all pairs $(I_1, I_3)$ of instances
for which there exists an instance $I_2$ of $\bs_2$ such that
$(I_1,I_2)$ belong to $\M_{12}$ and $(I_2,I_3)$ belong to
$\M_{23}$. It was shown in \cite{FKPT05} that the composition of an
arbitrary number of st-tgd mappings is
specified
by an SO tgd,
that SO tgds are closed under composition, and that
every SO tgd specifies the composition of a finite number of
st-tgd mappings.

\section{A Negative Result: SO tgds are not Enough}
\label{sec-neg}

As pointed out in the previous section, SO tgds were introduced in
\cite{FKPT05} to deal with the problem of composing schema mappings.
Thus, SO tgds are a natural starting point
for the study of languages for defining the composition of schema
mappings with target constraints, which is the goal of this paper.
Unfortunately, it can be
easily proved that this language is not rich enough to be able to
specify
the composition of some simple schema mappings with target constraints.
We now give an example.

\begin{example}\label{exa-no-sotgd}
Let $\M_{12} = (\bs_1, \bs_2, \Sigma_{12}, \Sigma_2)$ and $\M_{23} =
(\bs_2,
\bs_3, \Sigma_{23})$, where $\bs_1 = \{ P(\cdot, \cdot) \}$, $\bs_2 =
\{ R(\cdot, \cdot) \}$, $\bs_3 = \{ T(\cdot, \cdot) \}$ and
\begin{eqnarray*}
\Sigma_{12} & = & \{ P(x,y) \rightarrow R(x,y) \},\\
\Sigma_2 & = & \{ R(x,y) \wedge R(x,z) \rightarrow y  = z \},\\
\Sigma_{23} & = & \{ R(x,y) \rightarrow T(x,y) \}.
\end{eqnarray*}
Notice that $\Sigma_2$ consists of a key dependency over $\bs_2$.

Let $\M_{13} = \M_{12} \circ \M_{23}$.
We now show that $\M_{13}$ cannot be specified by an SO tgd $\sigma$.
Assume that it were; we shall derive a contradiction.
If $I_1$ is an arbitrary instance of $\bs_1$, then there is $I_3$ such
that
$(I_1, I_3) \in \M_{13}$ (for example, we could take $I_3$ to be the result
of chasing $I_1$ with $\sigma$ as in \cite{FKPT05}).
However, let $I_1$ be the instance of $\bs_1$ such that
$P^{I_1} = \{(1,2), (1,3) \}$.
Then
there is no instance $I_3$ of $\bs_3$
such that
$(I_1, I_3) \in \M_{12} \circ \M_{23}$, since $I_1$ does
not have any solutions
with respect to
$\M_{12}$.
\end{example}

From the previous example, we obtain the following proposition.
\begin{prop}\label{prop-exp-so-tgd}
There exist schema mappings $\M_{12} = (\bs_1, \bs_2, \Sigma_{12},
\Sigma_2)$ and $\M_{23} = (\bs_2, \bs_3, \Sigma_{23})$, where
$\Sigma_{12}$ and $\Sigma_{23}$ are sets of st-tgds and $\Sigma_2$ is
a set of key dependencies, such that $\M_{12} \circ \M_{23}$ cannot be
specified by an SO tgd.
\end{prop}

Proposition~\ref{prop-exp-so-tgd} does not rule out the possibility
that
the composition of $\M_{12}$ and $\M_{23}$ can be specified by using an
SO tgd together with some source and target constraints.
In fact,
if $\M_{12}$ and $\M_{23}$ are as in Example~\ref{exa-no-sotgd}, then
the composition $\M_{12} \circ \M_{23}$
can be
specified by a set of st-tgds together with some source constraints:
specifically,
$\M_{12} \circ \M_{23} = \M_{13}$,
where $\M_{13} = (\bs_1, \bs_3, \Sigma_1,
\Sigma_{13})$ and
\begin{eqnarray*}
\Sigma_1 & = & \{ P(x,y) \wedge P(x,z) \rightarrow y  = z \},\\
\Sigma_{13} & = & \{ P(x,y) \rightarrow T(x,y) \}.
\end{eqnarray*}
A natural question is then whether the language  of SO tgds together
with
source and target constraints is the right language for defining the
composition of schema mappings with source and target
constraints. Unfortunately, the following theorem shows that this is
not the case.
\begin{thm}\label{theo-exp-so-tgd-t}
There exist schema mappings $\M_{12} = (\bs_1, \bs_2, \Sigma_{12},
\Sigma_2)$ and $\M_{23} = (\bs_2, \bs_3$, $\Sigma_{23})$, where
$\Sigma_{12}$ and $\Sigma_{23}$ are sets of st-tgds and $\Sigma_2$ is
a set of key dependencies, such that $\M_{12} \circ \M_{23}$ cannot be
specified by any schema mapping of the form $(\bs_1, \bs_3,
\sigma_1, \sigma_{13}, \sigma_3)$, where
$\sigma_1$ is an arbitrary source constraint,
$\sigma_{13}$ is an SO tgd, and $\sigma_3$ is an arbitrary target
constraint.
\end{thm}
If we view a source constraint as a set of allowed source instances,
then when we say that $\sigma_1$ is an ``arbitrary source constraint''
in Theorem~\ref{theo-exp-so-tgd-t},
we mean that $\sigma_1$ allows an arbitrary set of source instances.
A similar comment applies to $\sigma_3$ being an ``arbitrary target
constraint''.

To prove this theorem, we use a notion of locality from \cite{ABFL04}.
Notions of locality \cite{FSV95,Gai82,Han65,fmt-book} have been widely
used to prove inexpressibility results for first-order logic (\FO)
and some
of its extensions.  The intuition underlying those notions of locality
is that \FO\ cannot express properties (such as
connectivity,
cyclicity, etc.) that involve nontrivial  recursive computations. The
setting of locality is as follows.  The {\em Gaifman graph} $\G(I)$ of
an instance $I$ of a schema $\bs$ is the graph whose nodes are the
elements of $\dom(I)$,
and such that there exists an edge between $a$ and $b$ in $\G(I)$ if
and only if $a$ and $b$ belong to the same tuple of a relation $R^I$,
for some $R\in \bs$. For example, if $I$ is an undirected graph, then
$\G(I)$ is $I$ itself.  The distance between two elements $a$ and $b$
in $I$ is considered to be
the distance between them in $\G(I)$. Given $a \in
\dom(I)$, the instance $N_d^I(a)$, called the {\em $d$-neighborhood of $a$
in $I$}, is defined as the restriction of $I$ to the elements at
distance at most $d$ from $a$, with $a$ treated as a distinguished
element (a constant in the vocabulary).

%
The notion of neighborhood of a point is used in \cite{ABFL04} to
introduce a notion of locality for data transformations.
Before we give this definition, we give the standard recursive
definition
of the {\em quantifier rank\/} $qr(\phi)$ of an \FO-formula $\phi$.
\begin{iteMize}{$\bullet$}\itemsep=1pt
\item
If $\varphi$ is quantifier-free, then $qr(\varphi) = 0$.
\item
$qr(\neg \varphi) = qr(\varphi)$
\item
$qr(\varphi_1 \land \varphi_2) = \max\{qr(\varphi_1), qr(\varphi_2)\}$
\item
$qr(\forall x \varphi) = 1 + qr(\varphi)$
\end{iteMize}
Following \cite{ABFL04}, we write
$N_d^I(a)
\equiv_k N_d^I(b)$ to mean that $N_d^I(a)$ and $N_d^I(b)$ agree
on all
\FO-sentences of quantifier rank at most $k$; that is, for every
\FO-sentence $\varphi$ of quantifier rank at most $k$, we have that
$N_d^I(a) \models \varphi$ if and only if $N_d^I(b) \models \varphi$.

\begin{defi}{\bf (\cite{ABFL04})}\ \
Given a source schema $\bs$ and a target schema $\bt$, a mapping
$\FF:\bs \rightarrow \bt$ is {\em locally consistent under
\FO-equivalence} if for every $r,\ell \geq 0$ there exist $d,k \geq 0$
such that, for every instance $I$ of $\bs$ and $a,b \in
\dom(I)$, if $N_{d}^I(a) \equiv_{k} N_{d}^I(b)$, then
\begin{enumerate}[(1)]\itemsep=1pt
\item $a \in \dom(\FF(I))$ if and only if $b \in \dom(\FF(I))$,
and

\item $N_r^{\FF(I)}(a) \equiv_\ell N_r^{\FF(I)}(b)$. \ebox
\end{enumerate}
\end{defi}

For a fixed schema mapping $(\bs,\bt,\Sigma_{st})$, we denote
by $\CUS$ the transformation from $\bs$ to  $\bt$, such
that $\CUS(I)$ is the
\emph{canonical universal solution}
for $I$, which is obtained by doing a naive chase of $I$ with
$\Sigma_{st}$.

\begin{prop}[\cite{ABFL04}]\label{prop-cus-st-tgd}
For every st-tgd mapping,
the transformation $\CUS$ is
locally consistent under \FO-equivalence.
\end{prop}

The previous proposition can be easily extended to the case of a
composition of a finite number of st-tgd  mappings.

\begin{lem}\label{lemma-cus-comp}
Let $n \geq 2$. For every $i \in [1,n-1]$, let $\M_i = (\bs_i,
\bs_{i+1}, \Sigma_{i \, i+1})$ be a schema mapping specified by a
set
$\Sigma_{i \, i+1}$ of st-tgds, and $\CUS^i$
be the canonical
universal
solution
transformation for $\M_i$. Assume that $\FF$ is the transformation from
$\bs_1$ to $\bs_n$ defined as:
\begin{eqnarray*}
\FF(I_1) & = & \CUS^{n-1}(\cdots (\CUS^2(\CUS^1(I_1))) \cdots),
\end{eqnarray*}
for every instance $I_1$ of $\bs_1$. Then $\FF$ is locally consistent
under \FO-equivalence.
\end{lem}
Lemma \ref{lemma-cus-comp} is one of the key components in the proof
of Theorem~\ref{theo-exp-so-tgd-t}.
We shall also utilize the next proposition
(Proposition~\ref{pro:chase-of-the-chase}) in
the proof
of Theorem~\ref{theo-exp-so-tgd-t}.
Proposition~\ref{pro:chase-of-the-chase} in
a generalization of Proposition~7.2
of \cite{Fa07}, which says that for the composition of two
st-tgd mappings, the ``chase of the chase'' is a universal solution.

\begin{prop}\label{pro:chase-of-the-chase}
Let $\M_1, \ldots, \M_k$ be schema mappings, each specified by st-tgds,
target egds, and target tgds.
Let $\M = \M_1 \circ \cdots \circ
\M_k$, and let $I$ be a source instance for $\M_1$.
Let $U$ be a result (if it exists) of chasing $I$ with $\M_1$,
then chasing
the result with $\M_2$, ..., and then chasing the result with $\M_k$.
Then $U$ is a universal solution for $I$ with respect to $\M$.
\end{prop}
\proof{
We use a simple trick from the proof of Proposition~7.2 in
\cite{Fa07}.
Define $\M'$ to be a schema mapping whose st-tgds consist of the st-tgds
of $\M_1$,
whose target egds consist of the union of the target egds of $\M_1,
\ldots, \M_k$,
and whose target tgds consist of all of the st-tgds of
$\M_2, \ldots, \M_k$, along with all of the target tgds of $\M_1,
\ldots, \M_k$.
If $I$ be a source instance for $\M_1$, and
$J$ is a target instance for $\M_k$, then it is easy to see that
$J$ is a solution for $I$ with respect to $\M$ if and only if
$J$ is a solution for $I$ with respect to $\M'$.
Hence,
$\M$ and $\M'$ are the same schema mapping semantically, in that they
consist of the same set of $(I,J)$ pairs.
If $I$ and $U$ are as in the statement of the proposition, then
it is easy to see that $U$ is a result of a chase of $I$ with
$\M'$.
So from Theorem~3.3 of \cite{FKMP05}, we have that $U$ is a universal
solution of $I$ with respect to $\M'$.
Since $\M$ and $\M'$ are the same schema mapping semantically,
it follows that
$U$ is a universal
solution of $I$ with respect to $\M$.\qed}

We need to say ``if it exists'' in the statement of
Proposition~\ref{pro:chase-of-the-chase}, since there are two
reasons that a result of the chase may not exist.
First, a target egd may try to equate two constants during a chase.
Second, target tgds might force an ``infinite chase''.
These problems do not arise for the composition of two st-tgd mappings,
the case considered in Proposition 7.2 of \cite{Fa07}.

\bigskip

\aproof{Theorem \ref{theo-exp-so-tgd-t}}{Let $\M_{12} = (\bs_1, \bs_2,
\Sigma_{12}, \Sigma_2)$ and $\M_{23} = (\bs_2, \bs_3, \Sigma_{23})$ be
schema mappings, where:
\begin{eqnarray*}
\bs_1 & = & \{ E(\cdot,\cdot), P_1(\cdot), Q_1(\cdot) \},\\
\bs_2 & = & \{ P_2(\cdot), Q_2(\cdot), R(\cdot,\cdot),
S(\cdot,\cdot)\},\\
\bs_3 & = & \{ V(\cdot) \},
\end{eqnarray*}
and
\begin{align*}
\Sigma_{12} \ = \ \{&P_1(x) \rightarrow P_2(x),\\
& Q_1(x) \rightarrow Q_2(x), \\
& E(x,y) \rightarrow \exists z_1 \exists z_2 \exists z_3\, (R(x,z_1)
\wedge R(y,z_2) \wedge S(z_1,z_3) \wedge S(z_2,z_3))\},\\
\Sigma_2 \ = \ \{& R(x,y) \wedge R(x,z) \rightarrow y = z,\\
& S(x,y) \wedge S(x,z) \rightarrow y = z,\\
& S(y,x) \wedge S(z,x) \rightarrow y = z\},\\
\Sigma_{23} \ = \ \{&P_2(x) \wedge R(x,z) \wedge R(y,z) \wedge Q_2(y)
\rightarrow V(x)\}.
\end{align*}
First, we show that $\M_{12} \circ \M_{23}$ cannot be
specified
by an
SO tgd. For the sake of contradiction, assume that $\sigma_{13}$ is an
SO tgd from $\bs_1$ to $\bs_3$ and that schema mapping $\M_{13}
= (\bs_1, \bs_3, \sigma_{13})$
is
the composition of $\M_{12}$
and $\M_{23}$, that is, $(I_1, I_3) \in \M_{12} \circ
\M_{23}$ if and only if $(I_1, I_3) \models \sigma_{13}$.

{From} Theorem 8.2 in \cite{FKPT05}, we know that every SO tgd
specifies
the composition of a finite number of
st-tgd mappings.
Thus, given that $\M_{13}$
is
the
composition of $\M_{12}$ and $\M_{23}$, we have that there exist
schema mappings $\M'_1 = (\bs'_1, \bs'_2, \Sigma'_{12})$, $\ldots$,
$\M'_{n-1} = (\bs'_{n-1}, \bs'_n, \Sigma'_{n-1\, n})$ such that $n
\geq 2$, $\bs'_1 = \bs_1$, $\bs'_n = \bs_3$, $\Sigma'_{i \, i+1}$ is a
set of st-tgds for every $i \in \{1, \ldots, n-1\}$, and $\M'_1 \circ
\ldots \circ \M'_{n-1}$
equals
the composition of $\M_{12}$ and
$\M_{23}$, that is, for
every pair  $(I_1, I_3) \in \inst(\bs_1) \times \inst(\bs_3)$:
\begin{eqnarray*}
(I_1,I_3) \in \M_{12} \circ \M_{23} & \Leftrightarrow &
(I_1, I_3) \in \M'_1 \circ \ldots \circ \M'_{n-1}.
\end{eqnarray*}
For every $i \in \{1, \ldots, n-1\}$, let $\CUS^i$ be the canonical
solution transformation for $\M'_i$, and assume that
$\FF :
\inst(\bs_1) \rightarrow \inst(\bs_3)$ is the transformation defined
as 
\begin{eqnarray*}
\FF(I_1) & = & \CUS^{n-1}(\cdots (\CUS^2(\CUS^1(I_1))) \cdots), 
\end{eqnarray*}
for every instance $I_1$ of $\bs_1$.  From Lemma \ref{lemma-cus-comp}, we
have that $\FF$ is locally consistent under \FO-equivalence. Thus, for
$r=1$ and $\ell = 1$, there exist $d, k \geq 0$ such that for every
instance $I_1$ of $\bs_1$ and for every $a,b \in \dom(I_1)$, if
$N^{I_1}_{d}(a) \equiv_{k} N^{I_1}_{d}(b)$ then (1) $a \in
\dom(\FF(I_1))$ if and only if $b \in \dom(\FF(I_1))$, and (2)
$N^{\FF(I_1)}_r(a)  \equiv_\ell N^{\FF(I_1)}_r(b)$.

Define an instance $I_1$ of $\bs_1$ with domain $\{a, a_1, \ldots, a_{d},
b, b_1, \ldots, b_{d}, c\}$ as follows: $P_1^{I_1} = \{a,b\}$,
$Q_1^{I_1} = \{ c \}$, and $E^{I_1}$ contains the
tuples represented by the following figure:
\begin{center}
\input{source.pstex_t}
\end{center}
Thus, $E^{I_1} = \{(a,a_1), (a_1,a_2), \ldots, (a_{d-1},a_d), (b,b_1),
(b_1, b_2), \ldots, b_{d-1},b_d)\}$.
As shown in the figure, $E^{I_1}$ is a union of two paths, one
containing $d+2$ elements with first element $a$ and last element $c$,
and another one containing $d+1$ elements with first element
$b$. Observe that $N^{I_1}_{d}(a) \equiv_k N^{I_1}_{d}(b)$ since
$N^{I_1}_{d}(a)$ is isomorphic to $N^{I_1}_{d}(b)$, with $a$ and $b$
treated as distinguished elements.

Let $I_3$ be the instance of $\bs_3$ that contains only the tuple $a$
in $V$ (that is, $V^{I_3} = \{a\}$). Next we show that $I_3$ is a
universal solution for $I_1$
with respect to
$\M_{12} \circ \M_{23}$. A universal
solution for $I_1$ in $\M_{12} \circ \M_{23}$ can be constructed by
first chasing with the set $\Sigma_{12}$ of st-tgds:
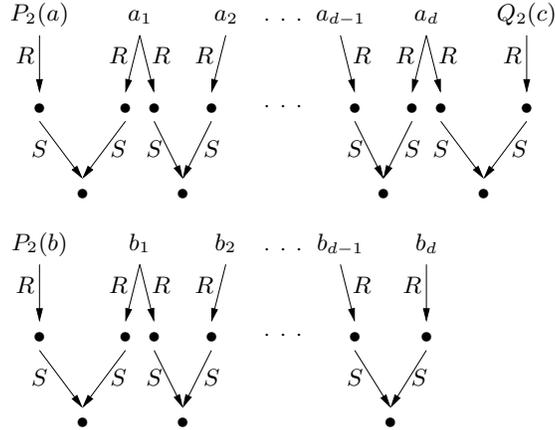
\begin{center}
\input{pre-can.pstex_t}
\end{center}
(where each element in the figure that is not in $\dom(I_1)$ is a
fresh null value, which is represented by a symbol $\bullet$ in the
figure), then chasing with the set $\Sigma_2$ of t-egds:
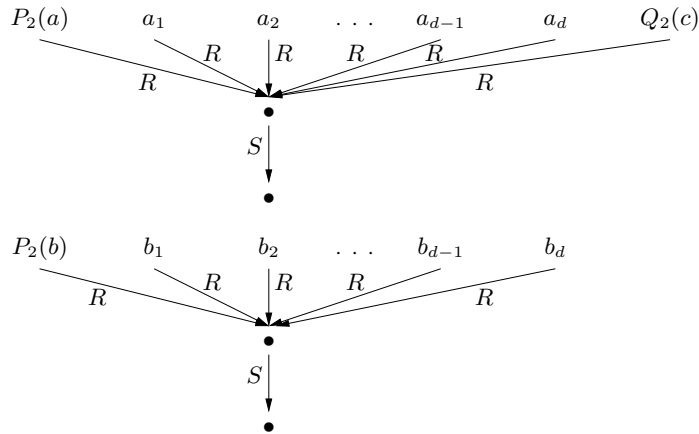
\begin{center}
\input{can.pstex_t}
\end{center}
and finally chasing with the set $\Sigma_{23}$ of st-tgds. The result
of this last step is $I_3$,  since after chasing with $\Sigma_{12}$ and
$\Sigma_2$, we have that $P_2$ contains elements $a$ and $b$, $Q_2$
contains element $c$, and $a$, $c$ is the only pair of elements for
which there exists an element $z$ such that $P_2(a)
\wedge R(a,z) \wedge R(c,z) \wedge Q_2(c)$ holds. We conclude
that $I_3$ is a universal solution for $I_1$.

By Proposition~\ref{pro:chase-of-the-chase}, we know that
$\FF(I_1)$ is a universal solution for $I_1$ with respect to $\M'_1
\circ
\cdots \circ \M'_{n-1}$.
Hence,
$\FF(I_1)$ is a universal
solution for $I_1$ with respect to $\M_{12} \circ \M_{23}$.
Again by Proposition~\ref{pro:chase-of-the-chase},
we know that $I_3$
is also a universal solution for $I_1$ with respect to $\M_{12} \circ
\M_{23}$.
Therefore, since $V^{I_3} = \{a\}$,
we conclude that $a \in V^{\FF(I_1)}$ and $b
\not\in V^{\FF(I_1)}$. Hence, we have that $a \in \dom(\FF(I_1))$ and
$b \not\in \dom(\FF(I_1))$, which contradicts the fact that $\FF$ is
locally consistent under \FO-equivalence and $N^{I_1}_{d}(a)
\equiv_{k} N^{I_1}_{d}(b)$. This concludes the proof that $\M_{12}
\circ \M_{23}$ cannot be
specified
by an SO tgd.

To conclude the proof of the theorem, we need to show that $\M_{12}
\circ \M_{23}$ cannot be
specified
by an SO tgd together with some
arbitrary source and target constraints. For the sake of contradiction, assume
that schema mapping $\M_{13}= (\bs_1, \bs_3, \gamma_1,
\gamma_{13}, \gamma_3)$
equals
the composition of $\M_{12}$ and
$\M_{23}$, where $\gamma_1$ is an arbitrary source constraint,
$\gamma_{13}$ is an SO tgd and $\gamma_3$ is an arbitrary target
constraint.  Given that $\M_{12} \circ \M_{23}$ cannot be
specified
by
an SO tgd, we have that
either
$\gamma_1$ is not trivial or $\gamma_3$ is not
trivial, where an arbitrary constraint is said to be trivial if it
allows all the possible instances. First assume that $\gamma_1$ is not
trivial, and let $I_1$ be an instance of $\bs_1$ such that $I_1$ does
not satisfy $\gamma_1$ ($I_1$ is not allowed by $\gamma_1$). Let
$\bot$ be a fresh null value and $I_2$ an instance of $\bs_2$ defined
as:
\begin{eqnarray*}
P_2^{I_2} & = & P_1^{I_1},\\
Q_2^{I_2} & = & Q_1^{I_1},\\
R^{I_2} & = & \{(a,\bot) \mid \text{there exists } b \in \dom(I_1)
\text{ such that } (a,b) \in E^{I_1} \text{ or } (b,a) \in
E^{I_1}\},\\
S^{I_2} & = & \{(\bot,\bot)\}.
\end{eqnarray*}
Furthermore, let $I_3$ be an instance of $\bs_3$ defined as $V^{I_3} =
P_2^{I_2}$. It is easy to see that $(I_1, I_2) \models \Sigma_{12}$,
$I_2 \models \Sigma_2$ and $(I_2, I_3) \models \Sigma_{23}$, which
implies that $(I_1,I_3) \in
\M_{12} \circ \M_{23}$.
Thus, given that $\M_{13}$
is
the composition of
$\M_{12}$ and $\M_{23}$, we have that $(I_1, I_3) \in
\M_{13}$.
We conclude that $I_1$ satisfies $\gamma_1$, which
contradicts our initial assumption.

Now suppose that $\gamma_3$ is not trivial, and let $I_3$ be an
instance of $\bs_3$ such that $I_3$ does not satisfy $\gamma_3$. Assume that
$I_1$, $I_2$ are the empty instances of $\bs_1$ and $\bs_2$,
respectively. It is easy to see that $(I_1, I_2) \models \Sigma_{12}$,
$I_2 \models
\Sigma_2$ and $(I_2, I_3) \models \Sigma_{23}$, which implies that
$(I_1,I_3) \in \M_{12} \circ \M_{23}$.
Thus, given that
$\M_{13}$
is
the composition of $\M_{12}$ and $\M_{23}$, we have
that $(I_1, I_3) \in \M_{13}$, and hence
$I_3$ satisfies
$\gamma_3$, which contradicts our initial assumption. This concludes
the proof of the theorem.}

%

\section{Source-to-target SO Dependencies}
\label{sec-st-so}

In Section \ref{sec-neg}, we showed that SO tgds, even with the
assistance of arbitrary source constraints and arbitrary target
constraints,  cannot always be used to specify the
composition of mappings with target constraints, even if only key
dependencies are
allowed as
target constraints of the mappings being composed.
 In this paper, we
define a richer class, called source-to-target SO dependencies
(st-SO dependencies).
This class of dependencies is the source-to-target restriction of the
class $\sk$ of dependencies introduced in \cite{NBM05}.
We
show that st-SO dependencies (together with appropriate target
constraints) are the right extension of SO tgds for the purpose
of expressing the composition of standard schema mappings.
The definition of st-SO dependencies is exactly like
the definition of SO tgds
in Definition~\ref{def:sotgd},
except that condition~3 is
changed to:

\begin{enumerate}[(1)]
\item[(3)]  Each $\psi_i$ is a conjunction of
\begin{iteMize}{$\bullet$}\itemsep=1pt
\item relational atomic formulas of
the form $T(t_1, \ldots, t_\ell)$, where $T$ is an $\ell$-ary relation
symbol of $\bt$ and $t_1$, $\ldots,$ $t_\ell$ are terms based on $\x_i$
and $\f$,
and

\item equality atoms of the form $t = t'$, where $t$ and $t'$ are
terms based on $\x_i$ and $\f$.
\end{iteMize}
\end{enumerate}

\noindent
It is sometimes convenient to rewrite an st-SO dependency
$\exists \f \, (\forall \x_1
(\varphi_1
\rightarrow \psi_1) \wedge \cdots \wedge \forall \x_n (\varphi_n
\rightarrow \psi_n))$
so that
each conclusion $\psi_i$ is either a conjunction of relational
atomic formulas or a single equality of terms
(this is possible because we can recursively replace
$\forall \x_i (\varphi_i
\rightarrow (\psi_i^1 \land \psi_i^2))$ by
$\forall \x_i (\varphi_i
\rightarrow \psi_i^1) \land
\forall \x_i (\varphi_i
\rightarrow \psi_i^2)$
without changing the meaning).
Let $\Phi$ be the result of such a rewriting.
If $\psi_i$ is a
conjunction of relational
atomic formulas, then we refer to
$\forall \x_i (\varphi_i \rightarrow \psi_i)$ as an
{\em SO tgd part\/} of $\Phi$,
and if $\psi_i$ is an equality $t = t'$, then we refer to
$\forall \x_i (\varphi_i \rightarrow \psi_i)$ as an
{\em SO egd part\/} of $\Phi$.

We adopt the same convention for the semantics of st-SO dependencies as
was given in Section~\ref{sec-prelim} for SO tgds, by assuming the
existence of a countably infinite universe that includes the active
domain.
As with SO tgds, it can be shown that the universe can be taken to be
finite but ``sufficiently large''.

We shall show that the composition of a finite number of standard schema
mappings
can be
specified by an st-SO dependency,
together with t-egds and a weakly acyclic set of t-tgds.
It is convenient to give these
second-order
schema mappings a name.
To emphasize the similarity of these second-order schema mappings with
the first-order case, we shall refer to these
second-order
schema mappings as
{\em SO-standard}.
Thus, an SO-standard schema mapping is one that is
specified by an st-SO dependency,
together with t-egds and a weakly acyclic set of t-tgds.

Note that st-SO dependencies, like SO tgds, are closed under
conjunction.
That is, the conjunction of two st-SO dependencies is equivalent to a
single st-SO dependency.
This is why we define an SO-standard schema mapping to have only one
st-SO dependency, not several.
Note also that every finite set of st-tgds can be expressed with an
SO tgd, and so with an st-SO dependency.
In particular, every standard schema mapping is an SO-standard schema
mapping.

\section{The Chase for st-SO Dependencies}
\label{sec-chase}

In \cite{FKPT05}, the well-known chase process is extended so that it
applies to an SO tgd $\Phi$.
If we define an {\em SO tgd part\/} of an SO tgd as we did for st-SO
dependencies, then the idea of the chase with SO tgds is
that each SO tgd part of $\Phi$ is treated
like a tgd (of course, the conclusion contains Skolem functions rather
than existential quantifiers).
In deciding whether the premise of the SO tgd part is instantiated
in the instance being chased, two terms
are treated as equal precisely if they are
syntactically identical.
So a premise containing the equality atom $f(x) = g(y)$
automatically fails to hold over an instance, and
a premise containing the equality atom $f(g(x)) = f(g(y))$
automatically fails to hold over an instance unless the instantiation of
$x$ equals the instantiation of $y$.

In this section, we discuss how the chase can be extended to apply to an
st-SO dependency.
We note that in \cite{NBM05},
a chase procedure for the dependencies studied there (which are like
ours but not necessarily source-to-target) was introduced.
However, their chase was not procedural, in that their chase
procedure says to set terms $t_1$ and $t_2$ to be equal when the
dependencies logically imply that $t_1 = t_2$.
Because of our source-to-target restriction, we are able to give an
explicit, polynomial-time procedure for equating terms.

For clarity, we keep the discussion here informal; it is not hard to
convert this into a formal version.
In chasing an instance $I$ with an st-SO
dependency
$\Phi$, we chase first with all of the SO egd parts of $\Phi$, and then
we chase with all of the SO tgd parts of $\Phi$.
We no longer
consider two terms
to be equal precisely if they are
syntactically identical, since an SO egd part may force, say, $f(0)$
and $g(1)$ to be equal, even though $f(0)$ and $g(1)$ are not
syntactically identical.

Given a source instance $I$ and an st-SO dependency $\Phi$, we now
describe
how to chase $I$ with the SO egd parts of $\Phi$.
Let $D$ be the active domain of $I$
(by our assumptions, $D$ consists of constants only).
Let $n$ be the maximal depth of nesting over all terms that appear in
$\Phi$.
Let $\f$ consist of the function symbols that appear in $\Phi$.
Let $T$ be the set of terms based on $D$ and $\f$
that have depth of nesting at most $n$.
This set $T$ is sometimes called the {\em Herbrand universe\/}
(with respect to
$D$ and $\f$) of depth $n$.
It is straightforward to see (by induction on depth) that
the size of $T$
is polynomial in the size of $D$, for a fixed choice
of $\Phi$.
We note that if we define $T'$ to be the subset of $T$ that
consists of all terms $t(\a)$, where
$t(\x)$ is a subterm of $\Phi$, and $\a$ is the result of replacing
members of $\x$ by values in $D$, then we could work just as well with
$T'$ as with $T$ in defining the chase.
However, the proofs are easier to give using $T$ instead of $T'$.

We now define a function $F$ with domain the members of $T$.
The values $F(t)$ are stored in a table that is
updated repeatedly during the chase process.
If $a$ is a member of $D$, then the initial value of $F(a)$ is $a$
itself (in fact, the value of $F(a)$ will never change for members $a$
of $D$).
If $t$ is a member of $T$ that is not in $D$ (so that
$t$ is of the form
$f(t_1, \ldots, t_k)$ for some function symbol $f$), then
$F(t)$ is initially taken to be a
new null value.
As we change $F$, we shall maintain the invariant that if
$f(t_1, \ldots, t_k)$ and
$f(t_1', \ldots, t_k')$ are members of $T$ where
$F(t_i) = F(t_i')$, for $1 \leq i \leq k$, then
$F(f(t_1, \ldots, t_k)) =
F(f(t_1', \ldots, t_k'))$.
This is certainly true initially, since $F$ is initially one-to-one on
members of $T$.


Let $N$ be the set of all of the new null values
(the values initially assigned to $F(t)$ when $t$ is not in $D$).
We create an ordering $\prec$
on $D \cup N$,
where the
members of
$D$ are an initial segment of the
ordering $\prec$,
followed by the
members of $N$.

We now begin chasing $I$ with the SO egd parts of $\Phi$, to change the
values of $F$.
Whenever $t$ is a member of $T$ such that
we replace a current value of $F(t)$ by a new value
during the chase process,
we will always replace the current value of $F(t)$ by a value that is
lower in the ordering $\prec$.
If $s_1(\y_1) = s_2(\y_2)$ is an equality in the premise of an SO egd
part of $\Phi$, then the equality $s_1(\e_1) = s_2(\e_2)$ evaluates to
``true'' where $\e_1$ and $\e_2$ consist of members of $D$,
precisely if the current value of $F(s_1(\e_1))$ equals the current
value of
$F(s_2(\e_2))$.
Each time an equality $t_1(\a) = t_2(\b)$ is forced (because of an
SO egd part with conclusion $t_1(\x) = t_2(\y)$),
and the current value of $F(t_1(\a))$ does not equal the current value
of $F(t_2(\b))$ we
proceed as follows.
Let $c_1$ be the smaller of these two values
and let $c_2$ be the larger of these two values
in our ordering $\prec$.
If $c_2$ is a constant, then the chase fails and halts.
Otherwise,
for every member $s$ of $T$ where the current value of $F(s)$ is
$c_2$, change the value so that the new value of $F(s)$ is $c_1$.
Note that under this change, the new value of
$F(t_1(\a))$ and the new value of
$F(t_2(\b))$
are the same (namely, $c_1$).

These changes in $F$ may propagate new changes in $F$, which we need to
make in order to maintain the invariant.
%
Assume that as a result of our changes in $F$ so far,
there are terms
$f(t_1, \ldots, t_k)$ and
$f(t_1', \ldots, t_k')$ in $T$ where
$F(t_i) = F(t_i')$, for $1 \leq i \leq k$, but
$F(f(t_1, \ldots, t_k))$ and
$F(f(t_1', \ldots, t_k'))$ are different.
As before, let
$c_1$ be the smaller of these two values
and let $c_2$ be the larger of these two values
in our ordering $\prec$.
If $c_2$ is a constant, then the chase fails and halts.
Otherwise,
for every member $s$ of $T$ where the current value of $F(s)$ is
$c_2$, change the value so that the new value of $F(s)$ is $c_1$.
Note that under this change, the new value of
$F(f(t_1, \ldots, t_k))$ and the new value of
$F(f(t_1', \ldots, t_k'))$ are the same (namely, $c_1$).
Continue this process until no more changes occur.
It is easy to see that we have maintained our invariant.
Continue chasing with SO egd parts until
no more changes occur.
Note that at most as many changes can occur as the size of $T$, since
every time a change occurs, there are strictly fewer values of $F(t)$ as
$t$ ranges over $T$.
This is the key reason why the chase runs in polynomial time.

Once $F$ has stabilized, so that no more changes are caused by chasing
with the SO egd parts of $\Phi$, then chase $I$ with the SO tgd parts
of $\Phi$.
If $s_1(\y_1) = s_2(\y_2)$ is an equality in the premise of an SO tgd
part of $\Phi$, then the equality $s_1(\e_1) = s_2(\e_2)$ evaluates to
``true'' where $\e_1$ and $\e_2$ consist of members of $I$,
precisely if $F(s_1(\e_1)) =  F(s_2(\e_2))$.
These chase steps produce the target relation $J$ that is taken to be
the result of
the chase (and we say that the chase {\em succeeds}).

We have the following theorem
about the chase process.

\begin{thm}\label{theo-chase}
Let $\Phi$ be a fixed st-SO dependency.
The chase of a ground instance $I$ with $\Phi$ runs in time polynomial
in the size of $I$.
The chase fails precisely if there is no solution for
$I$ with respect to $\Phi$.
If the chase succeeds, then it produces a universal solution
for $I$ with respect to $\Phi$.
\end{thm}

\proof{We first show that the chase of a ground instance $I$ with $\Phi$ runs
in time polynomial in the size of $I$ (when $\Phi$ is held fixed).
It is straightforward to show by induction on depth that the size of $T$
is polynomial in the size of $D$. As we noted, during the chase with the
SO egd parts, there are at most as many changes in the current value of
$F$ as the size of $T$.
So only polynomially many changes occur in the values of $F$.
For each such change, there is only a polynomial amount of work: the
time needed to chase each SO egd part and update F if needed along
with the
time to check the invariant and update F if needed.
Finally, since the SO tgd parts are source-to-target, it follows easily
that the final portion of the chase, that is, chasing with the SO tgd
parts, can also be done in polynomial time.
Therefore, the entire chase can be carried out in polynomial time.

Assume for now that there is $J$ such that $(I,J) \models \Phi$.
If the existentially quantified function symbols of $\Phi$ are given by
$\f$, then let
$\f^0$ denote the instantiation of $\f$ that shows that
$(I,J) \models \Phi$.
For each member $t$ of $T$, let $t^0$ be the value
obtained by replacing the function symbols $\f$ by $\f^0$.
By induction on the steps in the chase process, we can
show that
at all points during the chase process,
if the current value of $F(t_1)$ equals the current value of $F(t_2)$,
then necessarily $t_1^0 = t_2^0$.
Thus, whenever we set two values $F(t_1)$ and $F(t_2)$ equal during the
process of chasing with the SO egd parts, we are forced to do so.
This is clear when two values are made equal because of the
conclusion of an SO egd part.
It is also true when two values are made equal because of maintaining
the invariant, because the invariant is needed for the functions
in $\f^0$ to be
well-defined.
For example, assume that $f^0$ and $g^0$ are unary functions in
$\f^0$.
If $f^0(a) = b$, then necessarily $g^0(f^0(a)) = g^0(b)$,
and this is reflected by the requirement of the invariant that
$F(g(f(a)) = F(g(b))$.
Since the only time we make two values equal in the table for $F$ is
when we are forced to, it follows that if the chase process fails
for $I$
(because we try to set two constants to be equal), then there is no
solution for $I$ with respect to $\Phi$.
We now consider the other case, where the chase
succeeds
for $I$.

Assume
that the chase
succeeds
for $I$.
Let $n$ be the maximal depth of nesting of function symbols in $\Phi$.
Use (the final values
of) the table for $F$ to define our functions $\f^0$ on the Herbrand
universe of depth $n-1$.
For example, if $a$ is in the active domain $D$, and $F(f(a)) = c$,
then let
$f^0(a) = c$.
Similarly, if $a$ and $b$ are in $D$,
and
$F(h(g(a),b)) = d$,
then let
$h^0(g^0(a),b) = d$.
The invariant insures that the functions in $\f^0$ are well-defined on
the Herbrand universe of depth
$n-1$ (and the table then gives us the values of all members of
the Herbrand universe of depth $n$).
Our semantics requires the functions in $\f^0$ to be defined not just on
the Herbrand universe of depth $n-1$, but on the
entire universe.
If $f$ is a $k$-ary function symbol in $\f$, and if $c_1, \ldots, c_k$
are values such that
$f^0(c_1, \ldots, c_k)$ is not already determined by the rules we have
given, then let $f^0(c_1, \ldots, c_k)$ be arbitrary.
The key point is that $\Phi$ refers only to terms in the Herbrand
universe of depth n,
so what happens outside of the Herbrand
universe of depth n
is irrelevant, as far as satisfaction of
$\Phi$ is concerned.
The chase with the SO egd
parts force equalities among the values of the functions so that
$I$ (together with the choice of the functions) satisfies
the SO egd parts of $\Phi$.
If $J$ is the result of the chase, then
the chase
with the SO tgd parts force $(I,J)$ to satisfy the SO tgd parts of
$\Phi$.
Hence, $J$ is a solution for $I$ with respect to $\Phi$, as desired.

It is also clear that if $J'$ is an arbitrary
solution for
$I$
with respect to $\Phi$, then  up to a replacement (not necessarily
one-to-one)
of nulls in $J$ by other values (nulls or constants), every tuple of
every relation that appears in $J$ must appear in the corresponding
relation of $J'$ (since tuples are produced in the chase only if needed,
and equalities are forced in the chase only if needed).
But this means that there is a homomorphism from $J$ into $J'$, so $J$
is a universal solution, as desired.\qed}

\medskip


Because there is a polynomial-time chase for st-SO
dependencies, there is also a polyno\-mial-time chase for SO-standard
schema mappings:  first, chase with the st-SO dependency, and then with
the target dependencies.  The reason that chasing with the target
dependencies requires only polynomial time
is that the number of steps in this chase is polynomial,
because of the weak acyclicity assumption (Theorem 3.9 of \cite{FKMP05}).
We therefore can extend Theorem~\ref{theo-chase} to apply to SO-standard
schema mappings.
We state this in the following corollary.

\begin{corollary}\label{cor-chase}
Let $\M$ be an SO-standard schema mapping.
The chase of a ground instance $I$ with $\M$ runs in time polynomial
in the size of $I$.
The chase fails precisely if there is no solution for
$I$ with respect to $\M$.
If the chase succeeds, then it produces a universal solution
for $I$ with respect to $\M$.
\end{corollary}

\noindent
Note that in particular, Corollary~\ref{cor-chase} tells us that there
is a polynomial-time algorithm for determining, given a source instance
$I$, whether there is a solution for $I$, and if so, producing a
universal solution for $I$.

As shown in \cite{FKMP05}, we can use a universal solution
to obtain the certain
answers to unions of conjunctive queries in polynomial time.
We now recall the definition of the certain answers.
Let $\M = ({\bf S}, {\bf T}$, $\Sigma)$ be a schema mapping, and let
$q$ be a $k$-ary query posed
against the target schema $\bf T$.
Denote by
$q(J)$ the result of evaluating $q$ on a target instance $J$.
If $I$ is a source instance, then
the
{\em certain answers of q on I with respect to $\M$},
denoted
by $\certain{\M}{q}{I}$,
are the $k$-tuples $t$
such that, for every solution $J$ of $I$
with respect to
$\M$, we have that
$t \in q(J)$.
It should be noticed that if a source instance $I$ does not have any
solution with respect to the mapping $\M$, then $\certain{\M}{q}{I} =
\D^k$ (recall that $\D$ is the countably infinite domain from which
the entries of tuples are taken), as every $k$-tuple trivially
satisfies the previous
condition. In this case, we use the special symbol $\top$ to indicate
that every $k$-tuple belongs to $\certain{\M}{q}{I}$, that is, we say
that $\certain{\M}{q}{I} = \top$.
If $U$
is a
universal solution
for
$I$
with respect to
$\M$, and $q$ is a union of conjunctive queries, then
it is shown in \cite{FKMP05} that
$\certain{\M}{q}{I}$ equals $q(U)_{\downarrow}$, which is
the result of evaluating
$q$ on $U$ and then keeping only those tuples formed entirely of values
from $I$ (that is, tuples that do not contain nulls).
The equality  $\certain{\M}{q}{I} = q(U)_{\downarrow}$
holds for arbitrarily specified schema mappings $\M$
(as long as such a universal solution $U$ exists).
Corollary~\ref{cor-chase} therefore has the following corollary,
which is analogous to the same corollary in \cite{FKPT05} for
mappings specified by SO tgds.

\begin{corollary}\label{cor-universal}
Let $\M$ be an SO-standard schema mapping.
Let $q$ be a union of conjunctive queries over the target schema ${\bf T}$.
Then for every ground instance $I$ over ${\bf S}$,
the set $\certain{\M}{q}{I}$
can be computed in polynomial time (in the size of $I$).
\end{corollary}
\proof{
Assume that the arity of query $q$ is $k$, where $k \geq 0$. Then the
polynomial-time algorithm to compute $\certain{\M}{q}{I}$ works as
follows. It first checks (using the polynomial-time algorithm of
Corollary~\ref{cor-chase}) whether $I$ has a solution with respect to
$\M$. If not, then $\certain{\M}{q}{I} = \D^k$, and the algorithm
returns symbol $\top$ to indicate that every tuple with $k$ elements
belongs to $\certain{\M}{q}{I}$. Otherwise $I$ has at least one
solution with respect to $\M$, and the algorithm computes a universal
solution $U$ for $I$ as in Corollary~\ref{cor-chase}, and then it
returns $q(U)_{\downarrow}$ (recall that, as discussed above,
$\certain{\M}{q}{I} = q(U)_{\downarrow}$).\qed}

\section{A Positive Result: SO-Standard Schema Mappings are the Needed
Class}
\label{sec-pos}

\noindent
In this section, we show that SO-standard schema mappings (those
specified by an st-SO dependency, along with target
constraints consisting of t-egds and a weakly acyclic set of t-tgds)
exactly correspond to
the composition of standard schema mappings.

\subsection{Using SO-standard schema mappings to define compositions}
\label{sec-rep}
\noindent
Before we show that
the composition of an arbitrary number of
standard schema mappings is equivalent to an
SO-standard schema mapping,
we first show that
target constraints are needed (that is, st-SO dependencies by themselves
are not enough).
In fact, the next proposition says that
st-SO dependencies, without target
constraints,  are not capable of
specifying  even schema mappings specified by st-tgds and a set of key
dependencies.

\begin{prop}\label{prop-no-st-so}
There exists a schema mapping $\M_{12} = (\bs_1, \bs_2,
\Sigma_{12}, \Sigma_2)$, where $\Sigma_{12}$ is a set of st-tgds and
$\Sigma_2$ is a set of key dependencies, such that $\M_{12}$ cannot be
specified by an st-SO dependency.
\end{prop}
As we shall see, we get an easy proof of
Proposition~\ref{prop-no-st-so} by using the
following simple proposition, which
is analogous to the same result for st-tgds
\cite{Fa07}.

\begin{prop}\label{prop-union}
Let $\sigma_{12}$ be an st-SO dependency, let $I$ be a source instance,
and let $J$ be a target instance.
If
$(I,J) \models \sigma_{12}$ and $J \subseteq J'$, then $(I, J')
\models \sigma_{12}$.
\end{prop}

\aproof{Proposition \ref{prop-no-st-so}}{Let $\bs_1 = \{S(\cdot,
\cdot)\}$, $\bs_2 = \{T(\cdot,\cdot)\}$ and $\Sigma_{12} = \{S(x,y) \to
T(x,y)\}$, and assume that $\Sigma_2$ consists of the single key
dependency $T(x,y) \wedge T(x,z)  \to  y = z$.
By way of contradiction, assume that $\M_{12}$ can be specified
by an st-SO dependency $\sigma_{12}$.
Let $I =
\{S(1,2)\}$, $J = \{T(1,2)\}$ and $J' = \{T(1,2),T(1,3)\}$. Given that
$(I,J) \models \Sigma_{12} \cup \Sigma_2$, and $\sigma_{12}$ specifies
$\M_{12}$, we have that $(I,J) \models \sigma_{12}$.
So by
Proposition~\ref{prop-union}, we have that $(I, J') \models
\sigma_{12}$.
Since
$\sigma_{12}$ specifies $\M_{12}$, we therefore have that
$(I, J') \models \Sigma_{12} \cup \Sigma_2$.
But this is a contradiction,
since $J' \not\models \Sigma_2$.}

\medskip

\noindent
Let $\M_{12}$ and $\M_{23}$ be standard schema mappings.
The previous negative result implies that st-SO dependencies by
themselves
cannot necessarily specify the composition $\M_{12} \circ \M_{23}$.
Our next theorem, which we shall prove shortly,
implies that
$\M_{12} \circ \M_{23}$ is equivalent to an SO-standard schema mapping
$\M_{13}$.
In fact, it says that we can take the target constraints of $\M_{13}$ to
be the set $\Sigma_3$ of
target constraints of $\M_{23}$.
Intuitively, this theorem tells us that
st-SO dependencies are expressive enough to capture the
intermediate target constraints in a composition.

\begin{thm}\label{theo-exp-st-so}
Let $\M_{12} = (\bs_1, \bs_2, \Sigma_{12}, \Sigma_2)$ and $\M_{23} =
(\bs_2, \bs_3, \Sigma_{23}, \Sigma_3)$ be standard schema mappings
(so that
$\Sigma_{12}$, $\Sigma_{23}$ are sets of st-tgds, and $\Sigma_i$ $(i =
2,3)$ is the union of a set of t-egds and a weakly acyclic set of
t-tgds). Then there exists an st-SO dependency $\sigma_{13}$
such that the mapping $\M_{13} = (\bs_1, \bs_3,
\sigma_{13}, \Sigma_3)$ is equivalent to the composition
$\M_{12} \circ \M_{23}$.
\end{thm}

\noindent
In
Section~\ref{sec-comp-st-so},
we show that the composition of SO-standard schema mappings is also an
SO-standard schema mapping.
By combining this result
with Theorem \ref{theo-exp-st-so}
(and using the simple fact, noted earlier, that every standard schema
mapping is an SO-standard schema mapping),  we obtain our desired
result, namely,
that the composition of a finite number of standard schema mappings
is equivalent to an SO-standard schema mapping.

It is straightforward to show that Theorem~\ref{theo-exp-st-so}
is a consequence of the following proposition.

\begin{prop}\label{prop-exp-st-so}
Let $\M_{12}$ be a standard schema mapping, and let $\M_{23}$ be an
st-tgd mapping
(no target constraints).
Then the composition $\M_{12} \circ \M_{23}$ can be specified by an
st-SO dependency.
\end{prop}
As pointed out in Section \ref{sec-st-so}, the class of st-SO
dependencies corresponds to the source-to-target restriction of the
class of $\sk$ dependencies introduced in \cite{NBM05}. In fact,
Theorem \ref{theo-exp-st-so} and Proposition \ref{prop-exp-st-so}
were essentially established in \cite{NBM05} (see Theorems 6 and 9
and the paragraph after Theorem 10 in \cite{NBM05}), but they are
restated and clarified here for the sake of completeness.  We also
show here how Proposition \ref{prop-exp-st-so} is proved, which is a
straightforward adaptation of the proofs of Theorems 6 and 9 in
\cite{NBM05}, and the comments in the paragraph after Theorem 10 to
handle a weakly acyclic set of target tgds.

We now demonstrate, by example,
how an st-SO dependency $\sigma_{13}$
is obtained from $\M_{12}$ and $\M_{23}$ in Proposition
\ref{prop-exp-st-so}
(it will be clear
how to extend from the example to the general case).
Assume that $\bs_1 = \{A(\cdot,\cdot),
B(\cdot)\}$, $\bs_2 = \{C(\cdot,\cdot), D(\cdot,\cdot)\}$, $\bs_3 =
\{E(\cdot,\cdot)\}$. Furthermore, suppose that  $\Sigma_{12}$ consists
of the following st-tgds:
\begin{eqnarray}
\nonumber
A(x,y) & \to & C(x,y),\\
\label{eq-1-2}
B(x) & \to & \exists y \, C(x,y),
\end{eqnarray}
$\Sigma_2$ consists of the following t-tgds:
\begin{eqnarray}
\nonumber
C(x,y) \wedge C(y,z) & \to & C(z,x),\\
\label{eq-2-2}
C(x,y) & \to & \exists z \, D(x,z),\\
\nonumber
C(x,x) & \to & D(x,x),\\
\nonumber
D(x,y) & \to & D(y,x),
\end{eqnarray}
and $\Sigma_{23}$ consists of the st-tgd:
\begin{eqnarray}
\label{eq-3-1}
D(x,y) & \to & \exists z \, E(x,y,z).
\end{eqnarray}
To obtain $\sigma_{13}$, we first Skolemize
each dependency in $\Sigma_{12}$, $\Sigma_2$ and $\Sigma_{23}$ to
obtain the sets  $\E(\Sigma_{12})$, $\E(\Sigma_2)$ and
$\E(\Sigma_{23})$ of dependencies, respectively.
So we replace
\eqref{eq-1-2}, \eqref{eq-2-2} and \eqref{eq-3-1} by:
\begin{eqnarray*}
B(x) & \to & C(x,f(x)),\\
C(x,y) & \to & D(x,g(x,y)),\\
D(x,y) & \to & E(x,y,h(x,y)),
\end{eqnarray*}
respectively. Then for predicates $C$ and $D$, we introduce functions
$f_C$, $g_C$, $f_D$ and $g_D$, where $f_C$, $g_C$ have the same arity
as $C$,  and where $f_D$, $g_D$ have the same arity as $D$, and we
define
$\sigma_{13}$ as:
\begin{eqnarray*}
\exists f \exists g \exists h \exists f_C \exists g_C \exists f_D
\exists g_D \, \Psi,
\end{eqnarray*}
where $f$, $g$ and $h$ are the Skolem functions introduced above and
$\Psi$ is a conjunction of a set of dependencies defined as
follows. As predicate $C$ cannot be mentioned in $\Psi$, functions
$f_C$ and $g_C$ are used to replace it: the equality $f_C(\bar
a) = g_C(\bar a)$ is used to indicate that $C(\bar a)$ holds. Thus,
the first two conjuncts of $\Psi$ are generated from $\E(\Sigma_{12})$
by replacing $C(\bar x)$ by $f_C(\bar x) = g_C(\bar x)$:
\begin{eqnarray}
\nonumber
A(x,y) & \to & f_C(x,y) = g_C(x,y),\\
\label{eq-e12-2}
B(x) & \to & f_C(x,f(x)) = g_C(x,f(x)).
\end{eqnarray}
Similarly, functions $f_D$ and $g_D$ are used to replace predicate
$D$, and the dependencies in $\E(\Sigma_2)$ are used to generate the
following conjuncts of $\Psi$:
{\small
\begin{align}
\label{eq-e2-1}
\hspace{-7pt}\dom(x) \wedge \dom(y) \wedge \dom(z) \wedge f_C(x,y) = g_C(x,y)
& \wedge f_C(y,z) = g_C(y,z) \to f_C(z,x) = g_C(z,x),\\
\label{eq-e2-2}
\dom(x) \wedge \dom(y) \wedge f_C(x,y) = g_C(x,y) & \to f_D(x,g(x,y))
= g_D(x,g(x,y)),\\
\label{eq-e2-3}
\dom(x) \wedge f_C(x,x) = g_C(x,x) & \to f_D(x,x) = g_D(x,x),\\
\label{eq-e2-4}
\dom(x) \wedge \dom(y) \wedge f_D(x,y) = g_D(x,y) & \to f_D(y,x) = g_D(y,x),
\end{align}
{\normalsize where}} $\dom(\cdot)$ is a formula that defines the domain of the
instances of $\bs_1$, that is, $\dom(x)$ is $\exists y \, A(x,y) \vee
\exists z \, A(z,x) \vee B(x)$. This predicate is included in the
previous dependencies to satisfy the safety condition of st-SO
dependencies, namely, that every variable mentioned in a term has to be
mentioned in a source predicate.
We then use the standard approach for eliminating disjunctions in a
premise (for example, $\varphi_1 \lor \varphi_2 \to \psi$ can be
replaced by the two formulas
$\varphi_1 \to \psi$ and $\varphi_2 \to \psi$).

Notice that if an equality $f_C(a,f(a)) =
g_C(a,f(a))$ can be inferred by using dependency \eqref{eq-e12-2},
then we know that $C(a,f(a))$ holds. Thus, since $D(a,g(a,f(a)))$
can be obtained from $C(a,f(a))$ and the dependency
$C(x,y) \to D(x,g(x,y))$, it should be possible to infer that
$f_D(a,g(a,f(a))) = g_D(a,g(a,f(a)))$ holds by using the fact that
$f_C(a,f(a)) = g_C(a,f(a))$ holds and the dependencies in
$\Psi$. However, if $\dom(f(a))$ does not hold, then $f_C(a,f(a))
= g_C(a,f(a))$ does not satisfy the premise of dependency
\eqref{eq-e2-2} and, therefore, $f_D(a,g(a,f(a))) = g_D(a,g(a,f(a)))$
cannot be inferred by using this dependency. To overcome
this limitation, we also instantiate the above four dependencies with
the terms that appear in the tuples that are generated by repeatedly
applying the formulas in
$\E(\Sigma_2)$. More precisely, it is possible to infer that only
terms of the form $x$ and $f(y)$ need to be considered for the case of
predicate $C$ and, thus, dependencies \eqref{eq-e2-1}, \eqref{eq-e2-2}
and \eqref{eq-e2-3} are instantiated with all the possible combinations
of these types of terms.
For example, the following is one of the
conjuncts of $\Psi$ generated from formula \eqref{eq-e2-1}:
\begin{multline*}
\dom(x) \wedge \dom(y) \wedge \dom(z) \ \wedge \\
f_C(f(x),y) = g_C(f(x),y) \wedge f_C(y,f(z)) = g_C(y,f(z)) \ \to \\
f_C(f(z),f(x)) = g_C(f(z),f(x)),
\end{multline*}
while the following dependency is one of the conjuncts of $\Psi$
generated from formula \eqref{eq-e2-3}:
\begin{multline}
\dom(x) \wedge \dom(y) \wedge f_C(f(x),f(y)) = g_C(f(x),f(y)) \ \wedge\\
f(x) = f(y) \to f_D(f(x),f(y)) = g_D(f(x),f(y)).
\end{multline}
Notice that in the previous dependency we have included the equality
$f(x) = f(y)$, as it can be the case that $f(a) = f(b)$ holds for
distinct elements $a$ and $b$. Similarly, it is possible to infer that
only terms of the form $x$, $f(y)$, $g(x,y)$, $g(x,f(y))$, $g(f(x),y)$
and $g(f(x),f(y))$ need to be considered for the case of predicate
$D$. Thus, dependency \eqref{eq-e2-4} is instantiated with all the
possible combinations
of these types of terms.
For example, the
following is one of the conjuncts of $\Psi$ generated by this process:
\begin{multline*}
\dom(x) \wedge \dom(y) \wedge \dom(z) \wedge
f_D(f(x),g(f(y),z)) = g_D(f(x),g(f(y),z)) \ \to \\
f_D(g(f(y),z),f(x)) = g_D(g(f(y),z),f(x)).
\end{multline*}
Finally, the last conjuncts of $\Psi$ are generated from dependency
$D(x,y) \to E(x,y,h(x,y))$ as above. For example, the following
are two of these conjuncts:
\begin{align*}
& \dom(x) \wedge \dom(y) \wedge f_D(x,y) = g_D(x,y) \to E(x,y,h(x,y)),\\
& \dom(x) \wedge \dom(y) \wedge \dom(z) \wedge f_D(f(x),g(f(y),f(z)))
= g_D(f(x),g(f(y),f(z))) \ \to \\
& \hspace{200pt} E(f(x), g(f(y),f(z)), h(f(x), g(f(y),f(z)))).
\end{align*}
It is important to notice that the weak acyclicity of $\Sigma_2$
guarantees that the above process terminates.
That is, we need only consider terms up to a certain fixed depth of
nesting.
In particular,
in the above example, we
need to consider only
terms where the nesting
depth of functions is at most 2.

\begin{example}
We conclude this section by showing why weak acyclicity is necessary
to guarantee the termination of the above process.  Assume that $\M_{12} =
(\bs_1,
\bs_2, \Sigma_{12}, \Sigma_2)$ and $\M_{23} = (\bs_2, \bs_3,
\Sigma_{23})$, where $\bs_1 = \{A(\cdot,\cdot)\}$, $\bs_2 =
\{B(\cdot,\cdot)\}$, $\bs_3 = \{C(\cdot,\cdot)\}$, $\Sigma_{12}$
consists of the following st-tgd:
\begin{eqnarray*}
A(x,y) & \to & B(x,y),
\end{eqnarray*}
$\Sigma_2$ consists of the following t-tgd:
\begin{eqnarray}
\label{eq-nwa}
B(x,y) & \to & \exists z \, B(y,z),
\end{eqnarray}
and $\Sigma_{23}$ consists of the st-tgd:
\begin{eqnarray*}
B(x,y) & \to & C(x,y).
\end{eqnarray*}
Notice that $\M_{12}$ is not a standard schema mapping, as $\Sigma_2$
is not weakly acyclic.

In order to obtain an st-SO dependency $\sigma_{13}$ that
specifies
the
composition of $\M_{12}$ and $\M_{23}$, the above process first
Skolemizes
each dependency in $\Sigma_{12}$, $\Sigma_2$ and
$\Sigma_{23}$ to obtain the sets $\E(\Sigma_{12})$, $\E(\Sigma_2)$ and
$\E(\Sigma_{23})$ of dependencies, respectively. In particular,
the t-tgd \eqref{eq-nwa} is replaced by the dependency:
\begin{eqnarray}
\label{eq-nwa-s}
B(x,y) & \to & B(y,h(x,y)).
\end{eqnarray}
Then binary functions $f_B$ and $g_B$ are introduced, and
$\sigma_{13}$ is defined as $\exists h \exists f_B \exists g_B \, \Psi$,
where $\Psi$ is a conjunction of a set of dependencies defined as
follows. The first conjunct of $\Psi$ is generated from
$\E(\Sigma_{12})$ by replacing $B(x,y)$ by $f_B(x,y) = g_B(x,y)$:
\begin{eqnarray}
\label{eq-fg}
A(x,y) & \to & f_B(x,y) = g_B(x,y).
\end{eqnarray}
Then functions $f_B$ and $g_B$ are used to eliminate predicate $B$
from $\E(\Sigma_2)$. In particular, the following conjunct is included
in $\Psi$:
\begin{eqnarray}
\label{eq-fg-n}
\dom(x) \wedge \dom(y) \wedge f_B(x,y) = g_B(x,y) & \to &
f_B(y,h(x,y)) = g_B(y,h(x,y)),
\end{eqnarray}
where $\dom(\cdot)$ is a formula that defines the domain of the
instances of $\bs_1$, that is, $\dom(x)$ is $\exists u \, A(x,u) \vee
\exists v \, A(v,x)$. As mentioned above, predicate $\dom(\cdot)$ is
included in the previous dependency to satisfy the safety condition of
st-SO dependencies.

It should be noticed if $(a,b)$ is a tuple in $A$, one can infer
that $f_B(a,b) = g_B(a,b)$ holds by considering dependency
\eqref{eq-fg}, and then one can infer that $f_B(b,h(a,b)) =
g_B(b,h(a,b))$ holds by considering dependency \eqref{eq-fg-n}. By
definition of $\sigma_{13}$, this implies that $B(b,h(a,b))$ holds,
from which one concludes that $B(h(a,b),h(b,h(a,b)))$ also holds (from
dependency \eqref{eq-nwa-s}).  Thus, in this case it should be
possible to infer that
\begin{eqnarray}
\label{eq-eqh}
f_B(h(a,b),h(b,h(a,b))) = g_B(h(a,b),h(b,h(a,b)))
\end{eqnarray}
holds from the dependencies in $\Psi$. However, if $\dom(h(a,b))$ does not
hold, then one cannot infer equality \eqref{eq-eqh} from dependency
\eqref{eq-fg-n} and the fact that $f_B(b,h(a,b)) = g_B(b,h(a,b))$
holds. This forces
one
to instantiate dependency \eqref{eq-fg-n} with
the terms that appear in the tuples that are generated by repeatedly
applying \eqref{eq-nwa-s}. In particular, the following dependency is
included as a conjunct of $\Psi$ to be able to infer \eqref{eq-eqh}
from equality $f_B(b,h(a,b)) = g_B(b,h(a,b))$:
\begin{multline*}
\dom(x) \wedge \dom(y) \wedge f_B(x,h(x,y)) = g_B(x,h(x,y)) \ \to\\
f_B(h(x,y),h(x,h(x,y))) = g_B(h(x,y),h(x,h(x,y))).
\end{multline*}
The previous dependencies are used to deal with the terms where the
nesting depth of functions is at most 2. But given that $\Sigma_2$ is
not weakly acyclic, one also needs to deal with the terms where the
nesting depth of functions is 3, which forces one to include the
following dependency as a conjunct of $\Psi$:
\begin{multline*}
\dom(x) \wedge \dom(y) \wedge
f_B(h(x,y),h(x,h(x,y))) = g_B(h(x,y),h(x,h(x,y))) \ \to \\
f_B(h(x,h(x,y)),h(h(x,y),h(x,h(x,y)))) =
g_B(h(x,h(x,y)),h(h(x,y),h(x,h(x,y)))).
\end{multline*}
It is not difficult to see that the process does not terminate in this
case, as from the preceding dependency one needs to generate a formula
to deal with the terms where the nesting depth of functions is 4,
which in turn has to be used to generate a dependency to deal with
nesting depth 5, and so on.
\end{example}

\subsection{Composability of SO-standard schema mappings}
\label{sec-comp-st-so}
\noindent
The next theorem implies that the composition of SO-standard schema
mappings is an SO-standard schema mapping.
This is the final step we
need to show that the composition of
a finite number of standard schema mappings is given by an SO-standard
schema mapping.

\begin{thm}\label{theo-comp-st-so}
For every pair $\M_{12} = (\bs_1, \bs_2,
\sigma_{12}, \Sigma_2)$ and $\M_{23} = (\bs_2, \bs_3, \sigma_{23},
\Sigma_3)$  of schema mappings, where $\sigma_{12}$, $\sigma_{23}$
are st-SO dependencies
and $\Sigma_i$ $(i = 2,3)$ is the union of a set of t-egds and a
weakly acyclic set of t-tgds, there exists
an st-SO dependency $\sigma_{13}$ such that the schema
mapping $\M_{13} = (\bs_1, \bs_3, \sigma_{13}, \Sigma_3)$
is equivalent to the composition
$\M_{12} \circ \M_{23}$.
\end{thm}
Note that, just as in Theorem~\ref{theo-exp-st-so},
the set $\Sigma_3$ used in $\M_{23}$ is also used in
$\M_{13}$.
Theorem \ref{theo-comp-st-so} was
essentially established in \cite{NBM05} (see Theorems 6 and 9 and the
paragraph after Theorem 10 in \cite{NBM05}), since the class of st-SO
dependencies corresponds to the source-to-target restriction of the
class of $\sk$ dependencies introduced in \cite{NBM05}.

As pointed out in Section \ref{sec-rep}, the previous result is
fundamental to showing that
SO-standard schema mappings can define the composition of standard
schema mappings, since
from the combination of this result with Theorem \ref{theo-exp-st-so}
(and using the simple fact that every standard schema
mapping is an SO-standard schema mapping),
we obtain the following theorem as a consequence.

\begin{thm}\label{theo-comp}
The composition of a finite number of standard
schema mappings is equivalent to an SO-standard schema mapping.
\end{thm}

\subsection{SO-standard schema mappings
are exactly the needed class}
\label{sec-right}
\noindent
We have introduced st-SO dependencies (and SO-standard schema
mappings) because of Theorem~\ref{theo-comp}.
In
this section, we show that SO-standard schema mappings are exactly the
needed class,
since the converse of Theorem~\ref{theo-comp} also
holds. Specifically, we have the following theorem.

\begin{thm}\label{theo-right-so-ed1}
Every SO-standard schema mapping is equivalent to the
composition of a finite number of standard
schema mappings.
\end{thm}
This is proven by showing the following:

\begin{thm}\label{theo-right-so-ed}
Every schema mapping $\M = (\bs, \bt, \sigma_{st}$), where
$\sigma_{st}$ is an st-SO dependency,
is equivalent to the
composition of a finite number of schema mappings, each specified by
st-tgds and t-egds.
\end{thm}
Note that, somewhat surprisingly,
we do not need to make use of a weakly acyclic set of
t-tgds (or any t-tgds at all) in Theorem~\ref{theo-right-so-ed}.
In particular, let $\M_{12}$ and $\M_{23}$ be as in
Proposition~\ref{prop-exp-st-so} (where the
specification of $\M_{12}$ may make use of a weakly acyclic set of
t-tgds).
By Proposition~\ref{prop-exp-st-so}, the composition is given by a
schema mapping $\M_{13}$ specified by an
st-SO dependency; furthermore, by
Theorem~\ref{theo-right-so-ed}, we
know that $\M_{13}$ is
the composition of a finite number of schema mappings, each specified by
st-tgds and t-egds (no t-tgds).
So $\M_{12} \circ \M_{23}$ needs no t-tgds to specify it, even though
$\M_{12}$ makes use of t-tgds.

We now show how Theorem~\ref{theo-right-so-ed1} follows from
Theorem~\ref{theo-right-so-ed}.
Let $\M = (\bs, \bt, \sigma_{st}$, $\Sigma_t)$ be an SO-standard schema
mapping (where $\sigma_{st}$ is an st-SO dependency, and $\Sigma_t$ is
the union of a
set of t-egds and a weakly acyclic set of t-tgds).
Let  $\M' = (\bs, \bt, \sigma_{st}$), where we discard
$\Sigma_t$ from $\M$.
By Theorem~\ref{theo-right-so-ed}, where the role of
$\M$ is played by $\M'$, we know that there are schema mappings $\M_1,
\ldots, \M_k$, each specified by st-tgds and t-egds, such that
$\M' = \M_1 \circ \cdots \circ \M_k$.
Assume that $\M_k =  (\bs ' , \bt, \sigma_{st}, T_k$), with $T_k$
consisting only of t-egds.
Let $\M_k ' = (\bs ' , \bt, \sigma_{st}, T_k \cup \Sigma_t$).
Then $\M_1, \ldots, \M_{k-1}, \M_k'$ are standard schema mappings
($\M_k'$ is a standard schema
mapping,
since its only t-tgds are
those in
$\Sigma_t$).
Since
$(\bs, \bt, \sigma_{st}) = \M_1 \circ \cdots \circ \M_k$,
it follows easily that
$(\bs, \bt, \sigma_{st}$, $\Sigma_t) = \M_1 \circ \cdots \circ
\M_{k-1} \circ \M_k'$.
Thus,
$\M = \M_1 \circ \cdots \circ  \M_{k-1} \circ \M_k'$.

We now demonstrate, by example,
how Theorem~\ref{theo-right-so-ed} is proved (again, it will be clear
how to extend from the example to the general case).
Our proof is an extension of the proof
of Theorem~8.2 in \cite{FKPT05}, that
every SO tgd specifies the composition of a finite number of st-tgd
mappings.

Assume that $\bs = \{ S(\cdot) \}$, $\bt =
\{T(\cdot,\cdot)\}$, $\Sigma_t = \emptyset$ and $\sigma_{st}$ is the
following st-SO dependency:
\begin{eqnarray*}
\exists f \exists g \, [\forall x \, (S(x) \to T(f(g(x)), g(f(x))))
\ \wedge
\forall x \forall y \, (S(x) \wedge S(y) \wedge f(x) = f(y) \to g(x) =
g(y))].
\end{eqnarray*}
Next we construct schema mappings $\M_{12} = (\bs_1, \bs_2,
\Sigma_{12}, \Sigma_2)$, $\M_{23} = (\bs_2, \bs_3, \Sigma_{23},
\Sigma_3)$ and $\M_{34} = (\bs_3, \bs_4, \Sigma_{34})$ such that
(1) $\bs_1 = \bs$, (2) $\bs_4 = \bt$, (3) $\Sigma_{12}$, $\Sigma_{23}$
and $\Sigma_{34}$ are sets of st-tgds, (4) $\Sigma_2$ and $\Sigma_3$
are set of t-egds, and (5) the mapping specified by $\sigma_{st}$ is
equivalent to $\M_{12} \circ \M_{23} \circ \M_{34}$.

Define $\bs_2$ as $\{R_1(\cdot), F_1(\cdot,\cdot), G_1(\cdot,\cdot)\}$
and $\Sigma_{12}$ to consist of the following st-tgds:
\begin{eqnarray*}
S(x) & \to & R_1(x),\\
S(x) & \to & \exists y \, F_1(x,y),\\
S(x) & \to & \exists y \, G_1(x,y).
\end{eqnarray*}
Intuitively, we take $R_1$ to copy $S$, we take $F_1(x, y)$ to encode
$f(x) = y$, and we take $G_1(x, y)$ to encode $g(x) = y$. In
particular, the second and third dependencies have the effect of
guaranteeing that $f(x)$ and $g(x)$ are defined for every element $x$
in $S$, respectively.

Given that $\Sigma_{12}$ cannot guarantee that
$F_1$ and $G_1$ each define a single image for every element in $S$,
we let $\Sigma_2$ consist of the following t-egds:
\begin{eqnarray*}
F_1(x,y) \wedge F_1(x,z) & \to & y = z,\\
G_1(x,y) \wedge G_1(x,z) & \to & y = z,
\end{eqnarray*}
which
guarantee that $F_1$ and $G_1$ encode functions.
In the same way, define $\bs_3$ as $\{R_2(\cdot)$, $F_2(\cdot,\cdot),
G_2(\cdot,\cdot)\}$ and $\Sigma_{23}$ to consist of the following
st-tgds:
\begin{eqnarray*}
R_1(x) & \to & R_2(x),\\
F_1(x,y) & \to & F_2(x,y),\\
G_1(x,y) & \to & G_2(x,y),\\
F_1(x,y) & \to & \exists z \, G_2(y,z),\\
G_1(x,y) & \to & \exists z \, F_2(y,z).
\end{eqnarray*}
Intuitively, we take $R_2$ to copy $R_1$, $F_2$ to copy $F_1$, and
$G_2$ to copy $G_1$, and we include the fourth dependency to guarantee
that $g(y)$ is defined for all $y$ in the range of $f$, and we include
the fifth dependency to guarantee that $f(y)$ is defined for all $y$
in the range of $g$. Also as in the previous case, we include in
$\Sigma_3$ two t-egds that guarantee that $F_2$ and $G_2$ are indeed
functions:
\begin{eqnarray*}
F_2(x,y) \wedge F_2(x,z) & \to & y = z,\\
G_2(x,y) \wedge G_2(x,z) & \to & y = z.
\end{eqnarray*}
Given that at this point, we have predicates that
encode the values of all the terms that are used in $\sigma_{st}$, we
also include in $\Sigma_3$ dependencies that encode the conjuncts of
$\sigma_{st}$ of the form $\forall \bar x \, (\varphi \to t_1 =
t_2)$.
Thus,
in this case we include in $\Sigma_3$ the
following t-egd that encodes the conjunct $\forall x \forall y \,
(S(x) \wedge S(y) \wedge f(x) = f(y) \to g(x) = g(y))$:
\begin{eqnarray*}
R_2(x) \wedge R_2(y) \wedge F_2(x,z) \wedge
F_2(y,z) \wedge G_2(x,u) \wedge G_2(y,v) \to u = v.
\end{eqnarray*}
Finally, we use $R_2$, $F_2$ and $G_2$ to encode the remaining
conjuncts of $\sigma_{st}$, which indicate how to populate the target
relations of $\sigma_{st}$. Thus, we define $\Sigma_{34}$ to consist
of the following st-tgd:
\begin{eqnarray*}
R_2(x) \wedge G_2(x,y_1) \wedge F_2(y_1,y_2) \wedge
F_2(x,z_1) \wedge G_2(z_1,z_2) \to T(y_2,z_2).
\end{eqnarray*}
This concludes the demonstration by example of how to prove
Theorem~\ref{theo-right-so-ed}.
This demonstration gives, as a special
case
(when the st-SO
dependency is unnested) the following lemma  (where we note also the
number of schema mappings that are composed).

\begin{lem}\label{lem-onlytwo}
Every schema mapping $\M = (\bs, \bt, \sigma_{st}$), where
$\sigma_{st}$ is an unnested st-SO dependency,
is equivalent to the
composition of two schema mappings, each specified by
st-tgds and t-egds.
\end{lem}

\noindent
We note that
Theorem~\ref{theo-right-so-ed} follows immediately from
Lemma~\ref{lem-onlytwo} and the fact, as we show later, that
every st-SO dependency is equivalent to an unnested st-SO dependency,
Therefore, we really needed to prove only
Lemma~\ref{lem-onlytwo} (the unnested case) rather than the general case
that we dealt with in proving
Theorem~\ref{theo-right-so-ed}.

\section{Collapsing Results: Nesting is Not Necessary}
\label{sec-collap}

Recall that
we say that an st-SO dependency or SO tgd is {\em unnested\/} if its
depth of nesting is at most 1.
Thus, an unnested  st-SO dependency or SO tgd can contain terms like
$f(x)$, but not terms like
$f(g(x))$.
In this section, we present collapsing results about the depth of
nesting of function symbols in st-SO dependencies and SO tgds.
Specifically, we prove the following two theorems.

\begin{thm}\label{th-st-tgds-t-egds}
Every st-SO dependency is equivalent to an unnested st-SO dependency.
\end{thm}

\begin{thm}\label{th-st-tgds}
Every SO tgd is equivalent to an unnested SO tgd
\end{thm}
These two results, especially the second one, are the most technically
difficult results in the paper.
Both results are surprising, since the ``obvious'' way to try to denest,
which we now describe, does not work.
Consider for example the SO tgd
\begin{equation}\label{eq:first}
\exists f  \exists g \forall x \forall y (P(x,y) \land (f(g(x)) = y) \to
Q(x, y))
\end{equation}
The ``obvious'' way to denest~(\ref{eq:first})  is to introduce a new
variable $z$ and rewrite~(\ref{eq:first}) as
\begin{eqnarray}\label{eq:second}
\exists f  \exists g \forall x \forall y \forall z (P(x,y) \land
(g(x) = z) \land (f(z) = y) \to Q(x, y))
\end{eqnarray}
However, the formula~(\ref{eq:second}) is not an SO tgd, since it
violates
the safety condition (because the variable $z$ does not appear in
$P(x,y)$, the
only relational atomic formula in the premise of~(\ref{eq:second})).

It should be mentioned that in \cite{LS08}, Libkin and Sirangelo
introduce the second-order language of Skolemized STDs (SkSTDs), and
study some of its fundamental properties. In particular, it is shown
in \cite{LS08} that this language is closed under composition if the
premises of SkSTDs are restricted to be conjunctive
queries. Interestingly, this fragment of SkSTDs is similar to the
language of SO tgds but does not allow nesting of functions, which may
lead one to think that Theorem \ref{th-st-tgds} can be deduced from
the results in
\cite{LS08}. However, no safety condition is imposed on the premises
of SkSTDs in \cite{LS08} and, thus, nesting of functions is not needed
in this language as it can be eliminated in the ``obvious'' way shown
above. In fact, dependency \eqref{eq:second} is a valid constraint
according to \cite{LS08}.

Before giving the proofs of Theorems \ref{th-st-tgds-t-egds} and
\ref{th-st-tgds}, we present and discuss some corollaries of these
theorems.
\begin{corollary}\label{cor-unnestedok}
The composition of a finite number of st-tgd mappings can be specified
by an unnested SO tgd.
\end{corollary}
This is a strengthening of the result (Theorem 8.1 in \cite{FKPT05})
that the composition of a finite number of st-tgd mappings can be
specified by an SO tgd (thus,
Corollary~\ref{cor-unnestedok} says that we can replace
``SO tgd'' in Theorem 8.1 in \cite{FKPT05}
by ``unnested
SO tgd'').
Corollary~\ref{cor-unnestedok} follows immediately from the result we
just cited (Theorem
8.1 in
\cite{FKPT05}) and our Theorem~\ref{th-st-tgds}.
It was not even known before that
the composition of two st-tgd mappings can be specified by an unnested
SO tgd.
Thus, although it was shown in \cite{FKPT05} that each unnested SO tgd
specifies the composition of some pair of st-tgd mappings, the converse
was not shown.
In fact, for the composition of two st-tgd mappings, the composition
construction in
\cite{FKPT05} produces an SO tgd whose depth of nesting can be
2, not 1.

We feel that nested dependencies are difficult to understand
(just think about an equality like $f(g(x),h(f(x,y))) =
g(f(x,h(y)))$),
and probably also difficult to use in practice. On the other hand,
unnested dependencies seem to be more natural and readable.
For example,
it is easy to see that the ``nested mappings'' in \cite{FHHMPP06}
can be expressed by unnested SO tgds.
Corollary~\ref{cor-unnestedok} tells us that unnested SO tgds are also
expressive enough to specify the composition of an arbitrary number of
st-tgd mappings.

Theorem~\ref{th-st-tgds} has as another corollary the  following
collapsing result
about the number of compositions of st-tgd mappings.

\begin{corollary}\label{cor-compose}
The composition of a finite number of st-tgd mappings
is equivalent to the composition of two st-tgd mappings.
\end{corollary}
This follows from Corollary~\ref{cor-unnestedok} and the fact
(which is a special case of Theorem 8.4 of \cite{FKPT05}) that
a schema mapping specified by an  unnested SO tgd is
equivalent
to the composition of two st-tgd mappings.


The next two corollaries follow from Theorem~\ref{th-st-tgds-t-egds}
just as
Corollaries~\ref{cor-unnestedok} and~\ref{cor-compose}
follow from Theorem~\ref{th-st-tgds}.

\begin{corollary}\label{cor-unnestedok1}
The composition of a finite number of standard schema mappings
can be
specified by an unnested st-SO dependency, along with t-egds and a
weakly acyclic set of t-tgds.
\end{corollary}

\begin{corollary}\label{cor-compose1}
The composition of a finite number of standard schema  mappings
is equivalent to the composition of two standard schema mappings.
\end{corollary}
In fact, it follows from Corollary~\ref{cor-unnestedok1} and
Lemma~\ref{lem-onlytwo} that we can slightly strengthen
Corollary~\ref{cor-compose1} as follows.

\begin{corollary}\label{cor-compose2}
The composition of a finite number of standard schema  mappings
is equivalent to the composition $\M_1 \circ \M_2$
of two standard schema mappings $\M_1$ and $\M_2$, where the target
constraints of $\M_1$ are only t-egds (no t-tgds).
\end{corollary}
Corollary~\ref{cor-compose1} has a direct, almost trivial proof
that does not use our heavy machinery,
as we now show.
Let $\M_{12}$, $\M_{23}$, $\ldots$, $\M_{k-1\; k}$ be standard schema
mappings.
Define $\M_{12}'$ to have source schema the same as $\M_{12}$, target
schema equal to
the union of the target schemas of $\M_{12}$, $\ldots$, $\M_{k-2\;
k-1}$,
and constraints equal to the union of the constraints of $\M_{12}$,
$\ldots$,
$\M_{k-2\; k-1}$. Because all of the schemas are disjoint, it is easy to
see that
$\M_{12}'$ is a standard schema mapping (note that the st-tgds of
$\M_{23}$, $\ldots$, $\M_{k-2\; k-1}$ are now being treated as t-tgds of
$\M_{12}'$).  Then it is clear that
$$\M_{12} \circ \M_{23} \circ \ldots \circ \M_{k-1\; k} = \M_{12}'
\circ
\M_{k-1\; k} . $$
In contrast to Corollary~\ref{cor-compose1}, the reason that
Corollary~\ref{cor-compose} is quite unexpected is that there is no
obvious
way to deal with all of the st-tgds in the intermediate schema mappings.

Corollary~\ref{cor-unnestedok1}, unlike  Corollary~\ref{cor-compose1},
does not seem to have a simple
direct proof that avoids the machinery of
Theorem~\ref{th-st-tgds-t-egds}.
This is because our construction of the composition of two standard
schema mappings produces an st-SO dependency whose nesting depth can be
arbitrarily large.


%
%
%
%
%
%

Based on our collapsing results, there are
two alternative ways to deal with the composition of
multiple st-tgd mappings.
First,
by Corollary~\ref{cor-unnestedok},
we can replace this
composition by a single schema mapping, specified by an unnested SO tgd.
Second, by Corollary~\ref{cor-compose},
we can replace the composition by
the composition of only two st-tgd mappings.
Similarly, by using Corollaries~\ref{cor-unnestedok1} and~\ref{cor-compose1},
we have two alternative ways to deal with
the composition of
a large number of standard schema mappings.

We now provide the proofs of Theorems \ref{th-st-tgds-t-egds} and
\ref{th-st-tgds}.

\medskip

\aproof{Theorem \ref{th-st-tgds-t-egds}}{In this proof, we use the
following terminology. Given a term $t$,
recursively define the set of non-atomic sub-terms of $t$, denoted by
$\nat(t)$, and the list of variables of $t$, denoted by $\lv(t)$, as
follows: (1) if $t = x$, where $x$ is a variable, then $\nat(t) =
\emptyset$ and $\lv(t) = [x]$; (2) if $t = f(t_1, t_2, \ldots, t_n)$,
where $t_1$, $t_2$, $\ldots$, $t_n$ are terms, then
\begin{eqnarray*}
\nat(t) & = &  \{f(t_1, \ldots, t_n)\} \cup \bigcup_{i=1}^n \nat(t_i)
\end{eqnarray*}
and $\lv(t) =  \lv[t_1] \cdot \lv[t_2] \cdot \ldots \cdot \lv[t_n]$,
where $L_1 \cdot  L_2$ is the result of appending $L_2$ to $L_1$. For
example, if $t =
f(x,h(z,y,g(x)))$, then $\nat(t) = \{f(x,h(z,y,g(x)))$, $h(z,y,g(x)),
g(x)\}$ and $\lv(f(x,h(z,y,g(x)))) = [x,z,y,x]$. Moreover, consider a
term replacement $\sig(\cdot)$ that describes the {\em skeleton} of a
term. For example, if $t = f(x,g(y))$, then $\sig(t)$ is
$f(\_,g(\_))$, as this shows what are the functions that have been
included in $t$ and how they have been nested in this
term.%
\footnote{It should be noticed that a similar term replacement was
used in \cite{DGL00} to eliminate function expressions from a logic
program.  However, the term replacement used in \cite{DGL00}
considers only
terms without nesting of function
symbols.}
More
precisely, recursively define $\sig(\cdot)$ as follows: (1) $\sig(x) =
\_$ for every variable $x$, and (2) $\sig(f(t_1, \ldots, t_n)) =
f(\sig(t_1), \ldots, \sig(t_n))$, for every $n$-ary symbol $f$ and
terms $t_1$, $\ldots$, $t_n$. For example, $\sig(f(x,h(z,y,g(x)))) =
f(\_,h(\_,\_,g(\_)))$.  Finally, by considering function
$\sig(\cdot)$, define a second term replacement $\xi(\cdot)$ as
follows:
\begin{equation}
\label{eq-xi}
\xi(t) =
\begin{cases}
t & \text{if } t \text{ is a variable}\\
\xi_{\sig(t)}(x_1, \ldots, x_n) & \text{if } t \text{ is a non-atomic
term and } \lv(t) = [x_1, \ldots, x_n]
\end{cases}
\end{equation}
For example, if $t = f(x,h(z,y,g(x)))$, then $\xi(t) =
\xi_{f(\_,h(\_,\_,g(\_)))}(x,z,y,x)$.

Given an st-SO dependency $\sigma$ from a source schema $\bs$ to a
target schema $\bt$, next we show how to construct an
unnested st-SO dependency $\sigma^\star$ from $\bs$ to $\bt$ such that
$\sigma$ and $\sigma^\star$ are equivalent.
Let
$t_1$,
$\ldots$,
$t_\ell$ be the
non-atomic terms $t$ such that there exists an atomic formula
mentioned in $\sigma$ of the form either $t = t'$ or $t' = t$ or
$R(t_1, \ldots, t_{i-i}, t, t_{i+1}, \ldots, t_k)$ (where $k$ is the
arity of $R$ and $i \in \{1, \ldots, k\}$),
let $\mathcal{H}(\sigma) = \{t_1, \ldots, t_\ell\}$,
and let $\mathcal{ST}(\sigma)$ be the set of all non-atomic sub-terms of $t_1$,
$\ldots$, $t_\ell$, that is, $\mathcal{ST}(\sigma) = \bigcup_{i=1}^\ell
\nat(t_i)$.
Then define $\Xi(\sigma)$ as the following set of function symbols:
\begin{eqnarray*}
\Xi(\sigma) & = & \{ \xi_{\sig(t)} \mid t \in \mathcal{ST}(\sigma)\}.
\end{eqnarray*}
For example, if $\sigma$ is the following st-SO dependency:
\begin{eqnarray*}
\exists f \exists g \, [\forall x \forall y \, S(x,y) \to
T(x,f(x,g(y)),g(f(y,g(x))))],
\end{eqnarray*}
then
$\mathcal{H}(\sigma)$ is the set
$\{f(x,g(y)),g(f(y,g(x)))\}$. Thus, given that $\nat(f(x,g(y))) =
\{f(x,g(y)), g(y)\}$ and $\nat(g(f(y,g(x)))) = \{g(f(y,g(x))), f(y,g(x)), g(x)\}$,
we have that
$\mathcal{ST}(\sigma) = \{f(x,g(y)), g(y), g(f(y,g(x))), f(y,g(x)), g(x)\}$, and
\begin{eqnarray*}
\Xi(\sigma) & = & \{\xi_{f(\_,g(\_))},\ \xi_{g(\_)},\
\xi_{g(f(\_,g(\_)))}\},
\end{eqnarray*}
Note that two members of $\mathcal{ST}(\sigma)$, namely $g(y)$ and $g(x)$,
have the same skeleton $g(\_)$, as do
$f(x,g(y))$ and $f(y,g(x))$, which have the same skeleton
$f(\_,g(\_))$.
The set $\Xi(\sigma)$ plays a fundamental role in the definition of
the
st-SO dependency $\sigma^\star$. More precisely, assume
that $\Xi(\sigma) = \{\chi_1,
\chi_2, \ldots, \chi_m\}$. Then $\sigma^\star$ is defined as:
\begin{eqnarray*}
\exists \chi_1 \exists \chi_2 \cdots \exists \chi_m \, \Psi,
\end{eqnarray*}
where $\Psi$ is defined as the conjunction of the following
dependencies.
For every conjunct $\forall \bar x \, (\varphi \to
\psi)$ of $\sigma$,
the
st-SO dependency $\sigma^\star$ contains a conjunct
$\forall \bar x (\varphi' \to \psi')$, where (1) $\varphi'$ is
obtained from $\varphi$ by replacing every non-atomic term $t \in
\mathcal{ST}(\sigma)$ by $\xi(t)$, and (2) $\psi'$ is obtained from
$\psi$ by replacing every non-atomic term $t \in \mathcal{ST}(\sigma)$ by
$\xi(t)$. Furthermore, for every pair of (non-necessarily distinct)
terms $t, t' \in \mathcal{ST}(\sigma)$, if $t = f(t_1,
\ldots, t_n)$ and $t' = f(t'_1, \ldots, t'_n)$, the
following procedure is executed to obtain a set $\Gamma_{t,t'}$ of
dependencies,
and then $\bigwedge \Gamma_{t,t'}$ is included as a conjunct
of $\sigma^\star$. First, replace each occurrence of a variable in $t$
by a fresh variable to obtain a term $s = f(s_1,
\ldots, s_n)$, and replace each occurrence of a variable in $t'$ by a
fresh variable to obtain a term $s' = f(s'_1, \ldots, s'_n)$ (in
particular, $s$ and $s'$ have no variables in common). Assume that
$x_1$, $\ldots$, $x_p$ is the set of variables mentioned in $s$,
$s'$. Second, as in the proof of Proposition
\ref{prop-exp-st-so}, let $\dom(\cdot)$ be a formula that defines the
domain of the instances of $\bs$. Finally, let $\Gamma_{t,t'}$ be a
set of
dependencies
obtained from
the
dependency:
\begin{eqnarray}
\label{eq-disj}
\forall x_1 \cdots \forall x_p \, \bigg[\bigg(\bigwedge_{i=1}^p
\dom(x_i)\bigg) \wedge \bigg(\bigwedge_{j=1}^n \xi(s_i) =
\xi(s'_i)\bigg) & \to & \xi(s) = \xi(s')\bigg]
\end{eqnarray}
by repeatedly using the equivalences:
\begin{eqnarray}
((\alpha \vee \beta) \wedge \gamma) \to \delta & \equiv & ((\alpha
\wedge \gamma) \to \delta) \wedge ((\beta \wedge \gamma) \to
\delta), \label{eq:firsta}\\
(\exists x \, \alpha) \to \beta & \equiv & \forall x \, (\alpha \to \beta)
\ \ \ \ \ \text{if } x \text{ is not mentioned in } \beta,
\label{eq:seconda}
\end{eqnarray}
until all the
disjunctions and existential quantifications in the left-hand side of
\eqref{eq-disj} have
been eliminated.
\begin{example}
Let us give the intuition behind the definition of $\sigma^\star$
through an example. Assume that $\sigma$ is the following st-SO
dependency:
\begin{eqnarray}\label{eq:iseg}
\exists f \exists g \, \bigg[\forall x \, \bigg(A(x) \to (T(x,f(g(x)))
\wedge f(x) = g(x))\bigg) \ \wedge \ \forall x \, \bigg(B(x) \to
U(x,g(f(x)))\bigg)\bigg].
\end{eqnarray}
Then we have that $\mathcal{ST}(\sigma) = \{f(g(x)), g(x), g(f(x)), f(x)\}$ and
$\Xi(\sigma) = \{\xi_{f(g(\_))}, \xi_{g(\_)}, \xi_{g(f(\_))}$,
$\xi_{f(\_)}\}$. Intuitively, $\xi_{f(\_)}$ and $\xi_{g(\_)}$
are used to represent functions $f$ and $g$, respectively, and
$\xi_{f(g(\_))}$ and $\xi_{g(f(\_))}$ are used to represent the
composition functions $(g \circ f)$ and $(f \circ g)$,
respectively, thus eliminating the nesting of functions from
$\sigma$. More precisely,
the
st-SO dependency $\sigma^\star$ is defined as:
\begin{eqnarray*}
\exists \xi_{f(g(\_))} \exists \xi_{g(\_)} \exists \xi_{g(f(\_))}
\exists \xi_{f(\_)} \, \Psi,
\end{eqnarray*}
where $\Psi$ is defined as the conjunction of the following
dependencies.
First, given that $\forall x \, (A(x) \to (T(x,f(g(x)))
\wedge f(x) = g(x)))$ and $\forall x \, (B(x) \to U(x,g(f(x))))$ are
the conjuncts of $\sigma$, the following
dependencies
are conjuncts of $\Psi$:
\begin{eqnarray*}
&&\forall x \, (A(x) \to  (T(x,\xi_{f(g(\_))}(x)) \wedge \xi_{f(\_)}(x)
= \xi_{g(\_)}(x))),\\
&&\forall x \, (B(x) \to  U(x,\xi_{g(f(\_))}(x))).
\end{eqnarray*}
Furthermore, for every pair $t, t'$ of (non-necessarily distinct)
terms from $\mathcal{ST}(\sigma)$, if either $t = f(t_1)$ and $t' =
f(t'_1)$ or $t = g(t_1)$ and $t' = g(t'_1)$, then the following
conjuncts are included in $\Psi$. Assume that $t = t' =
f(g(x))$. First, each occurrence of a variable in these terms is
replaced by a fresh variable, generating the terms $s = f(g(u))$ and
$s' = f(g(v))$. Second, given
that the source schema consists of the unary predicates $A$ and $B$,
formula $\dom(x)$ is defined as $A(x) \vee B(x)$ (that is, $\dom(x)$
holds if $x$ is in the domain of a source instance). Finally, assuming
that $s_1 = g(u)$ and $s'_1 = g(v)$, let $\alpha$ be the following
dependency:
\begin{eqnarray*}
\forall u \forall v \, \bigg[\dom(u) \wedge \dom(v) \wedge \xi(s_1) =
\xi(s'_1) \ \to \ \xi(s) = \xi(s')\bigg],
\end{eqnarray*}
that is, $\alpha$ is:
\begin{eqnarray*}
\forall u \forall v \, \bigg[(A(u) \vee B(u)) \wedge
(A(v) \vee B(v)) \wedge \xi_{g(\_)}(u) = \xi_{g(\_)}(v) \ \to \
\xi_{f(g(\_))}(u) = \xi_{f(g(\_))}(v) \bigg].
\end{eqnarray*}
Then the set $\Gamma_{t,t'}$ consists of the following
dependencies:
\begin{eqnarray*}
\forall u \forall v \, \bigg[A(u) \wedge A(v) \wedge \xi_{g(\_)}(u) =
\xi_{g(\_)}(v) \ \to \ \xi_{f(g(\_))}(u) = \xi_{f(g(\_))}(v) \bigg],\\
\forall u \forall v \, \bigg[A(u) \wedge B(v) \wedge \xi_{g(\_)}(u) =
\xi_{g(\_)}(v) \ \to \ \xi_{f(g(\_))}(u) = \xi_{f(g(\_))}(v) \bigg],\\
\forall u \forall v \, \bigg[B(u) \wedge A(v) \wedge \xi_{g(\_)}(u) =
\xi_{g(\_)}(v) \ \to \ \xi_{f(g(\_))}(u) = \xi_{f(g(\_))}(v) \bigg],\\
\forall u \forall v \, \bigg[B(u) \wedge B(v) \wedge \xi_{g(\_)}(u) =
\xi_{g(\_)}(v) \ \to \ \xi_{f(g(\_))}(u) = \xi_{f(g(\_))}(v) \bigg],
\end{eqnarray*}
and each one of these four
dependencies
is a conjunct of $\sigma^\star$. It is
important to notice that these
dependencies
make explicit some
properties that are implicit in $\sigma$. Given that $f$ and $g$ are
function symbols in $\sigma$, we know that if $g(u) = g(v)$, then
$f(g(u)) = f(g(v))$. But this property does not immediately hold for
$\xi_{f(g(\_))}$ and $\xi_{g(\_)}$ and, thus, we have to include the
above four conjuncts into $\sigma^\star$ to enforce it. It should also
be noticed that
the formula
$\dom(\cdot)$ is included in the previous
dependencies
to satisfy the safety condition of st-SO
dependencies, namely that every variable mentioned in a term has to be
mentioned in a source predicate.

To give more intuition about the definition of dependency
$\sigma^\star$, we also consider the case $t = f(g(x))$ and $t' =
f(x)$. As above, we start by
replacing each occurrence of a variable in these terms by a fresh
variable, generating the terms $s = f(s_1)$ and $s' = f(s'_1)$, where
$s_1 = g(u)$ and $s'_1 = v$. Then we define a dependency $\beta$ as:
\begin{eqnarray*}
\beta & = & \forall u \forall v \, \bigg[(A(u) \vee B(u)) \wedge
(A(v) \vee B(v)) \wedge \xi_{g(\_)}(u) = v \ \to \
\xi_{f(g(\_))}(u) = \xi_{f(\_)}(v) \bigg],
\end{eqnarray*}
and, therefore, in this case the set $\Gamma_{t,t'}$ consists of the
following dependencies:
\begin{eqnarray*}
\forall u \forall v \, \bigg[A(u) \wedge A(v) \wedge \xi_{g(\_)}(u) =
v \ \to \ \xi_{f(g(\_))}(u) = \xi_{f(\_)}(v) \bigg],\\
\forall u \forall v \, \bigg[A(u) \wedge B(v) \wedge \xi_{g(\_)}(u) =
v \ \to \ \xi_{f(g(\_))}(u) = \xi_{f(\_)}(v) \bigg],\\
\forall u \forall v \, \bigg[B(u) \wedge A(v) \wedge \xi_{g(\_)}(u) =
v \ \to \ \xi_{f(g(\_))}(u) = \xi_{f(\_)}(v) \bigg],\\
\forall u \forall v \, \bigg[B(u) \wedge B(v) \wedge \xi_{g(\_)}(u) =
v \ \to \ \xi_{f(g(\_))}(u) = \xi_{f(\_)}(v) \bigg].
\end{eqnarray*}
As in the previous case, each one of the dependencies of $\Gamma_{t,t'}$
is a conjunct of $\sigma^\star$. It is important to notice that these
dependencies make explicit the fact that in $\sigma$, if $g(u) = v$,
then $f(g(u)) = f(v)$.

For the st-SO dependency~(\ref{eq:iseg}), we took $\dom(u)$ to be
$A(u) \lor B(u)$.
We then made use of~(\ref{eq:firsta}) to eliminate the disjunction in
$A(u) \lor B(u)$.
If the left-hand side of the first conjunct of~(\ref{eq:iseg}) had been
$P(x,y)$ instead of $A(x)$,
we would have taken $\dom(u)$ to be
$\exists w P(u,w) \lor \exists w P(w,u) \lor B(u)$.
We then would have made use not only
of~(\ref{eq:firsta}), but also~(\ref{eq:seconda}), to eliminate the
disjunctions and
existential quantifiers in
$\exists w P(u,w) \lor \exists w P(w,u) \lor B(u)$.

\end{example}
We now prove
that $\sigma \Leftrightarrow \sigma^\star$, that is,
that $\sigma$ and $\sigma^\star$ are equivalent.

($\Rightarrow$) If $(I,J) \models \sigma$, then it is straightforward
to prove that $(I,J) \models \sigma^\star$ (the interpretation of each
function symbol in $\Xi(\sigma)$ is defined from the corresponding
composition of the interpretations of the function symbols from
$\sigma$).

($\Leftarrow$) Assume that $\Xi(\sigma) = \{\chi_1, \ldots, \chi_m\}$
and that $(I,J) \models \sigma^\star$ with the instantiations
$\chi^0_1$, $\ldots$, $\chi^0_m$ of $\chi_1$, $\ldots$,
$\chi_m$. Moreover, assume that $f_1$, $\ldots$, $f_\ell$ are the
function symbols mentioned in $\sigma$. To show that $(I,J) \models
\sigma$, we first need to define from $\chi^0_1$, $\ldots$, $\chi^0_m$
the instantiations $f^0_1$, $\ldots$, $f^0_\ell$ of function symbols
$f_1$, $\ldots$, $f_\ell$, and then we have to show that $(I,J)$
satisfies all the conjuncts of $\sigma$ with these instantiations.

Given that $(I,J) \models \sigma^\star$, we
have by definition of satisfaction for st-SO dependencies that
there exists a countably infinite universe%
\footnote{As noted earlier, the universe can even be taken to
be finite, but we do not need this.}
$U$
such that (1) $U$ is the union
of $\dom(I) \cup \dom(J)$ and a set of nulls, and
(2) $(U;I,J)$
satisfies $\sigma^\star$ in the standard second-order logic sense. Assume
that $\bot$ is a fresh null value ($\bot \not\in U$) and that the arity of
function symbol $f_i$ is $k_i$ ($1 \leq i \leq \ell$). Then the domain of
each one of the functions $f^0_1$, $\ldots$, $f^0_\ell$ is defined to be
$U \cup \{ \bot \}$, and for every $(a_1, \ldots, a_{k_i}) \in (U \cup
\{ \bot  \})^{k_i}$, we define $f^0_i(a_1, \ldots, a_{k_i})$ as
follows. If there exist $f_i(t_1, \ldots, t_{k_i}) \in \mathcal{ST}(\sigma)$ and tuples $\bar b_1$, $\ldots$, $\bar b_{k_i}$ such that
for every $i \in \{1, \ldots, k_i\}$:
\begin{iteMize}{$\bullet$}\itemsep=1pt
\item $t_i$ is a variable, $\bar b_i = (a_i)$ and $a_i \in  \dom(I)$; or

\item $t_i$ is a non-atomic term, $a_i = \xi^0_{\sig(t_i)}(\bar b_i)$
and $\bar b_i \subseteq \dom(I)$ (that is, every element mentioned in
$\bar b_i$ is in $\dom(I)$);
\end{iteMize}
then $f^0_i(a_1, \ldots, a_{k_i})$ is defined as $\xi^0_{\sig(f_i(t_1,
\ldots, t_{k_i}))}(\bar b_1, \ldots, \bar b_{k_i})$. Otherwise,
$f^0_i(a_1, \ldots, a_{k_i})$ is defined as $\bot$ (in particular, if
$a_i = \bot$ for some $i \in \{1, \ldots, k_i\}$, then $f^0_i(a_1,
\ldots, a_{k_i}) = \bot$).

Before showing that all the conjuncts of $\sigma$ are satisfied by
$(I,J)$ under the instantiations $f^0_1$, $\ldots$, $f^0_\ell$ of
function symbols $f_1$, $\ldots$, $f_\ell$, we need to show that these
functions are well defined. That is, we have to show that if by using
the above definition, one has different ways of assigning a value to
$f^0_i(a_1, \ldots, a_{k_i})$, then all these ways assign the same value
to $f^0_i(a_1, \ldots, a_{k_i})$. In order to prove this, we need to
consider several cases. In this proof, we
consider only
one of these
cases, as the other ones can be handled in the same way. Assume that
for $i \in \{1, \ldots, \ell\}$ and elements $a_1$, $\ldots$,
$a_{k_i}$ from $U \cup \{\bot\}$, it holds that (1) $f_i(t_1, \ldots,
t_{k_i}) \in \mathcal{ST}(\sigma)$, (2) $a_i =
\xi^0_{\sig(t_i)}(\bar b_i)$ and $\bar b_i \subseteq \dom(I)$, for every $i
\in \{1, \ldots, k_i\}$, (3) $f_i(s_1, \ldots, s_{k_i}) \in
\mathcal{ST}(\sigma)$, and (4) $a_i = \xi^0_{\sig(s_i)}(\bar c_i)$ and $\bar
c_i \subseteq \dom(I)$, for every $i \in \{1, \ldots, k_i\}$. Then we
have to prove that:
\begin{eqnarray}
\label{eq-wd}
\xi^0_{\sig(f_i(t_1, \ldots, t_{k_i}))}(\bar b_1, \ldots, \bar b_{k_i}) & = &
\xi^0_{\sig(f_i(s_1, \ldots, s_{k_i}))}(\bar c_1, \ldots, \bar c_{k_i}).
\end{eqnarray}
Given a tuple $\bar x = (x_1, \ldots, x_p)$ of variables, let
$\dom(\bar x)$ be a shorthand for $\dom(x_1) \wedge \cdots \wedge
\dom(x_p)$. By definition of $\sigma^\star$ and the fact that
$(I,J) \models \sigma^\star$, we have that $(I,J)$ satisfies the
following instantiated dependency:
\begin{multline*}
\bigg(\bigwedge_{i=1}^{k_i} \dom(\bar b_i)\bigg) \wedge
\bigg(\bigwedge_{i=1}^{k_i} \dom(\bar c_i)\bigg) \wedge
\bigg(\bigwedge_{i=1}^{k_i} \xi_{\sig(t_i)}(\bar b_i) =
\xi_{\sig(s_i)}(\bar c_i)\bigg) \ \to\\ \xi_{\sig(f_i(t_1, \ldots,
t_{k_i}))}(\bar b_1, \ldots, \bar b_{k_i})  =   \xi_{\sig(f_i(s_1,
\ldots, s_{k_i}))}(\bar c_1, \ldots, \bar c_{k_i}).
\end{multline*}
Thus, we conclude that \eqref{eq-wd}
holds, since
for every $i \in \{1,
\ldots, k_i\}$, it holds that $\xi^0_{\sig(t_i)}(\bar b_i) = a_i =
\xi^0_{\sig(s_i)}(\bar c_i)$,  $\bar b_i \subseteq \dom(I)$ and  $\bar
c_i \subseteq \dom(I)$.

Now we move to the proof that all the conjuncts of $\sigma$ are
satisfied by $(I,J)$ under the instantiations $f^0_1$, $\ldots$,
$f^0_\ell$ of function symbols $f_1$, $\ldots$, $f_\ell$. In this
proof, we need the following claim, where we use the following
terminology. Given a non-atomic term $t = f_i(t_1, \ldots, t_{k_i})$
based on variables $x_1$, $\ldots$, $x_k$ and function symbols $f_1$,
$\ldots$, $f_\ell$, and
given
a variable substitution $\rho : \{x_1, \ldots,
x_k\} \to (U \cup
\{ \bot \})$, the evaluation of $\rho$ over $t$ is recursively defined
as $\rho(t) = f^0_i(\rho(t_1), \ldots,
\rho(t_{k_i}))$.

\begin{claim}\label{cla-ind}
Let $t \in \mathcal{ST}(\sigma)$ such that $\lv(t) = [x_1, \ldots,
x_k]$. Then for every variable substitution $\rho : \{x_1, \ldots,
x_k\} \to \dom(I)$, it holds that $\rho(t) = \xi^0_{\sig(t)}(\rho(x_1),
\ldots, \rho(x_k))$.
\end{claim}

\proof{By induction on the depth of nesting of functions in $t$.
\begin{iteMize}{$\bullet$}\itemsep=3pt
\item Base case: If the depth of nesting of functions in $t$ is 1,
then $t = f_i(x_1, \ldots, x_k)$ and $k = k_i$. Then we have that
$\rho(f_i(x_1, \ldots, x_k)) = f^0_i(\rho(x_1), \ldots, \rho(x_k))$. But
given that $\rho(x_j) \in
\dom(I)$ for every $j \in \{1, \ldots, k\}$, we have by definition of
$f^0_i$
that $f^0_i(\rho(x_1), \ldots, \rho(x_k)) = \xi^0_{\sig(f_i(x_1,
\ldots, x_k))}(\rho(x_1), \ldots, \rho(x_k))$. Thus, we conclude that
$\rho(t) = \xi^0_{\sig(t)}(\rho(x_1), \ldots, \rho(x_k))$.

\item Inductive step: Assume that the depth of nesting of functions in $t$ is
$p$, and that the property holds for every term with depth of nesting
of functions smaller than $p$. In this case, we have that $t =
f_i(t_1, \ldots, t_{k_i})$. Thus, we have that $\rho(t) =
f^0_i(\rho(t_1), \ldots, \rho(t_{k_i}))$. If $t_i$ is a variable,
then we have that $\rho(t_i) \in \dom(I)$ since $\rho : \{x_1, \ldots,
x_k\} \to \dom(I)$. On the other hand, if $t_i$ is a non-atomic term
such that $\lv(t_i) = [u_1, \ldots, u_q]$ (with $[u_1, \ldots, u_q]$ a
sub-list of $[x_1, \ldots, x_k]$, that is, $[u_1, \ldots, u_q]$
consisting of consecutive elements of $[x_1, \ldots, x_k]$), then given
that the depth of
nesting of functions in $t_i$ is smaller than $p$, we have by
induction hypothesis that $\rho(t_i) = \xi^0_{\sig(t_i)}(\rho(u_1),
\ldots, \rho(u_q))$. Thus, given that $\rho : \{x_1, \ldots,
x_k\} \to \dom(I)$, we have by definition of $f^0_i$ that
$f^0_i(\rho(t_1), \ldots, \rho(t_{k_i})) = \xi^0_{\sig(f_i(t_1,
\ldots, t_{k_i}))}(\rho(x_1), \ldots, \rho(x_k))$ and, therefore,
$\rho(t) =
\xi^0_{\sig(t)}(\rho(x_1), \ldots, \rho(x_k))$. This concludes the
proof of the claim.\qed
\end{iteMize}}

\medskip

\noindent
We finally have all the necessary ingredients to prove that $(I,J)
\models \sigma$. More precisely, we show next that all the
conjuncts of $\sigma$ are satisfied by $(I,J)$ under the
instantiations $f^0_1$, $\ldots$, $f^0_\ell$ of function symbols
$f_1$, $\ldots$, $f_\ell$. Let
\begin{eqnarray}
\label{eq-fin}
\forall x_1 \cdots \forall x_k \forall y_1 \cdots \forall y_m \, (
\varphi(x_1, \ldots, x_k, y_1, \ldots, y_m) \to
\psi(x_1, \ldots, x_k))
\end{eqnarray}
be a conjunct of
$\sigma$, and let $\rho$ be
a variable substitution with
domain  $\{x_1, \ldots, x_k, y_1, \ldots, y_m\}$ and range contained
in $\dom(I)$, and assume that $I \models
\varphi(\rho(x_1), \ldots, \rho(x_k), \rho(y_1), \ldots,
\rho(y_m))$ with the instantiations $f^0_1$, $\ldots$, $f^0_\ell$. Next we
show that $J \models \psi(\rho(x_1), \ldots,
\rho(x_k))$. Assume that
\begin{eqnarray*}
\forall x_1 \cdots \forall x_k \forall y_1 \cdots \forall y_m \, (
\varphi'(x_1, \ldots, x_k, y_1, \ldots, y_m) \to
\psi'(x_1, \ldots, x_k))
\end{eqnarray*}
is the conjunct of $\sigma^\star$ obtained from \eqref{eq-fin} by replacing
every non-atomic term $t \in \mathcal{ST}(\sigma)$ by $\xi(t)$. Then
given that the range of $\rho$ is contained in $\dom(I)$, we have by
Claim \ref{cla-ind} and definition of $\sigma^\star$ that $I \models
\varphi'(\rho(x_1), \ldots, \rho(x_k),
\rho(y_1), \ldots, \rho(y_m))$ with the instantiations $\chi^0_1$,
$\ldots$, $\chi^0_m$ of the function symbols $\chi_1$, $\ldots$,
$\chi_m$ (recall that $\Xi(\sigma) = \{\chi_1, \ldots,
\chi_m\}$). Thus, we conclude that $J \models
\psi'(\rho(x_1), \ldots, \rho(x_k))$ with the instantiations $\chi^0_1$,
$\ldots$, $\chi^0_m$ (as $(I,J)$ satisfies the conjuncts of
$\sigma^\star$ with these instantiations). Therefore, again by Claim
\ref{cla-ind} and definition of $\sigma^\star$, we have that $J
\models \psi(\rho(x_1), \ldots, \rho(x_k))$ with the instantiations
$f^0_1$, $\ldots$, $f^0_\ell$, which was to be shown. This concludes the
proof of the theorem.}

\medskip

\noindent
We now move to the proof of Theorem \ref{th-st-tgds}.
The following lemma will be used in this proof.
\begin{lem}\label{lem-2-un}
For every SO tgd $\sigma$ of nesting depth 2, there exists an unnested
SO tgd $\sigma^\star$ that is equivalent to $\sigma$.
\end{lem}

\proof{In this proof, we extensively used the terminology defined in
the proof of Theorem \ref{th-st-tgds-t-egds}. Besides, we say that a
term $t$ is an $i$-term if the depth of nesting of function symbols in
$t$ is $i$. For example, $f(x,y)$ is a 1-term while $g(f(x,y),z)$ is a
2-term.

Let $\sigma$ be an SO tgd from a source schema $\bs$ to a target
schema $\bt$. Assume that the depth of nesting of function symbols in
every term mentioned in $\sigma$ is at most 2, and that $f_1$,
$\ldots$, $f_\ell$ are the function symbols mentioned in
$\sigma$. Then define a set $\Theta$ of dependencies  as follows. For
every conjunct $\alpha$ of $\sigma$, we include the following
dependencies
as elements of $\Theta$. Assume that $t_1$, $\ldots$,
$t_m$ are the 1-terms $t$ for which there exists a 2-term $t'$
mentioned in $\alpha$ such that $t \in \nat(t')$. For example, if
$\alpha$ is the following
dependency:
\begin{eqnarray}\label{eq-so-tgd-n}
\forall x \forall y \, (S(x,y) \wedge f(x) = g(f(x),f(y)) \to
T(f(g(x,x)))),
\end{eqnarray}
then $f(x)$, $f(y)$ and $g(x,x)$ are the only 1-terms satisfying the
preceding condition. Furthermore, for every $i \in \{1, \ldots, m\}$,
define $T_i$ as the set  $\{x, f_1(\bar x_1), \ldots,
f_\ell(\bar x_\ell)\}$ of terms, where: (1) $x$ is a variable,
(2) each $\bar x_j$ ($1 \leq j \leq \ell)$ is a tuple of pairwise distinct variables,
(3) for every $j \in \{1, \ldots, \ell\}$, we have that
$x$ is not mentioned in $\bar x_j$,
and
(4) for every pair $j$, $k$ of distinct values in $\{1, \ldots,
\ell\}$, we have that
$\bar x_j$ and $\bar x_k$ do not have variables in
common.
Besides, assume that
for every pair $i$, $j$ of distinct values in
$\{1, \ldots, m\}$, we have that
$T_i$ and $T_j$ do not have
variables in common,
and for every $i \in \{1, \ldots, m\}$, we have that
$T_i$ and $\alpha$ do not have
variables in common.
For example, assuming
that $t_1 = f(x)$, $t_2 = f(y)$ and $t_3 = g(x,x)$ for the case of
conjunct \eqref{eq-so-tgd-n}, we have that:
\begin{eqnarray*}
T_1 & = & \{u_1, f(u_2), g(u_3,u_4)\},\\
T_2 & = & \{u_5, f(u_6), g(u_7,u_8)\},\\
T_3 & = & \{u_9, f(u_{10}), g(u_{11},u_{12})\},
\end{eqnarray*}
satisfy the preceding conditions.

Assume that
$\alpha$ is $\forall \bar x \, (\varphi \to \psi)$.
Then for
every $s_1 \in T_1$, $\ldots$, $s_m \in T_m$, define dependency
$\theta_{s_1,\ldots,s_m}$ as:
\begin{eqnarray*}
\forall \bar x \forall y_1 \cdots \forall y_n
\bigg[\bigg(\bigg(\bigwedge_{i=1}^n \dom(y_i)\bigg) \wedge
\bigg(\bigwedge_{j=1}^m \xi(t_j) = \xi(s_j)\bigg) \wedge
\varphi'\bigg) \ \to \ \psi'\bigg],
\end{eqnarray*}
where $\xi$ is defined as in the proof of Theorem
\ref{th-st-tgds-t-egds} (see \eqref{eq-xi}), $y_1$, $\ldots$, $y_n$
are the
variables mentioned in the
terms $s_1$, $\ldots$, $s_m$, and $\varphi'$, $\psi'$ are obtained
from $\varphi$ and $\psi$, respectively, by replacing as follows the
1-terms and 2-terms of these formulas. Every 1-term $t$ mentioned in
$\varphi$ (resp. $\psi$) is replaced by $\xi(t)$ in $\varphi'$
(resp. $\psi'$). Furthermore, every 2-term $t = f(t'_1, \ldots, t'_p)$
mentioned in $\varphi$ (resp. $\psi$) is replaced by $\xi(f(t''_1,
\ldots, t''_p))$ in $\varphi'$ (resp. $\psi'$), where $t''_i$ ($1 \leq
i \leq p$) is defined as
follows.
If $t_i'$ is a 1-term, and so $t_i'$ is $t_j$ for some $j$ in
$\{1, \ldots, m\}$, then let
$t_i''$ be $s_j$.
If $t_i'$ is a variable $v$, then let $t_i''$ be $v$.

Finally, for each dependency $\theta_{s_1, \ldots, s_m}$, let
$\Theta_{s_1, \ldots, s_m}$ be a set of dependencies obtained from
$\theta_{s_1, \ldots, s_m}$ by repeatedly using the equivalences:
\begin{eqnarray*}
((\alpha \vee \beta) \wedge \gamma) \to \delta & \equiv & ((\alpha
\wedge \gamma) \to \delta) \wedge ((\beta \wedge \gamma) \to
\delta),\\
(\exists x \, \alpha) \to \beta & \equiv & \forall x \, (\alpha \to \beta)
\ \ \ \ \ \text{if } x \text{ is not mentioned in } \beta,
\end{eqnarray*}
until all the disjunctions and existential quantifications in the
left-hand side of $\theta_{s_1, \ldots, s_m}$ have been
eliminated. Then all the dependencies in $\Theta_{s_1, \ldots, s_m}$
are included in $\Theta$.
\begin{example}\label{exa-conjunct}
Let us give the intuition behind the definition of $\Theta$
through an example. Assume that $\alpha$ is the following conjunct of
SO tgd $\sigma$:
\begin{eqnarray*}
\forall x \forall y \, (S(x,y) \wedge f(x) = g(f(x),f(y)) \to
T(f(g(x,x)))).
\end{eqnarray*}
Then, as mentioned above, we have that $t_1 = f(x)$, $t_2 = f(y)$,
$t_3 = g(x,x)$
are the
1-terms $t$ for which there exists a
2-term $t'$ in $\alpha$ such that $t \in \nat(t')$. Furthermore, as
also mentioned above, we can assume that:
\begin{eqnarray*}
T_1 & = & \{u_1, f(u_2), g(u_3,u_4)\},\\
T_2 & = & \{u_5, f(u_6), g(u_7,u_8)\},\\
T_3 & = & \{u_9, f(u_{10}), g(u_{11},u_{12})\}.
\end{eqnarray*}
Then for every $s_1 \in T_1$, $s_2 \in T_2$ and $s_3 \in T_3$, we have
to compute the formula $\theta_{s_1, s_2, s_3}$, and then to include
all the dependencies of $\Theta_{s_1, s_2, s_3}$ as elements of
$\Theta$. Assume that $s_1 = u_1$, $s_2 = g(u_7, u_8)$ and $s_3 =
f(u_{10})$. Then $\theta_{s_1,s_2,s_3}$ is the following dependency:
\begin{multline*}
\forall x \forall y \forall u_1 \forall u_7 \forall u_8 \forall u_{10}
\, \bigg[\bigg(S(x,y) \wedge \dom(u_1) \wedge \dom(u_7) \wedge
\dom(u_8) \wedge \dom(u_{10}) \ \wedge\\
\xi_{f(\_)}(x) = u_1 \wedge
\xi_{f(\_)}(y) = \xi_{g(\_,\_)}(u_7,u_8) \wedge \xi_{g(\_,\_)}(x,x) =
\xi_{f(\_)}(u_{10}) \ \wedge\\
\xi_{f(\_)}(x) = \xi_{g(\_,g(\_,\_))}(u_1,u_7,u_8)\bigg) \to
T(\xi_{f(f(\_))}(u_{10}))\bigg].
\end{multline*}
We note that equalities $\xi_{f(\_)}(x) = u_1$, $\xi_{f(\_)}(y) =
\xi_{g(\_,\_)}(u_7,u_8)$ and $\xi_{g(\_,\_)}(x,x) =
\xi_{f(\_)}(u_{10})$ correspond to $\xi(t_1) = \xi(s_1)$, $\xi(t_2) =
\xi(s_2)$ and $\xi(t_3) = \xi(s_3)$ in the definition of dependency
$\theta_{s_1, s_2, s_3}$, respectively. Furthermore, we note that
equality $\xi_{f(\_)}(x) = \xi_{g(\_,g(\_,\_))}(u_1,u_7,u_8)$ is
generated as follows from equality $f(x) = g(f(x),f(y))$ in
$\alpha$. The 1-term $f(x)$
in
the left-hand side of this equality is
replaced by $\xi(f(x)) = \xi_{f(\_)}(x)$ in $\theta_{s_1,s_2,s_3}$,
and the 2-term $g(f(x),f(y))$ is replaced by $\xi(g(u_1,g(u_7,u_8))) =
\xi_{g(\_,g(\_,\_))}(u_1, u_7, u_8)$ in $\theta_{s_1,s_2,s_3}$ (since
$t_1 = f(x)$, $s_1 = u_1$, $t_2 = f(y)$ and $s_2 =
g(u_7,u_8)$). Finally, we note that $T(\xi_{f(f(\_))}(u_{10}))$ is
obtained by replacing the 2-term $f(g(x,x))$ by $\xi(f(f(u_{10}))) =
\xi_{f(f(\_))}(u_{10})$ (since $t_3 = g(x,x)$ and $s_3 = f(u_{10})$).

Given that $\dom(x) = \exists u S(x,u) \vee \exists
v S(v,x)$ in this case, we have that the following dependency is one
of the elements of $\Theta_{s_1,s_2,s_3}$:
\begin{multline*}
\forall x \forall y \forall u_1 \forall u_7 \forall u_8 \forall u_{10}
\forall z_1 \forall z_7 \forall z_8 \forall z_{10}
\, \bigg[\bigg(S(x,y) \wedge S(u_1,z_1) \wedge S(z_7,u_7) \wedge
S(z_8,u_8) \ \wedge \\
S(u_{10},z_{10}) \wedge
\xi_{f(\_)}(x) = u_1  \wedge
\xi_{f(\_)}(y) = \xi_{g(\_,\_)}(u_7,u_8) \wedge \xi_{g(\_,\_)}(x,x) =
\xi_{f(\_)}(u_{10}) \ \wedge\\
\xi_{f(\_)}(x) = \xi_{g(\_,g(\_,\_))}(u_1,u_7,u_8)\bigg) \ \to \
T(\xi_{f(f(\_))}(u_{10}))\bigg].
\end{multline*}
As in the proof of Theorem \ref{th-st-tgds-t-egds}, function symbols
$\xi_{f(\_)}$, $\xi_{g(\_,\_)}$ are used to represent functions $f$
and $g$, respectively, and $\xi_{g(\_,g(\_,\_))}$, $\xi_{f(f(\_))}$
are used to represent functions $g(x,g(y,z))$ and $f(f(x))$,
respectively, thus eliminating the nesting of functions
from $\alpha$. It is important to notice that the preceding
dependency makes explicit some properties that are implicit in
$\alpha$. Given that $f$ and $g$ are function symbols in $\alpha$,
we know that if $f(x) = u_1$, $f(y) = g(u_7,u_8)$, $g(x,x) =
f(u_{10})$, $f(x) = g(f(x),f(y))$ and $T(f(g(x,x)))$ hold, then $f(x)
= g(u_1,g(u_7,u_8))$ and $T(f(f(u_{10})))$ also hold. But
this property is not immediately true for $\xi_{g(\_,g(\_,\_))}$ and
$\xi_{f(f(\_))}$, and, thus, we have to include the preceding
dependency in $\Theta$ to enforce it.
\end{example}
Assume that $\chi_1$, $\ldots$, $\chi_k$
are the
function
symbols mentioned in $\Theta$. Then unnested SO tgd $\sigma^\star$ is
defined as:
\begin{eqnarray*}
\exists \chi_1 \cdots \exists \chi_k \, \big(\bigwedge \Theta\big).
\end{eqnarray*}
Next we show that $\sigma$ and $\sigma^\star$ are equivalent.

($\Rightarrow$) If $(I,J) \models \sigma$, then it is straightforward
to prove that $(I,J) \models \sigma^\star$ (the interpretation of each
function symbol mentioned in $\Theta$ is defined from the
corresponding composition of the interpretations of the function
symbols from $\sigma$).

($\Leftarrow$) Assume that $(I,J) \models \sigma^\star$ with the
instantiations $\chi^0_1$, $\ldots$, $\chi^0_k$ of the function
symbols $\chi_1$, $\ldots$, $\chi_k$. To show that $(I,J) \models
\sigma$, we first need to define from $\chi^0_1$, $\ldots$, $\chi^0_k$
the instantiations $f^0_1$, $\ldots$, $f^0_\ell$ of the function
symbols $f_1$, $\ldots$, $f_\ell$, and then we have to show that
$(I,J)$ satisfies all the conjuncts of $\sigma$ with these
instantiations.

Let $\mathcal{FT}(I)$ be the set of all pairs $(f_i,\bar a)$ such that
(1) $i \in \{1, \ldots, \ell\}$, (2) $\bar a$ is a tuple of elements
from $\dom(I)$, and (3) the length of $\bar a$ is the same as the
arity of function symbol $f_i$. Furthermore, let $<$ be an arbitrary
linear order over $\mathcal{FT}(I)$, and define $\mathcal{DT}(J)$ as the
set of elements $a \in \dom(J)$ such that $a \in \dom(I)$ or $a
= \xi^0_{f_i(\_,\ldots,\_)}(\bar a)$ for some $i \in \{1, \ldots,
\ell\}$ and tuple $\bar a$ of elements from $\dom(I)$. The sets
$\mathcal{FT}(I)$ and $\mathcal{DT}(J)$ are used to define a
substitution $\kappa$,
which in turn is used to define the interpretations of function
symbols $f_1$, $\ldots$, $f_\ell$. More precisely, for every $a \in
\mathcal{DT}(J)$, define:
\begin{equation*}
\kappa(a) =
\begin{cases}
(\_,a) & \mbox{if } a \in \dom(I)\\
(f_j(\_,\ldots,\_), \bar b) &   \mbox{if } a \not\in \dom(I) \text{ and }
(f_j,\bar b) = {\displaystyle \min_{<} \{(f_k,\bar c) \in \mathcal{FT}(I)
\mid \xi^0_{f_k(\_,\ldots,\_)}(\bar c) = a\}}
\end{cases}
\end{equation*}
Note that in the second case in the definition of $\kappa(a)$, the set
over which the min is taken is nonempty, because
$a \in \mathcal{DT}(J)$ and $ a \not\in \dom(I)$.

Define
the instantiations $f^0_1$, $\ldots$, $f^0_\ell$ of
function symbols $f_1$, $\ldots$,
$f_\ell$ as follows.  Given that $(I,J) \models \sigma^\star$, we have
by definition of satisfaction for SO tgds that
there exists a countably infinite universe%
\footnote{Again, the universe can even be taken to be finite,
but we do not need this.}
$U$ such that (1) $U$ is the union of
$\dom(I) \cup \dom(J)$ and a set of nulls, and
(2) $(U;I,J)$
satisfies $\sigma^\star$ in the standard second-order logic
sense. Assume that $\bot$ is a fresh null value ($\bot \not\in U$) and
that the arity of function symbol $f_i$ is $k_i$ ($1 \leq i \leq \ell$). Then
the domain of each one of the functions $f^0_1$, $\ldots$, $f^0_\ell$ is
defined to be $U \cup \{ \bot \}$, and for every $(a_1, \ldots,
a_{k_i}) \in (U \cup \{ \bot  \})^{k_i}$, we define:
\begin{equation*}
f^0_i(a_1, \ldots, a_{k_i}) =
\begin{cases}
\xi^0_{f_i(w_1, \ldots, w_{k_i})}(\bar b_1, \ldots, \bar b_{k_i}) &
\text{if for every } i \in \{1, \ldots, k_i\}, \text{ it holds that}\\
& \hspace{2cm} a_i \in \mathcal{DT}(J) \text{ and } \kappa(a_i) = (w_i,
\bar b_i)\\
\bot & \text{otherwise}
\end{cases}
\end{equation*}
Notice that if $a_i = \bot$ in the definition above, for some $i
\in \{1, \ldots, k_i\}$, then $f^0_i(a_1, \ldots, a_{k_i}) = \bot$.

Next we show that $(I,J)$ satisfies%
\footnote{When we say that $(I,J)$ satisfies a formula, we mean that
$(U;I,J)$ satisfies the formula.  Similarly, when we say that
$I$ satisfies a formula, we mean that $(U;I)$ satisfies the formula, and
likewise for $J$ satisfying a formula.}
every conjunct of $\sigma$ with
the instantiations $f^0_1$, $\ldots$, $f^0_\ell$ of function symbols
$f_1$, $\ldots$, $f_\ell$. But before doing this, we give an example
that shows how the strategy of the proof works.
\begin{example}
Consider again conjunct \eqref{eq-so-tgd-n}. To prove that $(I,J)$
satisfies this conjunct under the preceding definition of the
functions in $\sigma$, we have to prove that if:
\begin{eqnarray*}
I & \models & S(a,b) \wedge f(a) = g(f(a),f(b)),
\end{eqnarray*}
then $J \models T(f(g(a,a)))$. In order to prove this, we first
need to figure out what the values of $f^0(a)$, $f^0(b)$, $g^0(f^0(a),
f^0(b))$ and $f^0(g^0(a,a))$ are. Given that $a, b \in
\dom(I)$, we have that $f^0(a) = \xi^0_{f(\_)}(a)$ and $f^0(b) =
\xi^0_{f(\_)}(b)$. The definition of $g^0(f^0(a),f^0(b))$ depends on whether
$\xi^0_{f(\_)}(a)$ and $\xi^0_{f(\_)}(b)$ belong to $\dom(I)$. Assume
that $\xi^0_{f(\_)}(a) = a_1$ and $\xi^0_{f(\_)}(b) = b_1$, where $a_1
\in \dom(I)$ and $b_1 \not\in \dom(I)$. Then by the preceding definition, we
have to compute $\kappa(a_1)$ and $\kappa(b_1)$ in order to compute
the value of $g^0(f^0(a),f^0(b))$. We have that $\kappa(a_1) =
(\_,a_1)$ since $a_1 \in \dom(I)$, and we assume
for this example
that $\kappa(b_1) =
(g(\_,\_),(c_1,c_2))$, where $c_1, c_2 \in \dom(I)$. That is, we
assume that $\xi^0_{g(\_,\_)}(c_1,c_2) = b_1$ and $(g,(c_1,c_2))$ is
the smallest element $(h,\bar d)$ in $\mathcal{FT}(I)$, according to the
linear order $<$, satisfying the condition $\xi^0_{h(\_, \ldots,
\_)}(\bar d) = b_1$. Thus, by the preceding definition, we have that:
\begin{eqnarray*}
g^0(f^0(a),f^0(b)) \ = \ g^0(a_1,b_1) \ = \
\xi^0_{g(\_,g(\_,\_))}(a_1,c_1,c_2).
\end{eqnarray*}
Finally, we also need to know what the value of $f^0(g^0(a,a))$ is. By the
preceding definition, we know that $g^0(a,a) =
\xi^0_{g(\_,\_)}(a,a)$. Assume that $\xi^0_{g(\_,\_)}(a,a) = d_1$ with
$d_1 \not\in \dom(I)$. Then, as in the previous case, we need to
compute $\kappa(d_1)$ in order to compute $f^0(g^0(a,a))$. Assume
for this example
that
$\kappa(d_1) = (f(\_),d_2)$, where $d_2 \in \dom(I)$. That is, assume
that $\xi_{f(\_)}(d_2) = d_1$ and $(f,d_2)$ is the smallest element
$(h,\bar d)$ in $\mathcal{FT}(I)$, according to the linear order $<$,
satisfying the condition $\xi^0_{h(\_, \ldots, \_)}(\bar d) =
d_1$. Thus, by the preceding definition, we have that:
\begin{eqnarray}\label{eq-alm-exa}
f^0(g^0(a,a)) \ = \ f^0(d_1) \ = \ \xi^0_{f(f(\_))}(d_2).
\end{eqnarray}
Therefore, from the previous discussion and the fact that that $I \models
f(a) = g(f(a),f(b))$, we conclude that:
\begin{multline*}
I \ \models \ S(a,b) \wedge \dom(a_1) \wedge \dom(c_1) \wedge
\dom(c_2) \wedge \dom(d_2) \wedge \xi_{f(\_)}(a) = a_1 \ \wedge \\
\xi_{f(\_)}(b) = \xi_{g(\_,\_)}(c_1,c_2) \wedge \xi_{g(\_,\_)}(a,a) =
\xi_{f(\_)}(d_2) \wedge
\xi_{f(\_)}(a) = \xi_{g(\_,g(\_,\_))}(a_1,c_1,c_2).
\end{multline*}
Hence, given that we assume that $(I,J) \models \sigma^\star$ and the
following dependency is one of the formulas $\theta_{s_1,s_2,s_3}$
(see Example \ref{exa-conjunct}):
\begin{multline*}
\forall x \forall y \forall u_1 \forall u_7 \forall u_8 \forall u_{10}
\, \bigg[\bigg(S(x,y) \wedge \dom(u_1) \wedge \dom(u_7) \wedge
\dom(u_8) \wedge \dom(u_{10}) \ \wedge\\
\xi_{f(\_)}(x) = u_1 \wedge
\xi_{f(\_)}(y) = \xi_{g(\_,\_)}(u_7,u_8) \wedge \xi_{g(\_,\_)}(x,x) =
\xi_{f(\_)}(u_{10}) \ \wedge\\
\xi_{f(\_)}(x) = \xi_{g(\_,g(\_,\_))}(u_1,u_7,u_8)\bigg) \ \to \
T(\xi_{f(f(\_))}(u_{10}))\bigg],
\end{multline*}
we conclude that $J \models T(\xi_{f(f(\_))}(d_2))$. But we know from
\eqref{eq-alm-exa} that $f^0(g^0(a,a)) = \xi^0_{f(f(\_))}(d_2)$ and, thus,
we have that $J \models T(f(g(a,a)))$, which was to be shown.
\end{example}
In general, we have to show that if $\forall \bar x \, (\varphi(\bar
x) \to \psi(\bar x))$ is a conjunct of $\sigma$ and $I \models
\varphi(\bar a)$ with the instantiations $f^0_1$, $\ldots$, $f^0_\ell$
of the function symbols $f_1$, $\ldots$, $f_\ell$, then $J \models
\psi(\bar a)$ with these instantiations. It is straightforward but
lengthy to generalize the strategy shown in the previous example to
this case. In particular, given that in the construction of
$\sigma^\star$ we consider all the possible cases for substitution
$\kappa$, the previous strategy can be applied in general. This
concludes the proof of the lemma.\qed}

\medskip

\aproof{Theorem \ref{th-st-tgds}}{The theorem is proved by induction
on the nesting depth $n$ of an SO tgd. If $n = 1$, then the property
trivially holds, and if $n = 2$, then the property holds by Lemma
\ref{lem-2-un}. Thus, let $\sigma$ be an SO tgd from a source schema $\bs$
to a target schema $\bt$, and assume that the nesting depth of
$\sigma$ is $n \geq 3$. Moreover, assume that the theorem holds for every SO
tgd $\sigma'$ of nesting depth $n' < n$.

By Theorem 8.4
in \cite{FKPT05}, we know that there exist schema
mappings $\M_1$, $\M_2$, $\M_3$, $\ldots$, $\M_{n+1}$ such that
$\sigma$
specifies
$\M_1 \circ \M_2 \circ \M_3 \circ \cdots
\circ \M_{n+1}$ and every mapping $\M_i$ ($1 \leq i \leq n+1$) is
specified by a set of st-tgds. For every $i \in \{1, \ldots, n+1\}$,
let $\sigma_i$ be an unnested SO tgd
that specifies
$\M_i$. We know, by
the definition of the algorithm Compose in \cite{FKPT05}, that there
exists an SO tgd $\sigma_{12}$
that specifies
the composition of the schema mappings specified by $\sigma_1$ and
$\sigma_2$ and whose nesting depth is at most 2. By Lemma
\ref{lem-2-un}, we have that there exists an unnested SO tgd
$\sigma_{12}^\star$ that is equivalent to $\sigma_{12}$. Thus, by
considering again the definition of the algorithm Compose in
\cite{FKPT05}, we have that there exists an SO tgd $\sigma_{13}$ such
that $\sigma_{13}$
specifies
the composition of the schema
mappings specified by $\sigma_{12}^\star$ and $\sigma_3$ and whose
nesting depth is at most 2. Hence, by considering again Lemma
\ref{lem-2-un}, we conclude that there exists an unnested SO tgd
$\sigma^\star_{13}$ that
specifies
the composition of the
schema mappings specified by $\sigma^\star_{12}$ and $\sigma_3$ and,
thus, also
specifies
$\M_1 \circ \M_2 \circ
\M_3$. Finally, by considering again the definition of algorithm
Compose in \cite{FKPT05}, we have that there exists an SO tgd
$\sigma'$ such that: (1) $\sigma'$
specifies
the composition of
the mappings specified by $\sigma^\star_{13}$, $\sigma_4$, $\ldots$,
$\sigma_{n+1}$; and (2) the depth of nesting of $\sigma'$ is at most
$n-1$ (since $\sigma^\star_{13}$, $\sigma_4$, $\ldots$, $\sigma_{n+1}$
are all unnested SO tgds). Therefore, we conclude that there exists an
SO tgd $\sigma'$ that is equivalent to $\sigma$ and whose depth of
nesting is at most $n-1$. But then by induction hypothesis, there
exists an unnested SO tgd $\sigma^\star$ that is equivalent to
$\sigma'$ and, hence, there exists an unnested SO tgd $\sigma^\star$
that is equivalent to $\sigma$. This concludes the proof of the
theorem.}

\section{Concluding Remarks}
\label{sec-concluding}

We have investigated the question of what language is needed to specify
the composition of schema mappings with target constraints.
In particular, we showed that {\em st-SO dependencies} (along with
appropriate target constraints) are exactly the right language for
specifying the
composition of
{\em standard schema mappings\/} (those specified by st-tgds, target
egds, and a weakly acyclic set of target tgds).
By contrast, we showed that SO tgds, even with arbitrary source and
target constraints, are not rich enough to be able to
specify in general the composition of two standard schema mappings.
In addition to their expressive power,
we also showed that st-SO dependencies
enjoy other desirable properties.  In particular, they have a
polynomial-time chase that generates a universal
solution,
which can be
used to find the certain answers to unions of conjunctive queries in
polynomial time.

We proved the surprising result that SO tgds and st-SO dependencies can
be denested:  that is, each such dependency is equivalent to another
dependency of that type with no nested function symbols.
These denesting results can be used to ``collapse'' multiple
compositions of schema mappings into the composition of two schema
mappings of that type.
In particular, we obtain the unexpected result that the composition of
an arbitrary number of st-tgd mappings is equivalent to the composition
of only two st-tgd mappings.

Our results gave us two ways to ``simplify'' the composition of an
arbitrary number of st-tgd mappings.
First, we could
replace the
composition by a single schema mapping, specified by an unnested SO tgd.
Second, we could replace the composition by
the composition of only two st-tgd schema mappings.
A similar comment applies to the composition of an arbitrary number of
standard schema mappings.

\section*{Acknowledgments}
The authors are grateful to Phokion Kolaitis for helpful discussions.
Marcelo Arenas was partially supported by FONDECYT grant 1090565.


\bibliographystyle{abbrv}
\bibliography{biblio}

\end{document}

%% file: source.pstex_t
\begin{picture}(0,0)%
\epsfig{file=source.pstex}%
\end{picture}%
\setlength{\unitlength}{2565sp}%
\begingroup\makeatletter\ifx\SetFigFont\undefined%
\gdef\SetFigFont#1#2#3#4#5{%
  \reset@font\fontsize{#1}{#2pt}%
  \fontfamily{#3}\fontseries{#4}\fontshape{#5}%
  \selectfont}%
\fi\endgroup%
\begin{picture}(4500,1332)(7651,-10099)
\put(8701,-10036){\makebox(0,0)[b]{\smash{\SetFigFont{10}{12.0}{\familydefault}{\mddefault}{\updefault}{\color[rgb]{0,0,0}$b_1$}%
}}}
\put(8101,-8911){\makebox(0,0)[b]{\smash{\SetFigFont{10}{12.0}{\familydefault}{\mddefault}{\updefault}{\color[rgb]{0,0,0}$E$}%
}}}
\put(9301,-8911){\makebox(0,0)[b]{\smash{\SetFigFont{10}{12.0}{\familydefault}{\mddefault}{\updefault}{\color[rgb]{0,0,0}$E$}%
}}}
\put(10501,-8911){\makebox(0,0)[b]{\smash{\SetFigFont{10}{12.0}{\familydefault}{\mddefault}{\updefault}{\color[rgb]{0,0,0}$E$}%
}}}
\put(11701,-8911){\makebox(0,0)[b]{\smash{\SetFigFont{10}{12.0}{\familydefault}{\mddefault}{\updefault}{\color[rgb]{0,0,0}$E$}%
}}}
\put(8101,-9811){\makebox(0,0)[b]{\smash{\SetFigFont{10}{12.0}{\familydefault}{\mddefault}{\updefault}{\color[rgb]{0,0,0}$E$}%
}}}
\put(9301,-9811){\makebox(0,0)[b]{\smash{\SetFigFont{10}{12.0}{\familydefault}{\mddefault}{\updefault}{\color[rgb]{0,0,0}$E$}%
}}}
\put(10501,-9811){\makebox(0,0)[b]{\smash{\SetFigFont{10}{12.0}{\familydefault}{\mddefault}{\updefault}{\color[rgb]{0,0,0}$E$}%
}}}
\put(7651,-9136){\makebox(0,0)[b]{\smash{\SetFigFont{10}{12.0}{\familydefault}{\mddefault}{\updefault}{\color[rgb]{0,0,0}$a$}%
}}}
\put(8701,-9136){\makebox(0,0)[b]{\smash{\SetFigFont{10}{12.0}{\familydefault}{\mddefault}{\updefault}{\color[rgb]{0,0,0}$a_1$}%
}}}
\put(11101,-9136){\makebox(0,0)[b]{\smash{\SetFigFont{10}{12.0}{\familydefault}{\mddefault}{\updefault}{\color[rgb]{0,0,0}$a_{d}$}%
}}}
\put(12151,-9136){\makebox(0,0)[b]{\smash{\SetFigFont{10}{12.0}{\familydefault}{\mddefault}{\updefault}{\color[rgb]{0,0,0}$c$}%
}}}
\put(11101,-10036){\makebox(0,0)[b]{\smash{\SetFigFont{10}{12.0}{\familydefault}{\mddefault}{\updefault}{\color[rgb]{0,0,0}$b_{d}$}%
}}}
\put(7651,-10036){\makebox(0,0)[b]{\smash{\SetFigFont{10}{12.0}{\familydefault}{\mddefault}{\updefault}{\color[rgb]{0,0,0}$b$}%
}}}
\end{picture}

%% file: pre-can.pstex_t
\begin{picture}(0,0)%
\epsfig{file=pre-can.pstex}%
\end{picture}%
\setlength{\unitlength}{2368sp}%
\begingroup\makeatletter\ifx\SetFigFont\undefined%
\gdef\SetFigFont#1#2#3#4#5{%
  \reset@font\fontsize{#1}{#2pt}%
  \fontfamily{#3}\fontseries{#4}\fontshape{#5}%
  \selectfont}%
\fi\endgroup%
\begin{picture}(5292,4497)(7501,-13390)
\put(8101,-13336){\makebox(0,0)[b]{\smash{\SetFigFont{9}{10.8}{\familydefault}{\mddefault}{\updefault}{\color[rgb]{0,0,0}$\bullet$}%
}}}
\put(8701,-9061){\makebox(0,0)[b]{\smash{\SetFigFont{9}{10.8}{\familydefault}{\mddefault}{\updefault}{\color[rgb]{0,0,0}$a_1$}%
}}}
\put(7651,-9061){\makebox(0,0)[b]{\smash{\SetFigFont{9}{10.8}{\familydefault}{\mddefault}{\updefault}{\color[rgb]{0,0,0}$P_2(a)$}%
}}}
\put(9376,-9511){\makebox(0,0)[b]{\smash{\SetFigFont{9}{10.8}{\familydefault}{\mddefault}{\updefault}{\color[rgb]{0,0,0}$R$}%
}}}
\put(7501,-9511){\makebox(0,0)[b]{\smash{\SetFigFont{9}{10.8}{\familydefault}{\mddefault}{\updefault}{\color[rgb]{0,0,0}$R$}%
}}}
\put(8476,-9511){\makebox(0,0)[b]{\smash{\SetFigFont{9}{10.8}{\familydefault}{\mddefault}{\updefault}{\color[rgb]{0,0,0}$R$}%
}}}
\put(8851,-10486){\makebox(0,0)[b]{\smash{\SetFigFont{9}{10.8}{\familydefault}{\mddefault}{\updefault}{\color[rgb]{0,0,0}$S$}%
}}}
\put(9451,-10486){\makebox(0,0)[b]{\smash{\SetFigFont{9}{10.8}{\familydefault}{\mddefault}{\updefault}{\color[rgb]{0,0,0}$S$}%
}}}
\put(8926,-9511){\makebox(0,0)[b]{\smash{\SetFigFont{9}{10.8}{\familydefault}{\mddefault}{\updefault}{\color[rgb]{0,0,0}$R$}%
}}}
\put(9601,-9061){\makebox(0,0)[b]{\smash{\SetFigFont{9}{10.8}{\familydefault}{\mddefault}{\updefault}{\color[rgb]{0,0,0}$a_2$}%
}}}
\put(11701,-9061){\makebox(0,0)[b]{\smash{\SetFigFont{9}{10.8}{\familydefault}{\mddefault}{\updefault}{\color[rgb]{0,0,0}$a_{d}$}%
}}}
\put(12751,-9061){\makebox(0,0)[b]{\smash{\SetFigFont{9}{10.8}{\familydefault}{\mddefault}{\updefault}{\color[rgb]{0,0,0}$Q_2(c)$}%
}}}
\put(11476,-9511){\makebox(0,0)[b]{\smash{\SetFigFont{9}{10.8}{\familydefault}{\mddefault}{\updefault}{\color[rgb]{0,0,0}$R$}%
}}}
\put(12601,-9511){\makebox(0,0)[b]{\smash{\SetFigFont{9}{10.8}{\familydefault}{\mddefault}{\updefault}{\color[rgb]{0,0,0}$R$}%
}}}
\put(10951,-10486){\makebox(0,0)[b]{\smash{\SetFigFont{9}{10.8}{\familydefault}{\mddefault}{\updefault}{\color[rgb]{0,0,0}$S$}%
}}}
\put(11551,-10486){\makebox(0,0)[b]{\smash{\SetFigFont{9}{10.8}{\familydefault}{\mddefault}{\updefault}{\color[rgb]{0,0,0}$S$}%
}}}
\put(11026,-9511){\makebox(0,0)[b]{\smash{\SetFigFont{9}{10.8}{\familydefault}{\mddefault}{\updefault}{\color[rgb]{0,0,0}$R$}%
}}}
\put(11926,-9511){\makebox(0,0)[b]{\smash{\SetFigFont{9}{10.8}{\familydefault}{\mddefault}{\updefault}{\color[rgb]{0,0,0}$R$}%
}}}
\put(10801,-9061){\makebox(0,0)[b]{\smash{\SetFigFont{9}{10.8}{\familydefault}{\mddefault}{\updefault}{\color[rgb]{0,0,0}$a_{d-1}$}%
}}}
\put(8701,-11461){\makebox(0,0)[b]{\smash{\SetFigFont{9}{10.8}{\familydefault}{\mddefault}{\updefault}{\color[rgb]{0,0,0}$b_1$}%
}}}
\put(7651,-11461){\makebox(0,0)[b]{\smash{\SetFigFont{9}{10.8}{\familydefault}{\mddefault}{\updefault}{\color[rgb]{0,0,0}$P_2(b)$}%
}}}
\put(8476,-11911){\makebox(0,0)[b]{\smash{\SetFigFont{9}{10.8}{\familydefault}{\mddefault}{\updefault}{\color[rgb]{0,0,0}$R$}%
}}}
\put(7501,-11911){\makebox(0,0)[b]{\smash{\SetFigFont{9}{10.8}{\familydefault}{\mddefault}{\updefault}{\color[rgb]{0,0,0}$R$}%
}}}
\put(9601,-11461){\makebox(0,0)[b]{\smash{\SetFigFont{9}{10.8}{\familydefault}{\mddefault}{\updefault}{\color[rgb]{0,0,0}$b_2$}%
}}}
\put(10801,-11461){\makebox(0,0)[b]{\smash{\SetFigFont{9}{10.8}{\familydefault}{\mddefault}{\updefault}{\color[rgb]{0,0,0}$b_{d-1}$}%
}}}
\put(8851,-12886){\makebox(0,0)[b]{\smash{\SetFigFont{9}{10.8}{\familydefault}{\mddefault}{\updefault}{\color[rgb]{0,0,0}$S$}%
}}}
\put(9451,-12886){\makebox(0,0)[b]{\smash{\SetFigFont{9}{10.8}{\familydefault}{\mddefault}{\updefault}{\color[rgb]{0,0,0}$S$}%
}}}
\put(11701,-11461){\makebox(0,0)[b]{\smash{\SetFigFont{9}{10.8}{\familydefault}{\mddefault}{\updefault}{\color[rgb]{0,0,0}$b_{d}$}%
}}}
\put(11551,-11911){\makebox(0,0)[b]{\smash{\SetFigFont{9}{10.8}{\familydefault}{\mddefault}{\updefault}{\color[rgb]{0,0,0}$R$}%
}}}
\put(11026,-11911){\makebox(0,0)[b]{\smash{\SetFigFont{9}{10.8}{\familydefault}{\mddefault}{\updefault}{\color[rgb]{0,0,0}$R$}%
}}}
\put(8926,-11911){\makebox(0,0)[b]{\smash{\SetFigFont{9}{10.8}{\familydefault}{\mddefault}{\updefault}{\color[rgb]{0,0,0}$R$}%
}}}
\put(9376,-11911){\makebox(0,0)[b]{\smash{\SetFigFont{9}{10.8}{\familydefault}{\mddefault}{\updefault}{\color[rgb]{0,0,0}$R$}%
}}}
\put(10951,-12886){\makebox(0,0)[b]{\smash{\SetFigFont{9}{10.8}{\familydefault}{\mddefault}{\updefault}{\color[rgb]{0,0,0}$S$}%
}}}
\put(7651,-12886){\makebox(0,0)[b]{\smash{\SetFigFont{9}{10.8}{\familydefault}{\mddefault}{\updefault}{\color[rgb]{0,0,0}$S$}%
}}}
\put(7651,-10486){\makebox(0,0)[b]{\smash{\SetFigFont{9}{10.8}{\familydefault}{\mddefault}{\updefault}{\color[rgb]{0,0,0}$S$}%
}}}
\put(11851,-10486){\makebox(0,0)[b]{\smash{\SetFigFont{9}{10.8}{\familydefault}{\mddefault}{\updefault}{\color[rgb]{0,0,0}$S$}%
}}}
\put(8476,-10486){\makebox(0,0)[b]{\smash{\SetFigFont{9}{10.8}{\familydefault}{\mddefault}{\updefault}{\color[rgb]{0,0,0}$S$}%
}}}
\put(8476,-12886){\makebox(0,0)[b]{\smash{\SetFigFont{9}{10.8}{\familydefault}{\mddefault}{\updefault}{\color[rgb]{0,0,0}$S$}%
}}}
\put(11626,-12886){\makebox(0,0)[b]{\smash{\SetFigFont{9}{10.8}{\familydefault}{\mddefault}{\updefault}{\color[rgb]{0,0,0}$S$}%
}}}
\put(12676,-10486){\makebox(0,0)[b]{\smash{\SetFigFont{9}{10.8}{\familydefault}{\mddefault}{\updefault}{\color[rgb]{0,0,0}$S$}%
}}}
\put(7651,-10036){\makebox(0,0)[b]{\smash{\SetFigFont{9}{10.8}{\familydefault}{\mddefault}{\updefault}{\color[rgb]{0,0,0}$\bullet$}%
}}}
\put(8551,-10036){\makebox(0,0)[b]{\smash{\SetFigFont{9}{10.8}{\familydefault}{\mddefault}{\updefault}{\color[rgb]{0,0,0}$\bullet$}%
}}}
\put(8851,-10036){\makebox(0,0)[b]{\smash{\SetFigFont{9}{10.8}{\familydefault}{\mddefault}{\updefault}{\color[rgb]{0,0,0}$\bullet$}%
}}}
\put(9451,-10036){\makebox(0,0)[b]{\smash{\SetFigFont{9}{10.8}{\familydefault}{\mddefault}{\updefault}{\color[rgb]{0,0,0}$\bullet$}%
}}}
\put(10951,-10036){\makebox(0,0)[b]{\smash{\SetFigFont{9}{10.8}{\familydefault}{\mddefault}{\updefault}{\color[rgb]{0,0,0}$\bullet$}%
}}}
\put(11551,-10036){\makebox(0,0)[b]{\smash{\SetFigFont{9}{10.8}{\familydefault}{\mddefault}{\updefault}{\color[rgb]{0,0,0}$\bullet$}%
}}}
\put(11851,-10036){\makebox(0,0)[b]{\smash{\SetFigFont{9}{10.8}{\familydefault}{\mddefault}{\updefault}{\color[rgb]{0,0,0}$\bullet$}%
}}}
\put(12751,-10036){\makebox(0,0)[b]{\smash{\SetFigFont{9}{10.8}{\familydefault}{\mddefault}{\updefault}{\color[rgb]{0,0,0}$\bullet$}%
}}}
\put(8101,-10936){\makebox(0,0)[b]{\smash{\SetFigFont{9}{10.8}{\familydefault}{\mddefault}{\updefault}{\color[rgb]{0,0,0}$\bullet$}%
}}}
\put(9151,-10936){\makebox(0,0)[b]{\smash{\SetFigFont{9}{10.8}{\familydefault}{\mddefault}{\updefault}{\color[rgb]{0,0,0}$\bullet$}%
}}}
\put(11251,-10936){\makebox(0,0)[b]{\smash{\SetFigFont{9}{10.8}{\familydefault}{\mddefault}{\updefault}{\color[rgb]{0,0,0}$\bullet$}%
}}}
\put(12301,-10936){\makebox(0,0)[b]{\smash{\SetFigFont{9}{10.8}{\familydefault}{\mddefault}{\updefault}{\color[rgb]{0,0,0}$\bullet$}%
}}}
\put(7651,-12436){\makebox(0,0)[b]{\smash{\SetFigFont{9}{10.8}{\familydefault}{\mddefault}{\updefault}{\color[rgb]{0,0,0}$\bullet$}%
}}}
\put(8551,-12436){\makebox(0,0)[b]{\smash{\SetFigFont{9}{10.8}{\familydefault}{\mddefault}{\updefault}{\color[rgb]{0,0,0}$\bullet$}%
}}}
\put(8851,-12436){\makebox(0,0)[b]{\smash{\SetFigFont{9}{10.8}{\familydefault}{\mddefault}{\updefault}{\color[rgb]{0,0,0}$\bullet$}%
}}}
\put(9451,-12436){\makebox(0,0)[b]{\smash{\SetFigFont{9}{10.8}{\familydefault}{\mddefault}{\updefault}{\color[rgb]{0,0,0}$\bullet$}%
}}}
\put(10951,-12436){\makebox(0,0)[b]{\smash{\SetFigFont{9}{10.8}{\familydefault}{\mddefault}{\updefault}{\color[rgb]{0,0,0}$\bullet$}%
}}}
\put(11701,-12436){\makebox(0,0)[b]{\smash{\SetFigFont{9}{10.8}{\familydefault}{\mddefault}{\updefault}{\color[rgb]{0,0,0}$\bullet$}%
}}}
\put(9151,-13336){\makebox(0,0)[b]{\smash{\SetFigFont{9}{10.8}{\familydefault}{\mddefault}{\updefault}{\color[rgb]{0,0,0}$\bullet$}%
}}}
\put(11326,-13336){\makebox(0,0)[b]{\smash{\SetFigFont{9}{10.8}{\familydefault}{\mddefault}{\updefault}{\color[rgb]{0,0,0}$\bullet$}%
}}}
\end{picture}

%% file: can.pstex_t
\begin{picture}(0,0)%
\epsfig{file=can.pstex}%
\end{picture}%
\setlength{\unitlength}{2368sp}%
\begingroup\makeatletter\ifx\SetFigFont\undefined%
\gdef\SetFigFont#1#2#3#4#5{%
  \reset@font\fontsize{#1}{#2pt}%
  \fontfamily{#3}\fontseries{#4}\fontshape{#5}%
  \selectfont}%
\fi\endgroup%
\begin{picture}(6624,4497)(7489,-13390)
\put(9901,-12436){\makebox(0,0)[b]{\smash{\SetFigFont{9}{10.8}{\familydefault}{\mddefault}{\updefault}{\color[rgb]{0,0,0}$\bullet$}%
}}}
\put(8701,-9061){\makebox(0,0)[b]{\smash{\SetFigFont{9}{10.8}{\familydefault}{\mddefault}{\updefault}{\color[rgb]{0,0,0}$a_1$}%
}}}
\put(9751,-10411){\makebox(0,0)[b]{\smash{\SetFigFont{9}{10.8}{\familydefault}{\mddefault}{\updefault}{\color[rgb]{0,0,0}$S$}%
}}}
\put(9901,-9061){\makebox(0,0)[b]{\smash{\SetFigFont{9}{10.8}{\familydefault}{\mddefault}{\updefault}{\color[rgb]{0,0,0}$a_2$}%
}}}
\put(12901,-9061){\makebox(0,0)[b]{\smash{\SetFigFont{9}{10.8}{\familydefault}{\mddefault}{\updefault}{\color[rgb]{0,0,0}$a_{d}$}%
}}}
\put(11701,-9061){\makebox(0,0)[b]{\smash{\SetFigFont{9}{10.8}{\familydefault}{\mddefault}{\updefault}{\color[rgb]{0,0,0}$a_{d-1}$}%
}}}
\put(14101,-9061){\makebox(0,0)[b]{\smash{\SetFigFont{9}{10.8}{\familydefault}{\mddefault}{\updefault}{\color[rgb]{0,0,0}$Q_2(c)$}%
}}}
\put(11626,-9436){\makebox(0,0)[b]{\smash{\SetFigFont{9}{10.8}{\familydefault}{\mddefault}{\updefault}{\color[rgb]{0,0,0}$R$}%
}}}
\put(10801,-9436){\makebox(0,0)[b]{\smash{\SetFigFont{9}{10.8}{\familydefault}{\mddefault}{\updefault}{\color[rgb]{0,0,0}$R$}%
}}}
\put(9301,-9436){\makebox(0,0)[b]{\smash{\SetFigFont{9}{10.8}{\familydefault}{\mddefault}{\updefault}{\color[rgb]{0,0,0}$R$}%
}}}
\put(10051,-9436){\makebox(0,0)[b]{\smash{\SetFigFont{9}{10.8}{\familydefault}{\mddefault}{\updefault}{\color[rgb]{0,0,0}$R$}%
}}}
\put(8626,-9736){\makebox(0,0)[b]{\smash{\SetFigFont{9}{10.8}{\familydefault}{\mddefault}{\updefault}{\color[rgb]{0,0,0}$R$}%
}}}
\put(12151,-9736){\makebox(0,0)[b]{\smash{\SetFigFont{9}{10.8}{\familydefault}{\mddefault}{\updefault}{\color[rgb]{0,0,0}$R$}%
}}}
\put(7501,-9061){\makebox(0,0)[b]{\smash{\SetFigFont{9}{10.8}{\familydefault}{\mddefault}{\updefault}{\color[rgb]{0,0,0}$P_2(a)$}%
}}}
\put(9751,-12811){\makebox(0,0)[b]{\smash{\SetFigFont{9}{10.8}{\familydefault}{\mddefault}{\updefault}{\color[rgb]{0,0,0}$S$}%
}}}
\put(8701,-11461){\makebox(0,0)[b]{\smash{\SetFigFont{9}{10.8}{\familydefault}{\mddefault}{\updefault}{\color[rgb]{0,0,0}$b_1$}%
}}}
\put(7501,-11461){\makebox(0,0)[b]{\smash{\SetFigFont{9}{10.8}{\familydefault}{\mddefault}{\updefault}{\color[rgb]{0,0,0}$P_2(b)$}%
}}}
\put(9901,-11461){\makebox(0,0)[b]{\smash{\SetFigFont{9}{10.8}{\familydefault}{\mddefault}{\updefault}{\color[rgb]{0,0,0}$b_2$}%
}}}
\put(11701,-11461){\makebox(0,0)[b]{\smash{\SetFigFont{9}{10.8}{\familydefault}{\mddefault}{\updefault}{\color[rgb]{0,0,0}$b_{d-1}$}%
}}}
\put(12901,-11461){\makebox(0,0)[b]{\smash{\SetFigFont{9}{10.8}{\familydefault}{\mddefault}{\updefault}{\color[rgb]{0,0,0}$b_{d}$}%
}}}
\put(12151,-11986){\makebox(0,0)[b]{\smash{\SetFigFont{9}{10.8}{\familydefault}{\mddefault}{\updefault}{\color[rgb]{0,0,0}$R$}%
}}}
\put(8101,-11986){\makebox(0,0)[b]{\smash{\SetFigFont{9}{10.8}{\familydefault}{\mddefault}{\updefault}{\color[rgb]{0,0,0}$R$}%
}}}
\put(10801,-11836){\makebox(0,0)[b]{\smash{\SetFigFont{9}{10.8}{\familydefault}{\mddefault}{\updefault}{\color[rgb]{0,0,0}$R$}%
}}}
\put(10051,-11836){\makebox(0,0)[b]{\smash{\SetFigFont{9}{10.8}{\familydefault}{\mddefault}{\updefault}{\color[rgb]{0,0,0}$R$}%
}}}
\put(9301,-11836){\makebox(0,0)[b]{\smash{\SetFigFont{9}{10.8}{\familydefault}{\mddefault}{\updefault}{\color[rgb]{0,0,0}$R$}%
}}}
\put(9901,-10036){\makebox(0,0)[b]{\smash{\SetFigFont{9}{10.8}{\familydefault}{\mddefault}{\updefault}{\color[rgb]{0,0,0}$\bullet$}%
}}}
\put(9901,-10936){\makebox(0,0)[b]{\smash{\SetFigFont{9}{10.8}{\familydefault}{\mddefault}{\updefault}{\color[rgb]{0,0,0}$\bullet$}%
}}}
\put(9901,-13336){\makebox(0,0)[b]{\smash{\SetFigFont{9}{10.8}{\familydefault}{\mddefault}{\updefault}{\color[rgb]{0,0,0}$\bullet$}%
}}}
\end{picture}

%% file: lmcs_comp.bbl
\begin{thebibliography}{10}

\bibitem{AK08}
F.~Afrati and N.~Kiourtis.
\newblock {Query Answering Using Views in the Presence of Dependencies}.
\newblock In {\em New Trends in Information Integration (NTII)}, pages 8--11,
  2008.

\bibitem{AK09}
F.~Afrati and P.~Kolaitis.
\newblock {Repair Checking in Inconsistent Databases: Algorithms and
  Complexity}.
\newblock In {\em International Conference on Database Theory (ICDT)}, pages
  31--41, 2009.

\bibitem{ALP08a}
F.~Afrati, C.~Li, and V.~Pavlaki.
\newblock {Data Exchange in the Presence of Arithmetic Comparisons}.
\newblock In {\em Extending Data Base Technology (EDBT)}, pages 487--498, 2008.

\bibitem{ALP08b}
F.~Afrati, C.~Li, and V.~Pavlaki.
\newblock {Data Exchange: Query Answering for Incomplete Data Sources}.
\newblock In {\em 3rd International Conference on Scalable Information
  Systems}, 2009.

\bibitem{ABFL04}
M.~Arenas, P.~Barcel\'o, R.~Fagin, and L.~Libkin.
\newblock {Locally Consistent Transformations and Query Answering in Data
  Exchange}.
\newblock In {\em 23rd ACM Symposium on Principles of Database Systems (PODS)},
  pages 229--240, 2004.

\bibitem{AFN10}
M.~Arenas, R.~Fagin, and A.~Nash.
\newblock {Composition with Target Constraints}.
\newblock In {\em 13th International Conference on Database Theory (ICDT)},
  pages 129--142, 2010.

\bibitem{APR08}
M.~Arenas, J.~P\'{e}rez, and C.~Riveros.
\newblock {The Recovery of a Schema Mapping: Bringing Back Exchanged Data}.
\newblock In {\em 27th ACM Symposium on Principles of Database Systems (PODS)},
  pages 13--22, 2008.

\bibitem{Ba09}
P.~Barcel\'{o}.
\newblock {Logical Foundations of Relational Data Exchange}.
\newblock {\em SIGMOD Record}, 38(1):49--58, 2009.

\bibitem{Bernstein03}
P.~A. Bernstein.
\newblock {Applying Model Management to Classical Meta-Data Problems}.
\newblock In {\em Conference on Innovative Data Systems Research (CIDR)}, pages
  209--220, 2003.

\bibitem{CFP84}
M.~Casanova, R.~Fagin, and C.~Papadimitriou.
\newblock {Inclusion Dependencies and their Interaction with Functional
  Dependencies}.
\newblock {\em J. Computer and System Sciences}, 20(1):29--59, 1984.

\bibitem{CDO09}
B.~Cautis, A.~Deutsch, and N.~Onose.
\newblock {Querying Data Sources that Export Infinite Sets of Views}.
\newblock In {\em International Conference on Database Theory (ICDT)}, pages
  84--97, 2009.

\bibitem{CG09}
R.~Chirkova and M.~Genesereth.
\newblock {Equivalence of {SQL} Queries in Presence of Embedded Dependencies}.
\newblock In {\em 28th ACM Symposium on Principles of Database Systems (PODS)},
  pages 217--226, 2009.

\bibitem{CK86}
S.~S. Cosmadakis and P.~C. Kanellakis.
\newblock {Functional and Inclusion Dependencies: A Graph Theoretic Approach}.
\newblock In {\em {Advances in Computing Research}}, volume~3, pages 163--184.
  1986.

\bibitem{DNR08}
A.~Deutsch, A.~Nash, and J.~Remmel.
\newblock {The Chase Revisited}.
\newblock In {\em 27th ACM Symposium on Principles of Database Systems (PODS)},
  pages 149--158, 2008.

\bibitem{DPT06}
A.~Deutsch, L.~Popa, and V.~Tannen.
\newblock {Query Reformulation with Constraints}.
\newblock {\em SIGMOD Record}, 35(1):65--73, 2006.

\bibitem{DT03}
A.~Deutsch and V.~Tannen.
\newblock {Reformulation of XML Queries and Constraints}.
\newblock In {\em International Conference on Database Theory (ICDT)}, pages
  225--241, 2003.

\bibitem{DGL00}
O.~M. Duschka, M.~R. Genesereth, and A.~Y. Levy.
\newblock {Recursive Query Plans for Data Integration}.
\newblock {\em J. Log. Program.}, 43(1):49--73, 2000.

\bibitem{En01}
H.~B. Enderton.
\newblock {\em {A Mathematical Introduction to Logic: Second Edition}}.
\newblock Academic Press, 2001.

\bibitem{Fa07}
R.~Fagin.
\newblock {{Inverting Schema Mappings}}.
\newblock {\em ACM Trans. Database Syst.}, 32(4), 2007.

\bibitem{FHHMPV09}
R.~Fagin, L.~Haas, M.~Hernandez, R.~Miller, L.~Popa, and Y.~Velegrakis.
\newblock {Clio: Schema Mapping Creation and Data Exchange}.
\newblock In A.~Borgida, V.~Chaudhri, P.~Giorgini, and E.~Yu, editors, {\em
  Conceptual Modeling: Foundations and Applications, Essays in Honor of {John
  Mylopoulos}}, volume 5600 of {\em Lecture Notes in Computer Science}, pages
  198--236. Springer-Verlag, 2009.

\bibitem{FKNP08}
R.~Fagin, P.~Kolaitis, A.~Nash, and L.~Popa.
\newblock {Towards a Theory of Schema-Mapping Optimization}.
\newblock In {\em 27th ACM Symposium on Principles of Database Systems (PODS)},
  pages 33--42, 2008.

\bibitem{FKMP05}
R.~Fagin, P.~G. Kolaitis, R.~J. Miller, and L.~Popa.
\newblock {Data Exchange: Semantics and Query Answering}.
\newblock {\em Theoretical Computer Science}, 336:89--124, 2005.
\newblock Preliminary version in {\em International Conference on Database
  Theory (ICDT)}, pages 207--224, 2003.

\bibitem{FKPT05}
R.~Fagin, P.~G. Kolaitis, L.~Popa, and W.~C. Tan.
\newblock {Composing Schema Mappings: Second-order Dependencies to the Rescue}.
\newblock {\em ACM Trans. Database Syst.}, 30(4):994--1055, 2005.

\bibitem{FSV95}
R.~Fagin, L.~Stockmeyer, and M.~Vardi.
\newblock {On Monadic NP vs Monadic coNP}.
\newblock {\em Information and Computation}, 120(1):78--92, 1995.

\bibitem{FHHMPP06}
A.~Fuxman, M.~A. Hern{\'a}ndez, C.~T.~H. Ho, R.~J. Miller, P.~Papotti, and
  L.~Popa.
\newblock {Nested Mappings: Schema Mapping Reloaded}.
\newblock In {\em Very Large Data Bases (VLDB)}, pages 67--78, 2006.

\bibitem{FKMT06}
A.~Fuxman, P.~Kolaitis, R.~Miller, and W.-C. Tan.
\newblock {Peer Data Exchange}.
\newblock {\em ACM Transactions on Database Systems}, 31(4):1454--1498, 2006.

\bibitem{Gai82}
H.~Gaifman.
\newblock {On Local and Non-local Properties}.
\newblock In {\em Herbrand Symposium Logic Colloquium, North Holland}, pages
  105--135, 1982.

\bibitem{GLLR07}
G.~D. Giacomo, D.~Lembo, M.~Lenzerini, and R.~Rosati.
\newblock {On Reconciling Data Exchange, Data Integration, and Peer Data
  Management}.
\newblock In {\em 26th ACM Symposium on Principles of Database Systems (PODS)},
  pages 133--142, 2007.

\bibitem{GN08}
G.~Gottlob and A.~Nash.
\newblock {Efficient Core Computation in Data Exchange}.
\newblock {\em Journal of the ACM}, 55(2), 2008.

\bibitem{GS08}
G.~Gottlob and S.~Szeider.
\newblock {Fixed-Parameter Algorithms For Artificial Intelligence, Constraint
  Satisfaction and Database Problems}.
\newblock {\em Computer Journal}, 51(3):303--325, 2008.

\bibitem{GKIT07}
T.~Green, G.~Karvounarakis, Z.~Ives, and V.~Tannen.
\newblock {Update Exchange with Mappings and Provenance}.
\newblock In {\em International Conference on Very Large Data Bases (VLDB)},
  pages 675--686, 2007.

\bibitem{Han65}
W.~P. Hanf.
\newblock {Model-theoretic Methods in the Study of Elementary Logic}.
\newblock In {\em The Theory of Models; Addison, Henkin, and Tarski, eds.,
  North Holland}, pages 132--145, 1965.

\bibitem{HS07}
A.~Hernich and N.~Schweikardt.
\newblock {{CWA}-solutions for Data Exchange Settings with Target
  Dependencies}.
\newblock In {\em 26th ACM Symposium on Principles of Database Systems (PODS)},
  pages 113--122, 2007.

\bibitem{KT08}
G.~Karvounaraki and V.~Tannen.
\newblock {Conjunctive Queries and Mappings with Unequalities}.
\newblock Technical Report MS-CIS-08-37, University of Pennsylvania, 2008.

\bibitem{Ko05}
P.~G. Kolaitis.
\newblock {Schema Mappings, Data Exchange, and Metadata Management}.
\newblock In {\em 24th ACM Symposium on Principles of Database Systems (PODS)},
  pages 61--75, 2005.

\bibitem{Len02}
M.~Lenzerini.
\newblock {Data Integration: A Theoretical Perspective}.
\newblock In {\em 21st ACM Symposium on Principles of Database Systems (PODS)},
  pages 233--246, 2002.

\bibitem{fmt-book}
L.~Libkin.
\newblock {\em {Elements of Finite Model Theory}}.
\newblock Springer-Verlag, 1st edition, 2004.

\bibitem{LS08}
L.~Libkin and C.~Sirangelo.
\newblock {Data Exchange and Schema Mappings in Open and Closed Worlds}.
\newblock In {\em 27th ACM Symposium on Principles of Database Systems (PODS)},
  pages 139--148, 2008.

\bibitem{MH03}
J.~Madhavan and A.~Y. Halevy.
\newblock {Composing Mappings Among Data Sources}.
\newblock In {\em International Conference on Very Large Data Bases (VLDB)},
  pages 572--583, 2003.

\bibitem{Me08}
M.~Meier.
\newblock {Towards Rule-Based Minimization of {RDF} Graphs under Constraints}.
\newblock In {\em Web Reasoning and Rule Systems}, volume 5341 of {\em Lecture
  Notes in Computer Science}, pages 89--103. Springer-Verlag, 2008.

\bibitem{Me04}
S.~Melnik.
\newblock {\em {Generic Model Management: Concepts and Algorithms}}, volume
  2967 of {\em Lecture Notes in Computer Science}.
\newblock Springer, 2004.

\bibitem{NBM05}
A.~Nash, P.~A. Bernstein, and S.~Melnik.
\newblock {Composition of Mappings Given by Embedded Dependencies}.
\newblock In {\em 24th ACM Symposium on Principles of Database Systems (PODS)},
  pages 172--183, 2005.

\bibitem{PT09}
P.~Papotti and R.~Torlone.
\newblock {Schema Exchange: Generic Mappings for Transformaing Data and
  Metadata}.
\newblock {\em Data and Knowledge Engineering}, 68(7):665--682, 2009.

\bibitem{tK09}
B.~ten Cate and P.~Kolaitis.
\newblock {Structural Characterizations of Schema-Mapping Languages}.
\newblock In {\em International Conference on Database Theory (ICDT)}, pages
  63--72, 2009.

\bibitem{VSS02}
P.~Vassiliadis, A.~Simitsis, and S.~Skiadopoulos.
\newblock {On the Logical Modeling of ETL Processes}.
\newblock In {\em International Conference on Advanced Information Systems
  Engineering (CAiSE)}, pages 782--786, 2002.

\end{thebibliography}
